\documentclass[a4paper,12pt]{article}
\usepackage{jheppub}

\usepackage[T1]{fontenc}
\usepackage[utf8]{inputenc}
\usepackage{graphicx,cancel,float}
\usepackage{amssymb}
\usepackage{textcomp}
\usepackage{amsmath}
\usepackage{tabularx}
\usepackage{bm}
\usepackage{times}
\usepackage{color}
\usepackage{slashed}
\usepackage{multirow}
\usepackage{verbatim}
\usepackage{epsfig,epstopdf}
\usepackage{subfigure}

\allowdisplaybreaks[4]

\def\lsim{\mathrel{\raise.3ex\hbox{$<$\kern-.75em\lower1ex\hbox{$\sim$}}}}
\def\gsim{\mathrel{\raise.3ex\hbox{$>$\kern-.75em\lower1ex\hbox{$\sim$}}}}

\definecolor{orange}{rgb}{1,0.5,0}

\newcommand{\minigraph}[5][0.25in]{\begin{minipage}{#2}\begin{center}\includegraphics[width=#2]{#5}\\\vspace{#3}\hspace{#1}{\footnotesize #4}\end{center}\end{minipage}}

\preprint{}
\subheader{\it \today}

\title{Searching for heavy neutral lepton and lepton number violation through VBS at high-energy muon colliders}

\author[a]{Tong Li}
\emailAdd{litong@nankai.edu.cn}
\affiliation[a]{
School of Physics, Nankai University, Tianjin 300071, China}

\author[a,b]{Chang-Yuan Yao}
\emailAdd{yaocy@nankai.edu.cn}
\affiliation[b]{
Deutsches Elektronen-Synchrotron DESY, Notkestr. 85, 22607 Hamburg, Germany}

\author[a]{Man Yuan}
\emailAdd{yuanman@mail.nankai.edu.cn}

\preprint{DESY-23-092}
\abstract{
High-energy muon collider can play as an emitter of electroweak gauge bosons and thus leads to substantial vector boson scattering (VBS) processes. In this work, we investigate the production of heavy neutral lepton (HNL) $N$ and lepton number violation (LNV) signature through VBS at high-energy muon colliders. VBS induces LNV processes $W^\pm Z/\gamma\to \ell^\pm N \to \ell^\pm \ell^\pm W^\mp\to \ell^\pm \ell^\pm q\bar{q}'$ with an on-shell HNL $N$ at $\mu^+\mu^-$ colliders. In analogy to neutrinoless double-beta decay with the HNL in t-channel, the LNV signature $W^+W^+\to \ell^+\ell^+$ can also happen via VBS at same-sign muon collider. They provide clean and robust LNV signatures to tell the nature of Majorana HNLs and thus have more advantageous benefits than direct $\mu\mu$ annihilation. We analyze the potential of searching for Majorana HNL and obtain the exclusion limits on mixing $V_{\ell N}$. Based on this same-sign lepton signature, we also obtain the sensitivity of muon collider to the Weinberg operator.
}

\begin{document}

\maketitle
\setcounter{page}{2}

\newpage

\section{Introduction}
\label{sec:Intro}

Neutrino oscillation experiments provide clear and compelling evidence that neutrinos have non-zero, but very small masses. The Standard Model (SM) includes only left-handed neutrino fields in the lepton doublets, and therefore predicts neutrino masses of exactly zero. Although one can certainly introduce right-handed (RH) neutrino field, the Yukawa coupling has to be tuned to a very small constant $y_\nu\lesssim 10^{-13}$ to accommodate the observed neutrino mass. A more economical way to generate neutrino mass in the SM content is through the so-called ``Weinberg operator''~\cite{Weinberg:1979sa}
\begin{eqnarray}
\ell_L\ell_L HH\;,
\end{eqnarray}
where $\ell_L$ and $H$ stand for the SM left-handed lepton doublet and the Higgs doublet, respectively. The price we pay here is to introduce a high-dimensional (dimension-5) operator and the violation of global lepton number symmetry. The minimal ultraviolet (UV) realization of this dimension-5 operator is the Type I Seesaw mechanism~\cite{Minkowski:1977sc,Yanagida:1979as,GellMann:1980vs,Glashow:1979nm,Mohapatra:1979ia,Shrock:1980ct,Schechter:1980gr}. In the minimal Type I Seesaw, the ``sterile neutrinos'' have the nature of Majorana fermions as they transform as singlet under the SM gauge group. The RH neutrinos possess a Majorana mass term $M_R$ and interact with SM leptons through a Yukawa coupling $y_\nu$. The light neutrino masses are given by $m_\nu\sim y_\nu^2 v^2/M_R$ with $v$ being the vacuum expectation value (vev) of SM Higgs. The heavy Majorana neutrinos can also arise in the Type III Seesaw mechanism with triplet fermionic multiplet~\cite{Foot:1988aq}. The eigenstates of RH neutrinos are usually referred as heavy neutral leptons (HNLs) which will be denoted as $N$ below. After mass mixing, one obtains an important mixing matrix $V_{\ell N}$ transiting heavy neutrinos to charged leptons in the mixed mass-flavor basis~\cite{Atre:2009rg}.

If $y_\nu$ is close to the electron Yukawa coupling $y_e\sim 10^{-6}$ in Type I Seesaw, the heavy Majorana neutrino receives a mass at TeV level and can be searched through lepton number violation (LNV) signature at high-energy colliders. This low-scale neutral lepton and the mixing matrix $V_{\ell N}$ have been searched at colliders. See Ref.~\cite{Cai:2017mow} and references therein for a review of LNV searches. The mass of HNL has been excluded up to about 1 TeV depending on the value of $|V_{\ell N}|^2\lesssim 0.1$~\cite{CMS:2018jxx,CMS:2018iaf,CMS:2022rqc,ATLAS:2023tkz}. The future FCC-hh can push the exclusion limit up to a few TeV~\cite{Antusch:2016ejd,Pascoli:2018heg,Liu:2023gpt}. One needs a machine with higher energy and integrated
luminosity to further explore the measurement of $V_{\ell N}$ for the high mass region of HNL $N$. High-energy muon colliders have recently received much attention in the community~\cite{MICE:2019jkl,Delahaye:2019omf, Bartosik:2020xwr,Han:2020uid,Yin:2020afe,AlAli:2021let,Bose:2022obr,Maltoni:2022bqs,Narain:2022qud,Li:2022kkc,Li:2023ksw,Chowdhury:2023imd}. The high-energy options
of muon collider can have multi-TeV center-of-mass (c.m.) energies and high integrated luminosity
scaling with energy quadratically~\cite{Delahaye:2019omf}
\begin{equation}
\label{eq:lumi}
\mathcal{L}=\left(\frac{\sqrt{s}}{10\ {\rm TeV}}\right)^2 \times 10~\textrm{ab}^{-1}\;.
\end{equation}
It thus provides an excellent opportunity to produce and discover HNLs in Type I and III Seesaw mechanisms.

Recently, there arose quite a few studies of searching for HNL at $\mu^+\mu^-$ colliders~\cite{Mekala:2023diu,Kwok:2023dck,Li:2023tbx}.
They proposed that an HNL $N_\ell$ can be produced together with a light ``active'' neutrino $\nu_\ell$ from $\mu^+\mu^-$ collision with c.m. energy $\sqrt{s}=3$ or 10 TeV. The HNL then decays into a charged lepton $\ell^\pm$ and a $W$ gauge boson with hadronic decay. The production mode then becomes
\begin{eqnarray}
\mu^+\mu^-\to N_\ell \bar{\nu}_\ell\to \ell^\pm \bar{\nu}_\ell q\bar{q}'\;.
\label{eq:Nnu}
\end{eqnarray}
Despite of large production cross section of $\sigma(\mu^+\mu^-\to N_\mu \bar{\nu}_\mu)\sim\mathcal{O}(10)$ pb through a $W$ exchange in t-channel~\cite{Li:2023tbx}, this signal channel cannot tell whether the HNL is Majorana fermion because of the missing neutrino in final states. One has to define a forward-backward symmetry in $N_\ell$'s decay pattern~\cite{Li:2023tbx} or other kinematic quantity~\cite{Kwok:2023dck} to distinguish Majorana and Dirac neutrinos.
An alternative approach to search for Majorana neutrino is to consider the inverse $0\nu\beta\beta$-like channel $\mu^+\mu^+\to W^+ W^+$~\cite{Yang:2023ojm,Jiang:2023mte} which however relies on a same-sign muon collider.

In this work, we propose a clear way to search for heavy Majorana neutrino through LNV signature at $\mu^+\mu^-$ colliders.
At high-energy muon colliders, the initial muon beams substantially emit electroweak (EW) gauge bosons under an approximately unbroken SM gauge symmetry. The gauge bosons
are associated with very energetic muons or muon-neutrinos in the forward region with respect to the beam. The EW gauge bosons behave like initial state partons and lead to vector boson scattering (VBS) processes~\cite{Costantini:2020stv,AlAli:2021let,Han:2021udl,BuarqueFranzosi:2021wrv,Ruiz:2021tdt}.
The VBS becomes an increasingly important mode as colliding energies go higher, due to the logarithmic enhancement from
gauge boson radiation. Thus, a LNV signal can be produced through VBS
\begin{eqnarray}
W^\pm Z/\gamma\to \ell^\pm N \to \ell^\pm \ell^\pm W^\mp\to \ell^\pm \ell^\pm q\bar{q}'\;.
\end{eqnarray}
This signal is composed of same-sign charged leptons in final states and can thus clearly tell the Majorana nature of heavy neutrinos. It is exactly the smoking-gun signature searching for Majorana neutrino via Drell-Yan production $pp\to W^\ast\to \ell^\pm N$~\cite{Han:2006ip,Atre:2009rg} or $W\gamma$ VBS~\cite{Dev:2013wba,Alva:2014gxa} at LHC. It can serve as a robust channel to identify LNV and Majorana neutrino suppose a discovery of HNL through the channel in Eq.~(\ref{eq:Nnu}). We will utilize the leading-order framework of electroweak parton distribution functions (EW PDFs)~\cite{Han:2021udl,Ruiz:2021tdt} to calculate the HNL production through VBS. The detector simulation of LNV signal and SM backgrounds will be performed to analyze the search potentials on heavy Majorana neutrino and obtain the exclusion limit on mixing $V_{\ell N}$.

In analogy to neutrinoless double-beta ($0\nu\beta\beta$) decay, the VBS process can also induce LNV signature with the HNL in t-channel~\cite{Dicus:1991fk,Han:2006ip,Atre:2009rg,Fuks:2020att,Fuks:2020zbm,CMS:2022rqc,Schubert:2022lcp}.
This VBS process happens at same-sign muon collider (e.g., $\mu^+ \mu^+$ collider) and leads to the signal of same-sign charged leptons
\begin{eqnarray}
W^+ W^+ \to \ell^+ \ell^+\;.
\end{eqnarray}
This process provides a clean LNV signal compared with the inverse process $\mu^+\mu^+\to W^+ W^+$ as there are no missing neutrinos and the suppression from two $W$ bosons' decay branching fractions. We also analyze this LNV channel and explore the exclusion limit on mixing $V_{\ell N}$. It can help to probe all charged lepton flavors and corresponding mixing elements rather than only $V_{\mu N}$ through $\mu^+\mu^+\to W^+ W^+$.
In the absence of HNL, it was shown that this LNV process can also help to probe the Weinberg operator~\cite{Fuks:2020zbm}. We also evaluate the sensitivity of muon collider to the scale of Weinberg operator.

This paper is organized as follows. In Sec.~\ref{sec:Seesaw}, we first outline the HNLs in canonical Type I and Type III Seesaw models.
Then we simulate the production of HNL and LNV signature via VBF processes at high-energy $\mu^+\mu^-$ colliders in Sec.~\ref{sec:HNL}.
The results of projected sensitivity bounds for $V_{\ell N}$ are also given. In Sec.~\ref{sec:HNLt}, we also analyze the LNV signature with the HNL in t-channel through VBS at same-sign muon collider.
Finally, in Sec.~\ref{sec:Con} we summarize our conclusions.

\section{Heavy neutral leptons in Seesaw mechanisms}
\label{sec:Seesaw}

The HNLs can be realized in canonical Type I and Type III Seesaw mechanisms.
The neutrino Yukawa interactions in Type I Seesaw are
\begin{eqnarray}
-\mathcal{L}_Y^I=Y^D_\nu\bar{\ell}_L\tilde{H}N_R+{\frac{1}{2}}\overline{(N^c)_L} M_R N_R +\text{h.c.},
\end{eqnarray}
where $\tilde{H}=i\sigma_2 H^\ast$, $N_R$ denotes RH neutrino, and $M_R$ is the RH neutrino mass matrix.
Once $H$ receives the SM vev $\langle H\rangle=v/\sqrt{2}$, the Dirac mass term becomes $m_D=Y^D_\nu v/\sqrt{2}$ and the diagonalization of the neutrino mass matrix is
\begin{eqnarray}
\mathbb{N}^\dagger
\left(
  \begin{array}{cc}
    0 & m_D \\
    m_{D}^{T} & M_R \\
  \end{array}
\right) \mathbb{N}^\ast &=&
\left(
  \begin{array}{cc}
    m_\nu^{\rm diag} & 0 \\
    0 & M_N^{\rm diag}  \\
  \end{array}
\right),
\label{eq:type1NuMixMatrix}
\end{eqnarray}
with the transformation of mass eigenstates as
\begin{eqnarray}
\left(
  \begin{array}{c}
    \nu_L \\
    (N^c)_L \\
  \end{array}
\right) = \mathbb{N}\left(
  \begin{array}{c}
    \nu_L \\
    (N^c)_L \\
  \end{array}
\right)_{mass}, \ \ \ \mathbb{N}= \left(
  \begin{array}{cc}
    U & V \\
    X & Y \\
  \end{array}
\right)\;.
\label{eq:nuMixDefs}
\end{eqnarray}
The mass eigensvalues of physical neutrinos are $m_\nu^{\rm diag}={\rm diag}(m_1,m_2,m_3)$ and $M_N^{\rm diag}={\rm diag}(M_1,\cdots)$.
With another matrix $E$ diagonalizing the charged lepton mass matrix, we have
\begin{eqnarray}
E^\dagger U\equiv U_{\rm PMNS}\;,~~~E^\dagger V\equiv V_{\ell N}\;,
\end{eqnarray}
where $U_{\rm PMNS}$ is the approximate Pontecorvo-Maki-Nakagawa-Sakata (PMNS) neutrino mass mixing matrix. In particular, the matrix $V_{\ell N}$ transits heavy neutrinos to charged leptons~\cite{Atre:2009rg}.
The active neutrino states are decomposed into a general number of massive eigenstates
\begin{eqnarray}
\nu_{\ell} = \sum_{m=1}^3 (U_{\rm PMNS})_{\ell m}\nu_{m} + \sum_{m'=1} (V_{\ell N})_{\ell m'} N_{m'}^c \;,
\end{eqnarray}
where we assume that the HNL mass eigenstates are denoted as $N_{m'}$.

From Eq.~(\ref{eq:type1NuMixMatrix}), one can derive a relationship between diagonalized neutrino mass matrices and mixing matrices
\begin{eqnarray}
U^\ast_{\rm PMNS}m_\nu^{\rm diag} U^\dagger_{\rm PMNS}+V^\ast_{\ell
N}M_N^{\rm diag}V^\dagger_{\ell N}=0 \ .
\label{typei}
\end{eqnarray}
In the mixed mass-flavor basis, the matrix $V_{\ell N}$ determines the interactions between HNL and SM EW gauge bosons such as
\begin{eqnarray}
\mathcal{L}_{\rm Type-I}&\supset& -{\frac{g}{\sqrt{2}}}W_\mu^- \sum_{\ell=e}^{\tau} \Big(\sum_{m=1}^3\bar{\ell} (U_{\rm PMNS})_{\ell m} \gamma^\mu P_L \nu_m + \sum_{m'=1}\bar{\ell} (V_{\ell N})_{\ell m'} \gamma^\mu P_L N^c_{m'} \Big) +\text{h.c.} \nonumber \\
&&-\frac{g}{2\cos\theta_W} Z_\mu \sum_{\ell=e}^{\tau} \Big(\sum_{m=1}^3\bar{\nu}_\ell (U_{\rm PMNS})_{\ell m} \gamma^\mu P_L \nu_{m}+\sum_{m'=1}\bar{\nu}_\ell (V_{\ell N})_{\ell m'} \gamma^\mu P_L N^c_{m'} \Big) + \text{h.c.}\;.\nonumber \\
\end{eqnarray}
The partial width of one HNL $N$ decay into charged lepton becomes
\begin{eqnarray}
\Gamma(N\to \ell^\pm W^\mp)&\equiv& \Gamma(N\to \ell^+ W^-) = \Gamma(N\to \ell^- W^+) \nonumber \\
&=& \frac{G_F}{8\sqrt{2}\pi}|V_{\ell N}|^2 m_{N}(m_{N}^2+2m_W^2)\left(1-\frac{m_W^2}{m_{N}^2}\right)^2,
\label{partialN}
\end{eqnarray}
and an asymptotic behavior holds when $m_{N}\gg m_W, m_Z, m_h$.
\begin{eqnarray}
\Gamma(N\to \sum_\ell \ell^\pm W^\mp)\approx \Gamma(N\to \sum_\nu \nu Z+\bar{\nu}Z)\approx \Gamma(N\to \sum_\nu \nu h+\bar{\nu}h)\;,
\end{eqnarray}
where $\ell=e,\mu,\tau$.

The Type III Seesaw introduces SU$(2)_L$ triplet leptons $\Sigma_L$ with zero hypercharge in addition to the SM fields~\cite{Foot:1988aq}
\begin{eqnarray}
&&\Sigma_L = \left(
  \begin{array}{cc}
    \Sigma_L^0/\sqrt{2} & \Sigma^+_L \\
    \Sigma_L^- & -\Sigma_L^0/\sqrt{2} \\
  \end{array}
\right) \ .
\end{eqnarray}
The Type III Seesaw Lagrangian is given by
\begin{equation}
 \mathcal{L}_{\rm Type-III} =
 {\rm Tr}\left[\overline{\Sigma_L}i\cancel{D}\Sigma_L\right]
-\left(\frac{1}{2}{\rm Tr}\left[\overline{\Sigma_L^c}M_\Sigma \Sigma_L\right]+Y_\Sigma \overline{\ell_L}\tilde{H}\Sigma_L^c+\text{h.c.}\right)\;.
\label{eq:typeIIILag}
\end{equation}
The Yukawa term in Eq.~(\ref{eq:typeIIILag}) induces both the neutrino mass mixing and a mass mixing between the charged SM leptons and the charged triplet leptons
\begin{eqnarray}
\left(
  \begin{array}{cc}
    \overline{\nu_L^c} & \overline{\Sigma_L^{0c}} \\
  \end{array}
\right) \left(
  \begin{array}{cc}
    0 & Y_\Sigma^Tv_0/2\sqrt{2} \\
    Y_\Sigma v_0/2\sqrt{2} & M_\Sigma/2 \\
  \end{array}
\right) \left(
  \begin{array}{c}
    \nu_L \\
    \Sigma_L^0 \\
  \end{array}
\right)
+
\left(
  \begin{array}{cc}
    \overline{\ell_R} & \overline{\Sigma_L^{+c}} \\
  \end{array}
\right) \left(
  \begin{array}{cc}
    m_{\ell} & 0 \\
    Y_\Sigma v_0 & M_\Sigma \\
  \end{array}
\right) \left(
  \begin{array}{c}
    \ell_L \\
    \Sigma_L^- \\
  \end{array}
\right)+\text{h.c.}\;.\nonumber \\
\end{eqnarray}
After diagonalizing mass matrices by introducing unitary matrices, one obtains mass eigenvalues for neutrinos and charged leptons.
The mixing between the SM charged leptons and triplet leptons follows the same relation in Eq.~(\ref{typei}).
The interactions between HNL and SM EW gauge bosons are
also the same as those in Type I Seesaw.

In this section, we outline the theoretical motivations of the mixing matrix $V_{\ell N}$ between heavy neutrinos and charged leptons, i.e., Type I and Type III Seesaw models. In fact, the mixing parameter would scale like $V_{\ell N}^2\sim m_\nu/m_N$ and should be very small. We must state that we will assume the mixing angle $V_{\ell N}$ and the HNL mass $m_N$ as independent parameters in the following collider studies.

\section{Heavy neutral lepton and lepton number violation through VBS at $\mu^+\mu^-$ collider}
\label{sec:HNL}

In this section, we investigate the search for Majorana HNL through VBS processes at future high-energy muon colliders. We simulate the LNV signatures
from the decay of on-shell Majorana neutral lepton and obtain the sensitivity to the mixing parameters between HNL and SM charged leptons.

\subsection{The VBS production of HNL and LNV signal}

\begin{figure}[h!]
\begin{center}
\minigraph{6cm}{-0.05in}{(a)}{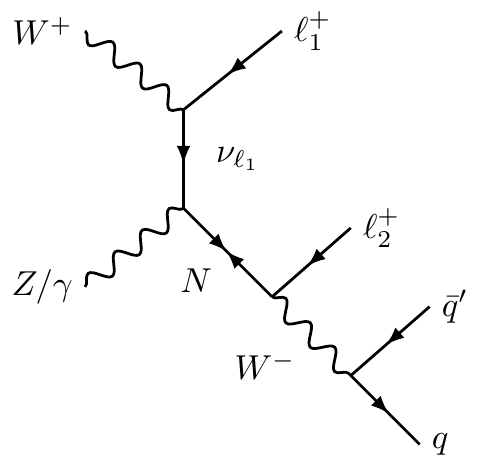}
\hspace{1cm}
\minigraph{6cm}{-0.05in}{(b)}{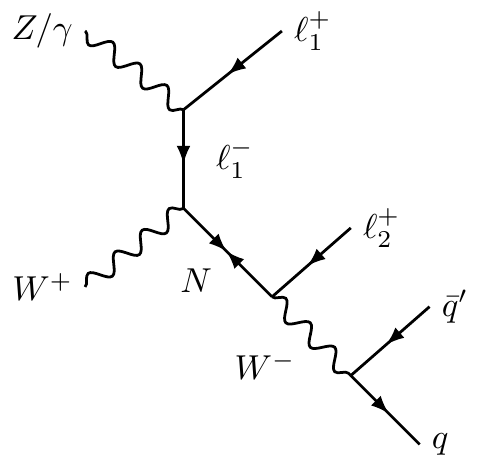}\\
\vspace{1cm}
\minigraph{9cm}{-0.05in}{(c)}{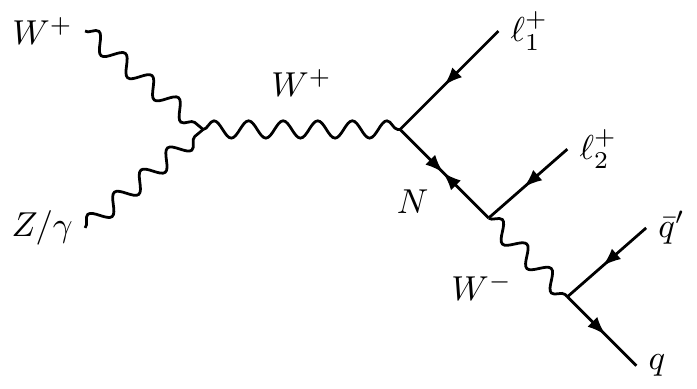}
\end{center}
\caption{The Feynman diagrams of processes $V_i~V_j\to \ell^+_1 N  \to \ell^+_1 \ell^+_2 q~\bar{q}'$ at the muon collider.
}
\label{fig-FD1}
\end{figure}

We propose to produce an HNL accompanied by a SM charged lepton $\ell_1^{\pm} N$. Due to charge conservation, this combination can only be produced through VBS processes at muon colliders. After considering the decay of HNL $N$, a LNV signature with $\Delta L =2$ can emerge
\begin{equation}
V_i~V_j \to \ell_1^{\pm}N \to \ell_1^{\pm} \ell_2^{\pm}q~\bar{q}' \;,
\label{eqn-VV1}
\end{equation}
where $\ell_1,\ell_2=e,\mu,\tau$ and $V_i,V_j$ denote SM EW gauge bosons. The decay channel of Majorana HNL is considered to be $N \to \ell_2^{\pm} W^{\mp} $ and the $W$ boson has subsequent decay into $q~\bar{q}'$. The Feynman diagrams of above LNV production are shown in Fig.~\ref{fig-FD1}. The panels (a) and (b) illustrate t-channel processes via exchanging a SM lepton and the panel (c) represents an s-channel process mediated by a virtual $W$ boson.

\begin{figure}[h!]
\begin{center}
\minigraph{7.2cm}{-0.05in}{}{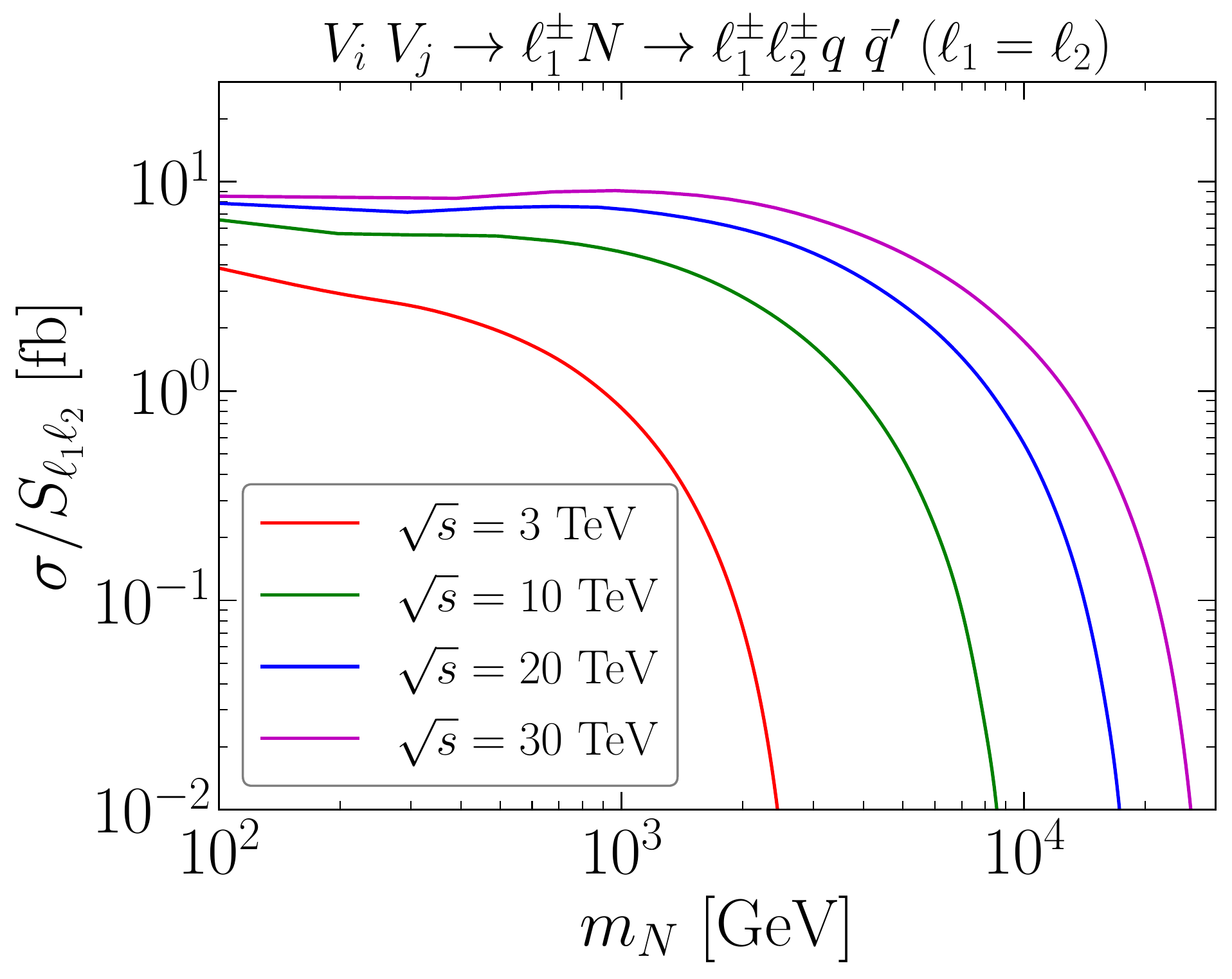}
\minigraph{7.2cm}{-0.05in}{}{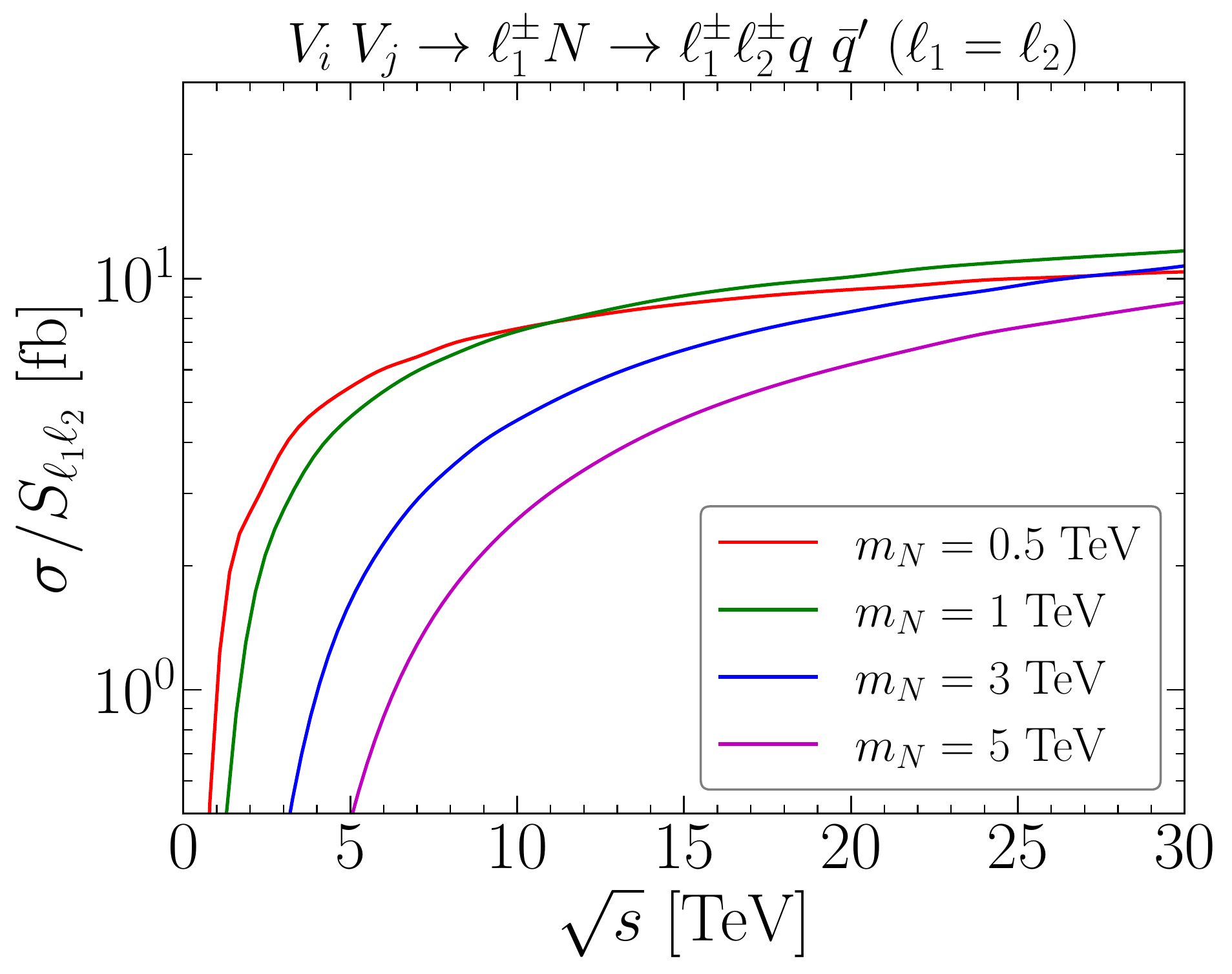}
\end{center}
\caption{The parameter-independent cross section $\sigma_0=\sigma/S_{\ell_1\ell_2}$ for the VBS process $V_i~V_j \to \ell_1^{\pm} N \to \ell_1^{\pm} \ell_2^{\pm} q~\bar{q}'$ with $\ell_1=\ell_2$, as a function of the HNL mass $m_N$ (left) or c.m. energy $\sqrt{s}$ (right) at muon colliders. The benchmark c.m. energy of the muon collider are $\sqrt{s}=3,~10,~20$ and 30 TeV, and the benchmark HNL masses are $m_N=0.5,~1,~3$ and 5 TeV.
}
\label{fig-xsc1}
\end{figure}

For the initial gauge boson partons $V_i$ and $V_j$, the VBS production cross section can be factorized as the product of the parton luminosity $d\mathcal{L}_{ij}/d\tau$ and the partonic cross section
\begin{eqnarray}
\sigma(\ell^+\ell^-\to F+X)&=&\int^1_{\tau_0}d\tau \sum_{ij}\frac{d\mathcal{L}_{ij}}{d\tau} \sigma(V_i V_j\to F)\;,\nonumber \\
\frac{d\mathcal{L}_{ij}}{d\tau}&=&\frac{1}{1+\delta_{ij}}\int^1_\tau \frac{d\xi}{\xi}\Big[f_i(\xi,Q^2)f_j(\frac{\tau}{\xi},Q^2)+(i\leftrightarrow j)\Big]\;,
\end{eqnarray}
where $F (X)$ denotes an exclusive final state (the underlying remnants), $f_{i(j)}(\xi,Q^2)$ is the EW PDF for vector $V_{i(j)}$ with $Q=\sqrt{\hat{s}}/2$ being the factorization scale, $\tau_0=m_F^2/s$ and $\tau=\hat{s}/s$.
By summing over all gauge boson initial states, one can obtain the total cross section of the ``inclusive'' production processes.
The scattering amplitude of the LNV process in Eq.~\eqref{eqn-VV1} is proportional to $|V_{\ell_1N}V_{\ell_2N}|$ and the corresponding cross sections can be expressed as
\begin{align}
\sigma ~(V_i~V_j \to \ell_1^{\pm}N \to \ell_1^{\pm} \ell_2^{\pm} q\bar{q}' )
& \approx \sigma ~(V_i~V_j \to \ell_1^{\pm}N ) \times {\rm BR} (N \to \ell_2^{\pm} q\bar{q}' )\times (2-\delta_{\ell_1\ell_2})  \nonumber \\
& \equiv  \frac{|V_{\ell_1 N}V_{\ell_2 N}|^2}{\sum_{\ell = e, \mu , \tau  } |V_{\ell N}|^2} \times \sigma_0 \times (2-\delta_{\ell_1\ell_2}) \;,
\label{eq:VBSxsec}
\end{align}
where $\sigma_0$ represents the part of the cross section that is independent of the mixing parameters.
We then define a new parameter which includes all the information of the mixing parameters~\cite{Atre:2009rg}
\begin{equation}
S_{\ell_1\ell_2} \equiv \frac{|V_{\ell_1 N}V_{\ell_2 N}|^2}{\sum_{\ell = e, \mu , \tau  } |V_{\ell N}|^2} \; .
\end{equation}
The parameter-independent cross section $\sigma_0=\sigma/S_{\ell_1\ell_2}$ for $\ell_1=\ell_2$ are shown in Fig.~\ref{fig-xsc1}. The cross section of $\ell_1\neq \ell_2$ case is two times that shown here because of the exchange $\ell_1\leftrightarrow \ell_2$ in Eq.~(\ref{eq:VBSxsec}).
If including the charge conjugation, another factor of two should be multiplied.
For the following simulation of LNV signatures, we choose three cases of lepton flavor combinations in the final states with $\ell_1,\ell_2=e,\mu$
\begin{equation}
S_{\mu \mu} = \frac{|V_{\mu N}|^4}{\sum_{\ell = e, \mu , \tau  } |V_{\ell N}|^2},~
S_{e e} = \frac{|V_{e N}|^4}{\sum_{\ell = e, \mu , \tau  } |V_{\ell N}|^2},~
S_{e \mu} = \frac{|V_{e N}V_{\mu N}|^2}{\sum_{\ell = e, \mu , \tau  } |V_{\ell N}|^2} \; .
\end{equation}
Next we consider the production of one single HNL denoted as $N$.

\subsection{The search for LNV signal at $\mu^+\mu^-$ collider}
\label{III-B}

We employ the Universal FeynRules Output (UFO)~\cite{Degrande:2011ua} files of the model ``HeavyN''~\cite{Degrande:2016aje,Pascoli:2018heg} and pass the model files to MadGraph5\_aMC@NLO~\cite{Alwall:2014hca} to generate the above LNV signal events at parton-level.
The latest version of MadGraph5 has included the leading order framework of EW PDFs for the computation of the VBS processes at muon colliders~\cite{Ruiz:2021tdt}.
We then use PYTHIA 8~\cite{Sjostrand:2014zea, Bierlich:2022pfr} for the parton shower of parton-level events. The detector card of the muon collider in Delphes 3~\cite{deFavereau:2013fsa} is utilized to include the detector effects.

Our LNV signal is composed of two same-sign charged leptons plus jet(s), as shown in Fig.~\ref{fig-FD1}. We consider the following SM backgrounds
\begin{align}
& {\rm B_1}:~~ V~V \to W^{\pm} W^{\pm} W^{\mp} \to \ell^{\pm} \ell^{\pm}\mathop{\nu_{\ell}}\limits^{(-)}\mathop{\nu_{\ell}}\limits^{(-)} q ~\Bar{q}' \nonumber \;,\\
& {\rm B_2}:~~ V~V \to t ~\overline{t} ~W^{\pm} \to b~W^{+} \overline{b}~W^- W^{\pm} \to
b ~\Bar{b}~ \ell^{\pm} \ell^{\pm}  \mathop{\nu_{\ell}}\limits^{(-)}\mathop{\nu_{\ell}}\limits^{(-)}+X \;,
\label{eqn-bkg1}
\end{align}
where $X$ denotes the decay products of the opposite-sign $W$ boson in ${\rm B_2}$ background, i.e., $X=\ell^\mp \mathop{\nu_{\ell}}\limits^{(-)}$ or $q\bar{q}'$.
For ${\rm B_1}$ background, the intermediate states are three $W$ bosons which are produced through VBS process. The two same-sign $W$ bosons leptonically decay into a pair of same-sign leptons, and the other $W$ decays into dijet. Thus, the final states of ${\rm B_1}$ consist of two same-sign charged leptons, two jets and missing energy.
In general, $b~(\Bar{b})$ quark can be mistakenly identified as a light jet in the detector of the collider. Thus we also consider the ${\rm B_2}$ background in which a pair of $t~\Bar{t}$ quarks associated with a $W$ boson are produced by VBS process. The $t~(\Bar{t})$ quark decays into the $b~(\Bar{b})$ quark and the $W^{\pm}$ boson. The same-sign $W$ bosons again decay into the lepton pairs and the other $W$ with opposite charge may decay into all possible final states. The events with $X=\ell^\mp \mathop{\nu_{\ell}}\limits^{(-)}$ can be efficiently reduced by imposing a soft $p_T$ cut on the opposite-sign lepton. Here we include them for a conservative background estimate.

\begin{figure}[h!]
\begin{center}
      \minigraph{3.6cm}{-0.05in}{}{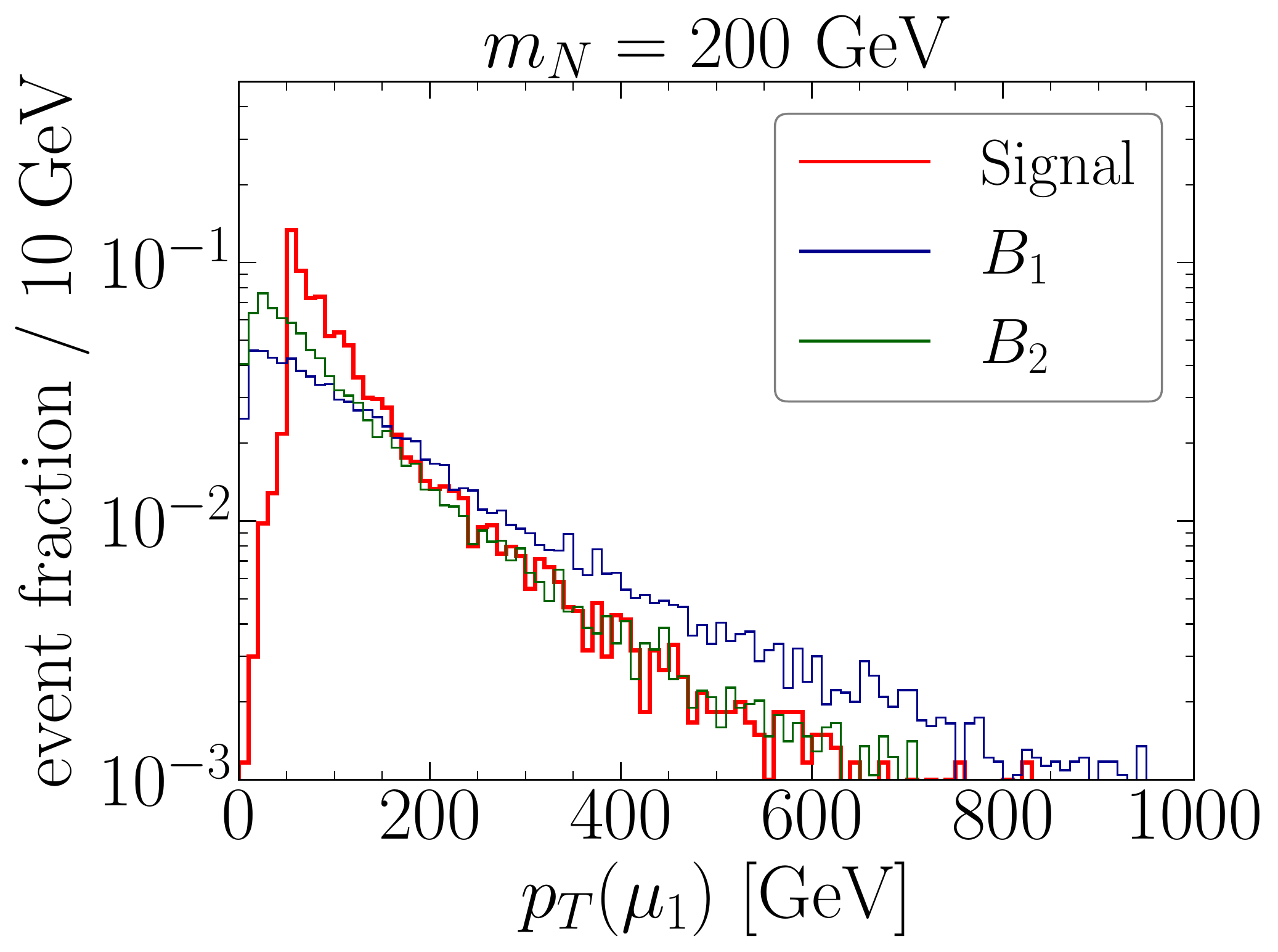}
      \minigraph{3.6cm}{-0.05in}{}{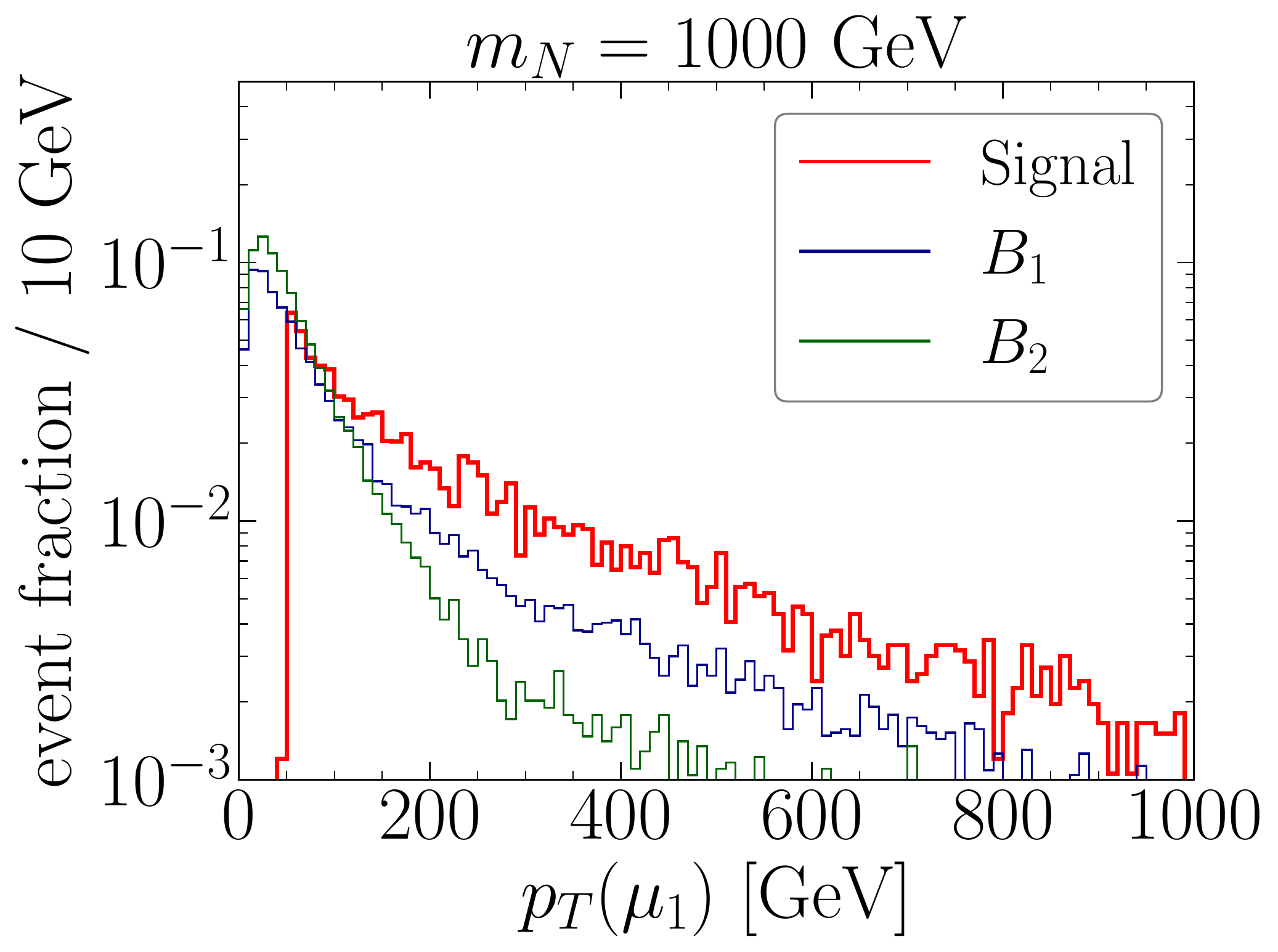}
      \minigraph{3.6cm}{-0.05in}{}{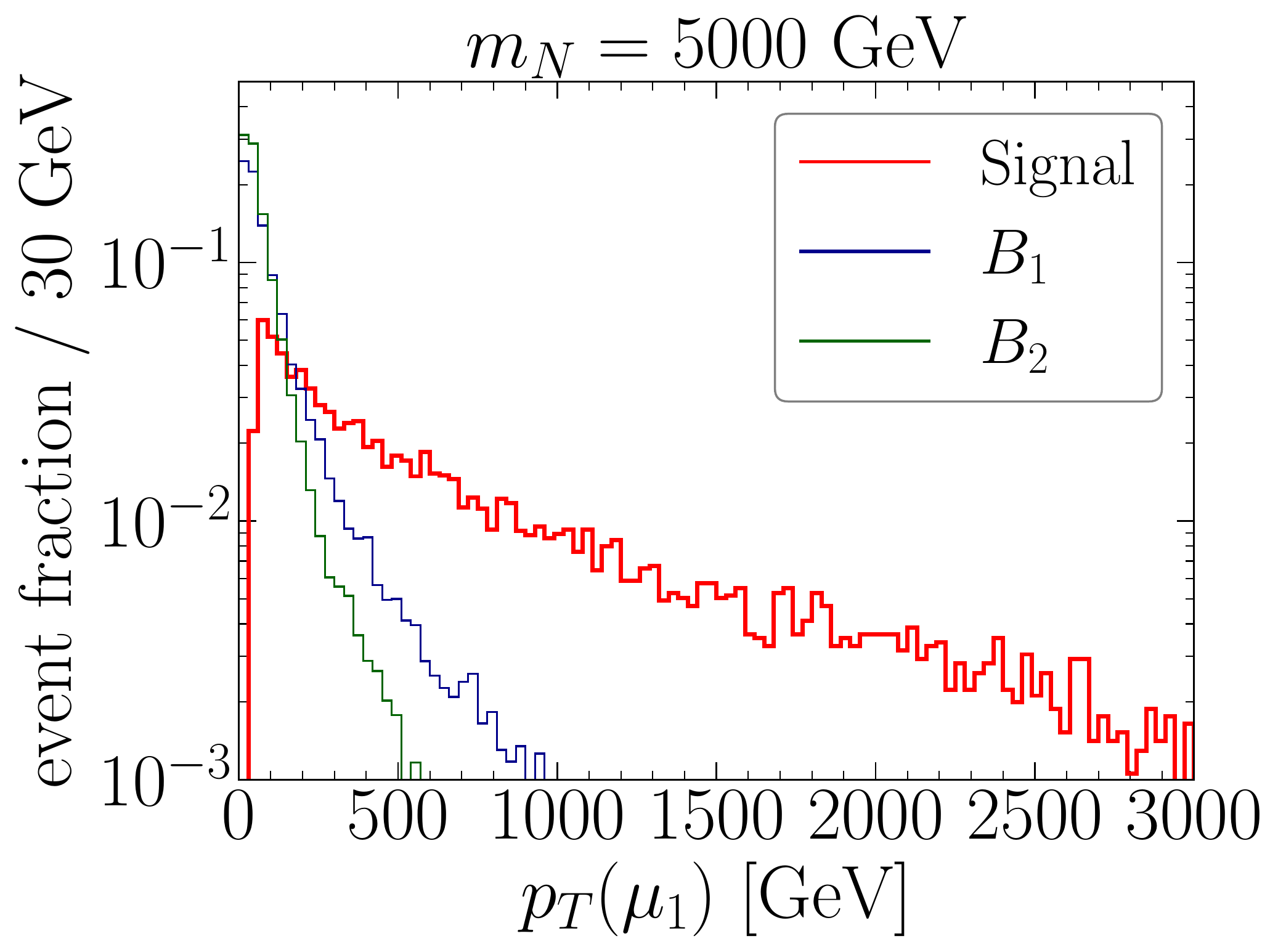}
      \minigraph{3.6cm}{-0.05in}{}{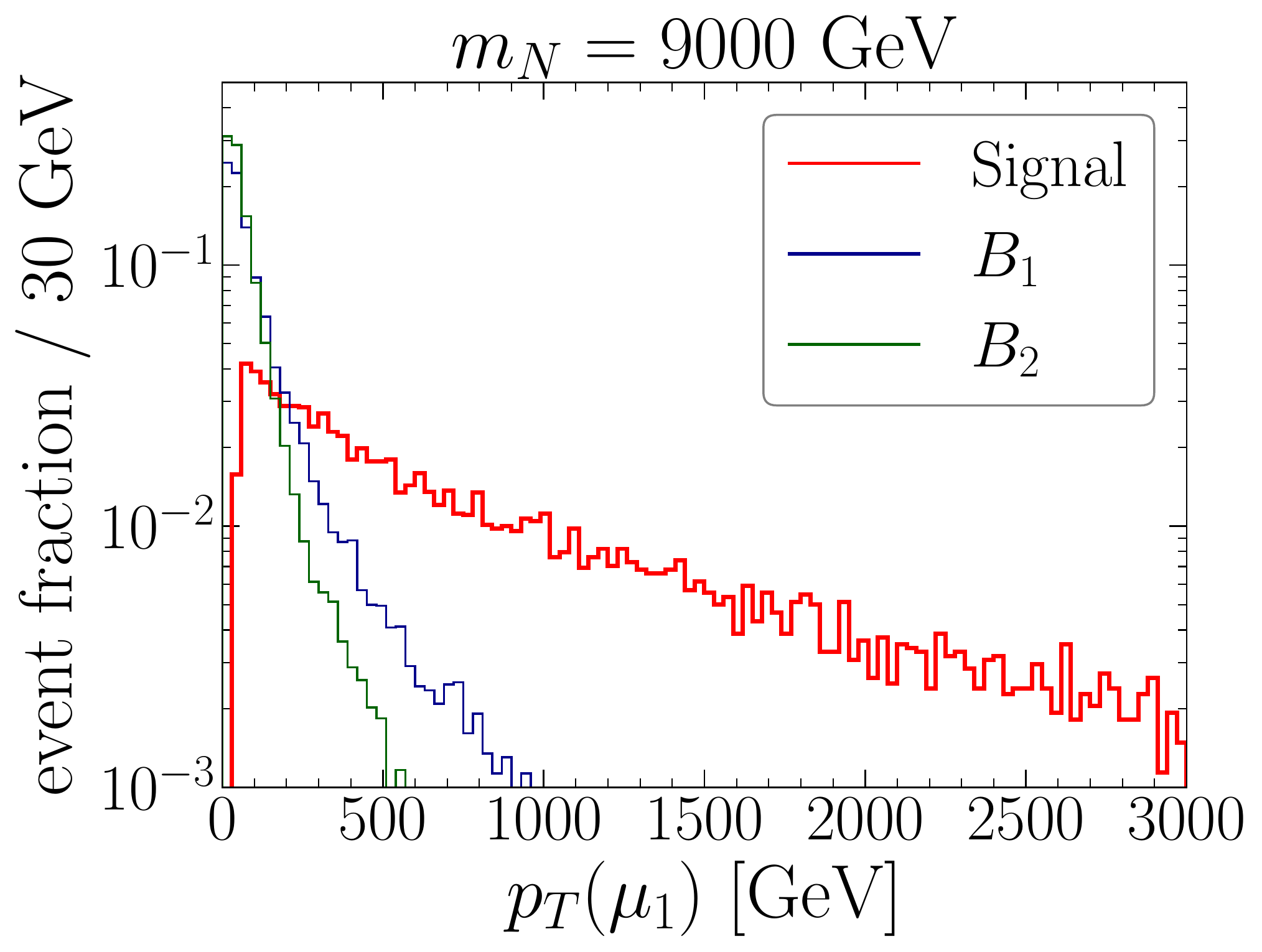}\\
      \minigraph{3.6cm}{-0.05in}{}{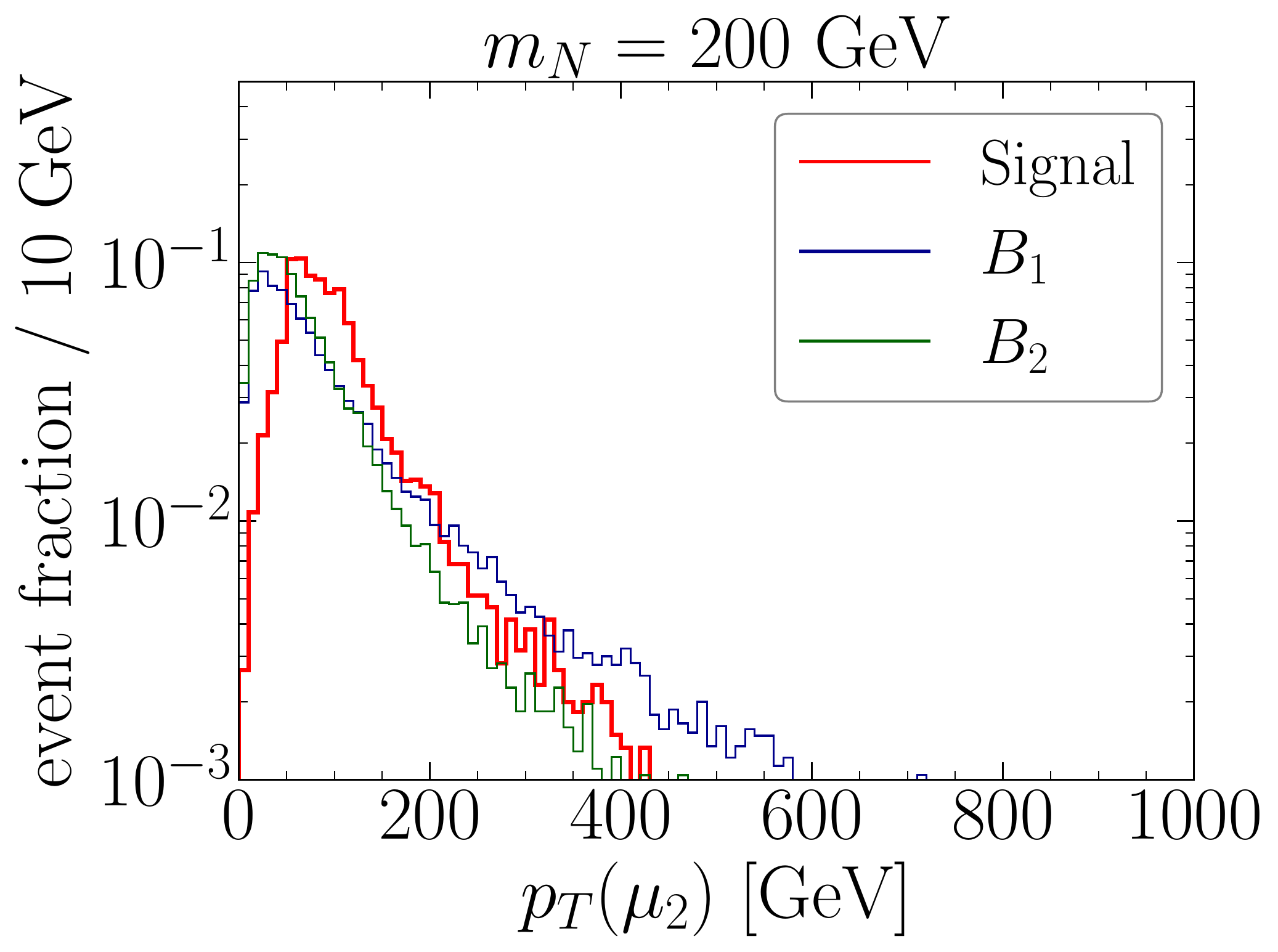}
      \minigraph{3.6cm}{-0.05in}{}{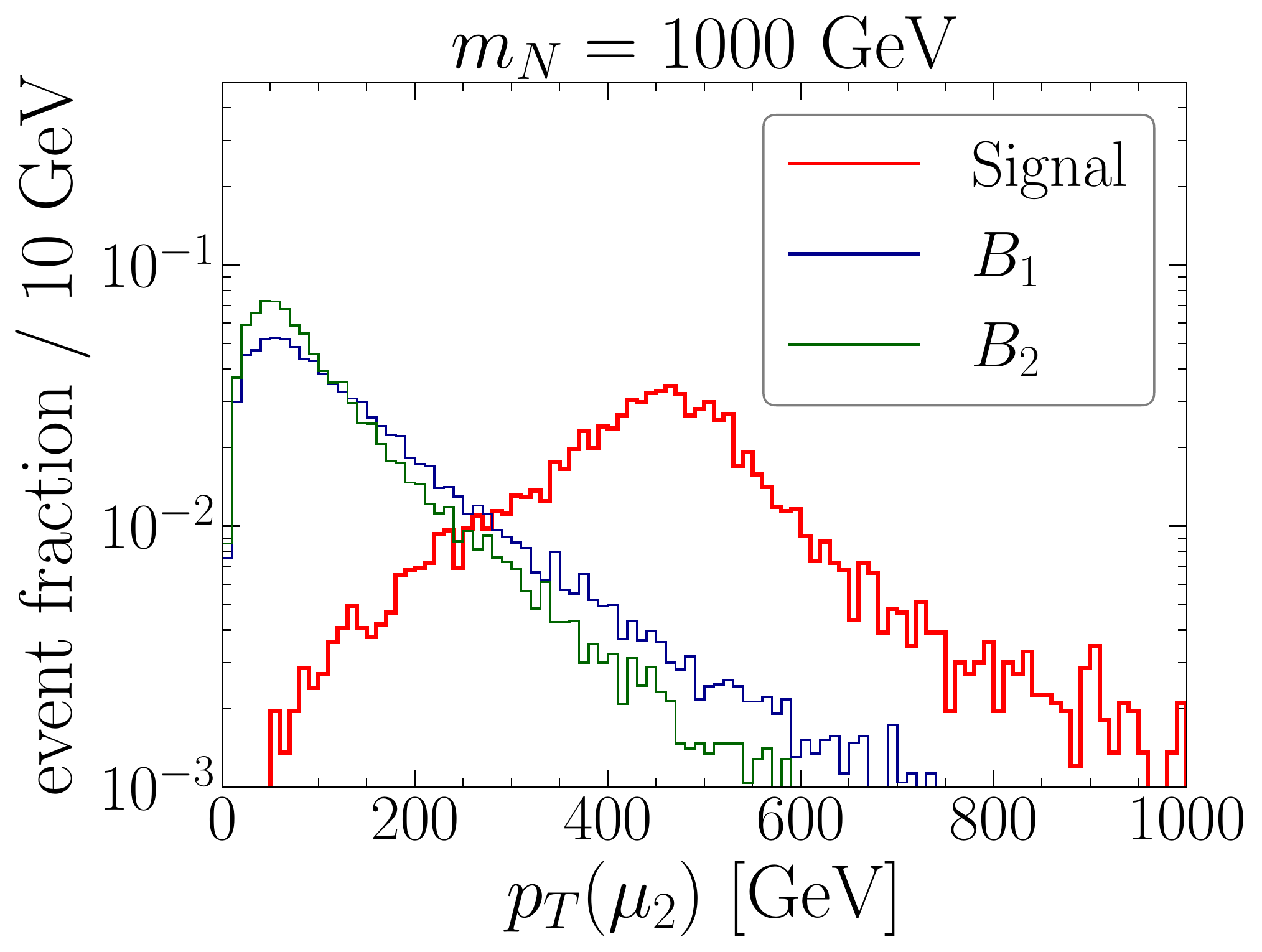}
      \minigraph{3.6cm}{-0.05in}{}{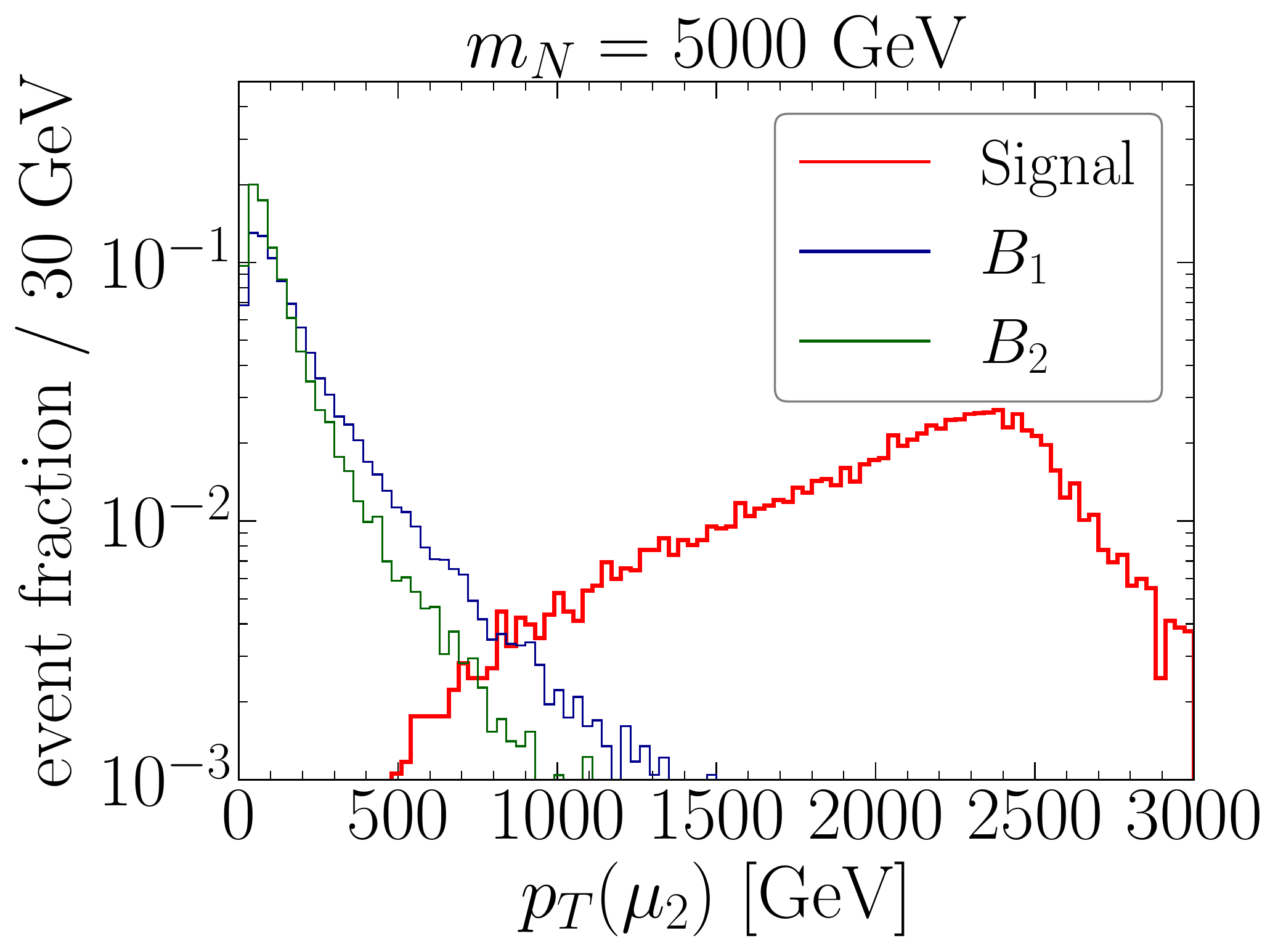}
      \minigraph{3.6cm}{-0.05in}{}{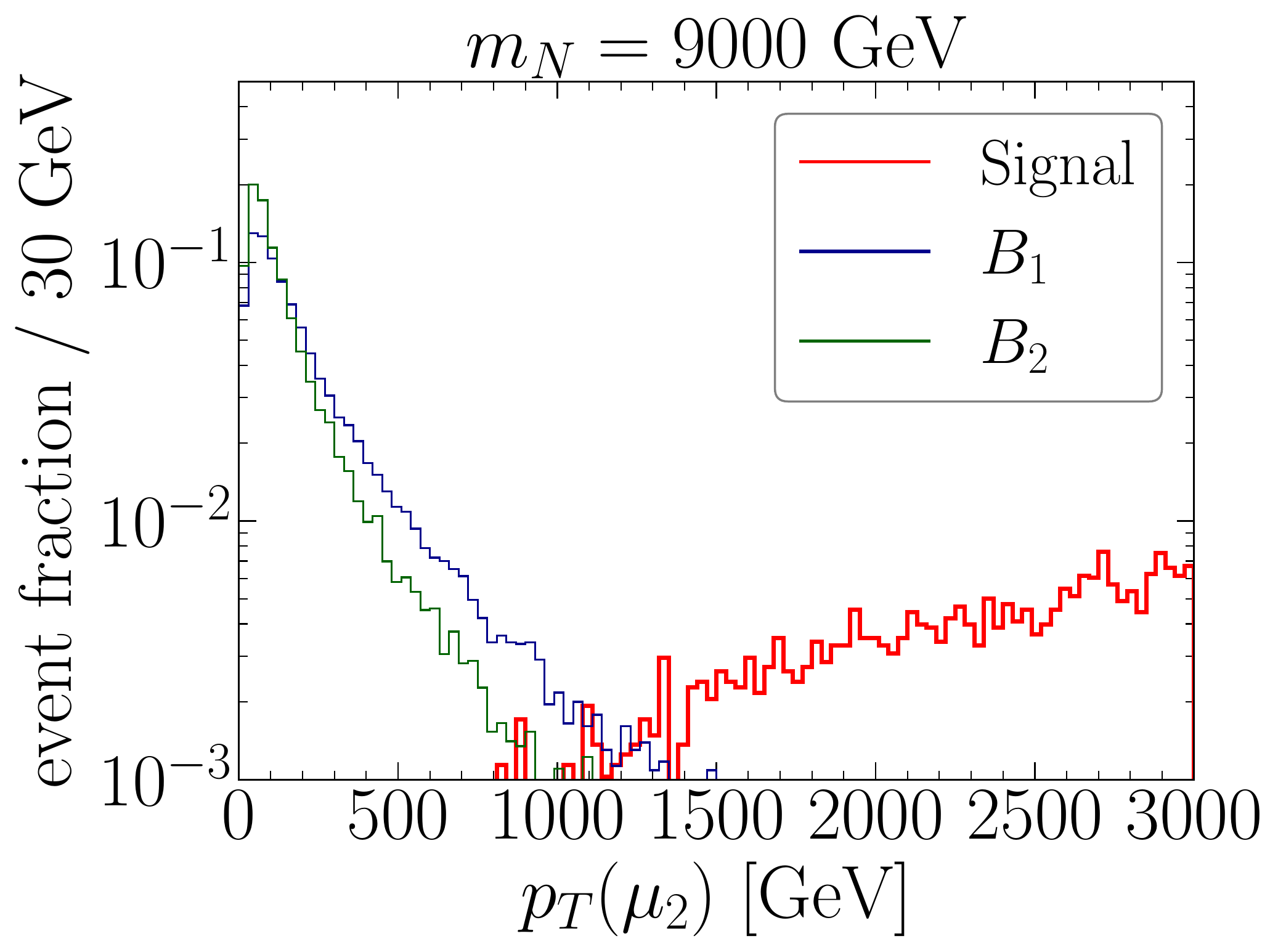}
\end{center}
\caption{The normalized distributions of the transverse momentum $p_T(\mu_1)$ (top) and $p_T(\mu_2)$ (bottom) are presented for the signal and the backgrounds for $\sqrt{s}=$ 30 TeV.
The benchmark masses of HNL from left to right are $m_N=200,~1000,~5000$ and $9000$ GeV.
}
\label{fig-SB1}
\end{figure}

\begin{figure}[h!]
\begin{center}
      \minigraph{3.6cm}{-0.05in}{}{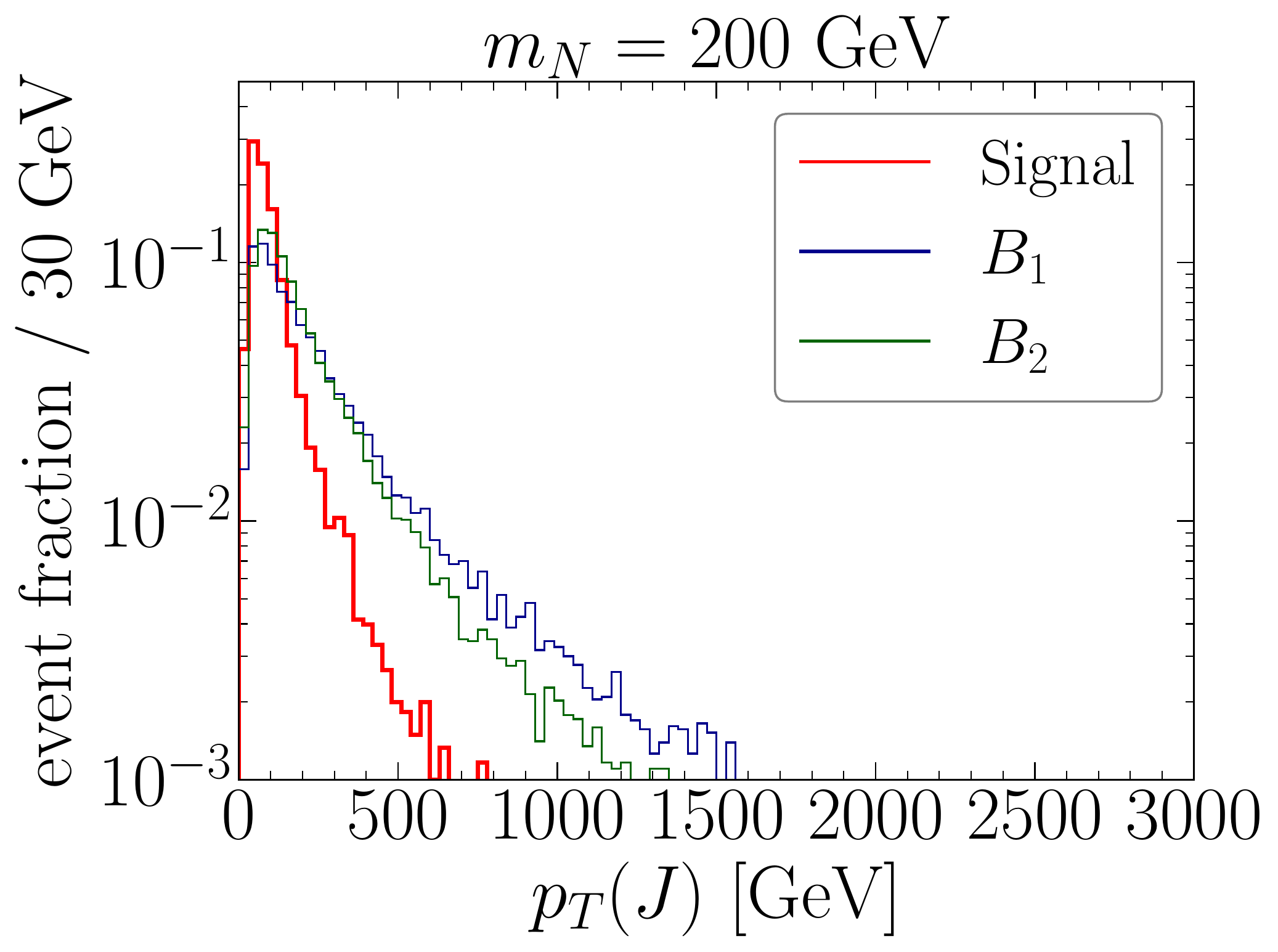}
      \minigraph{3.6cm}{-0.05in}{}{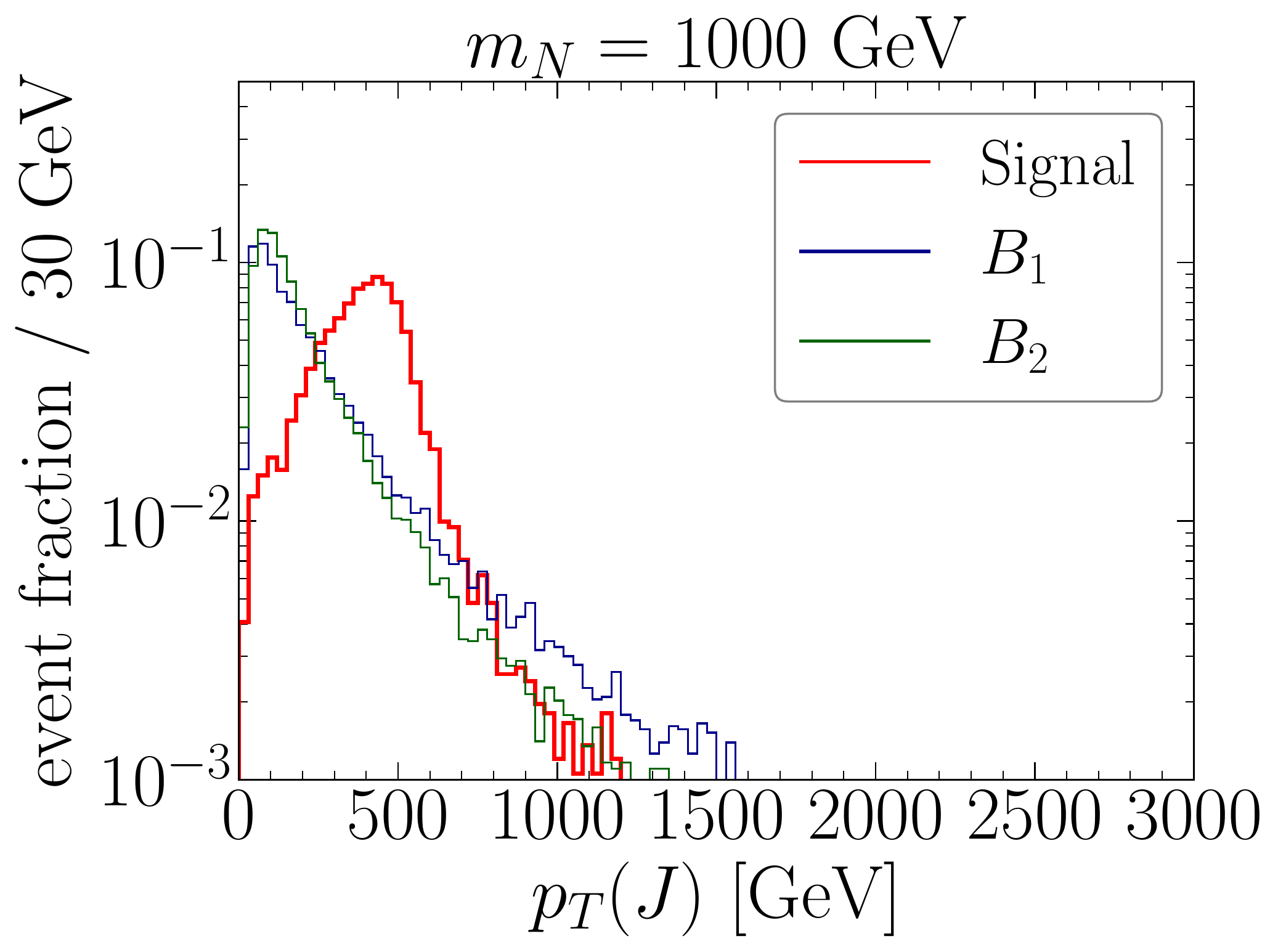}
      \minigraph{3.6cm}{-0.05in}{}{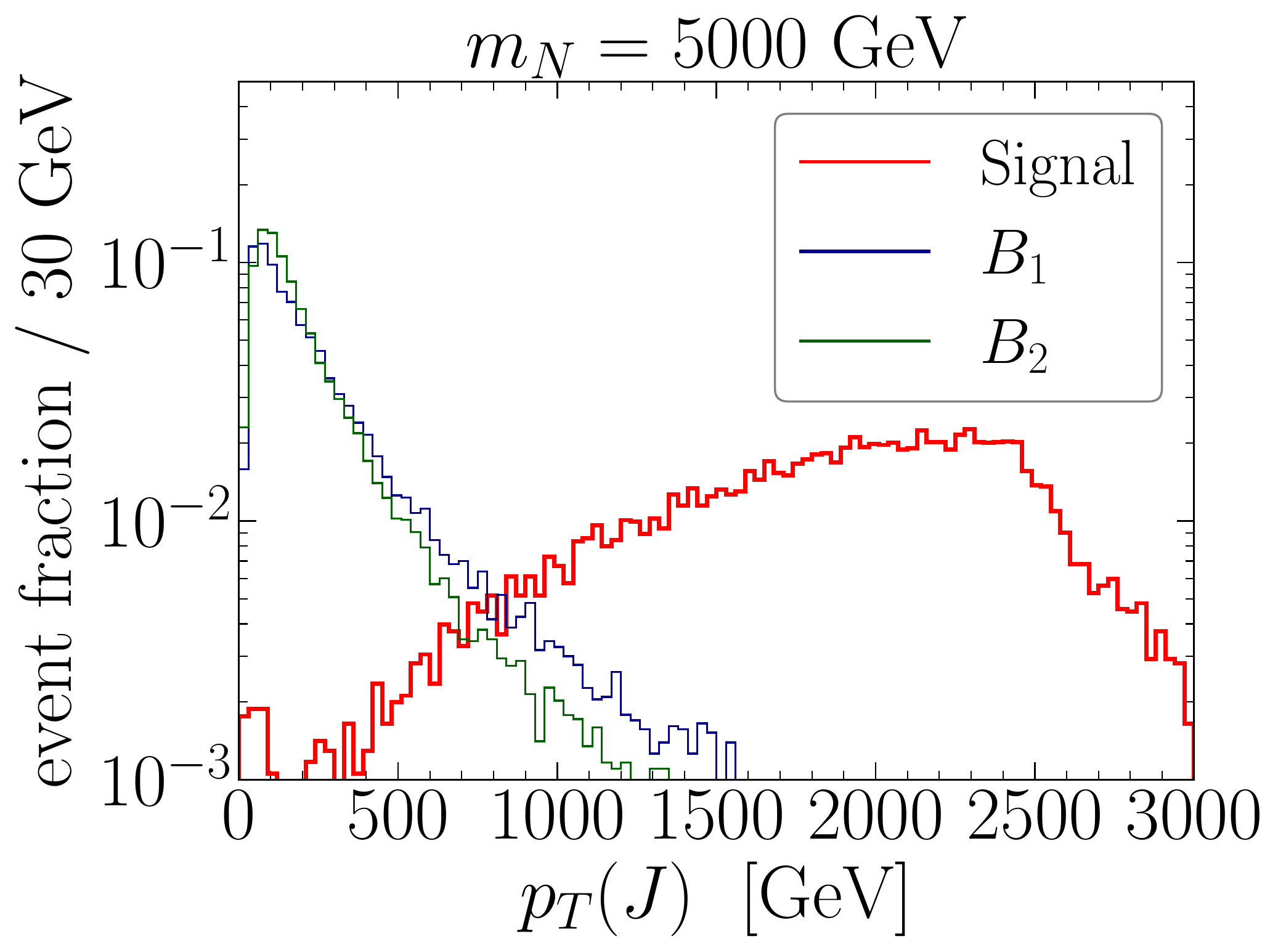}
      \minigraph{3.6cm}{-0.05in}{}{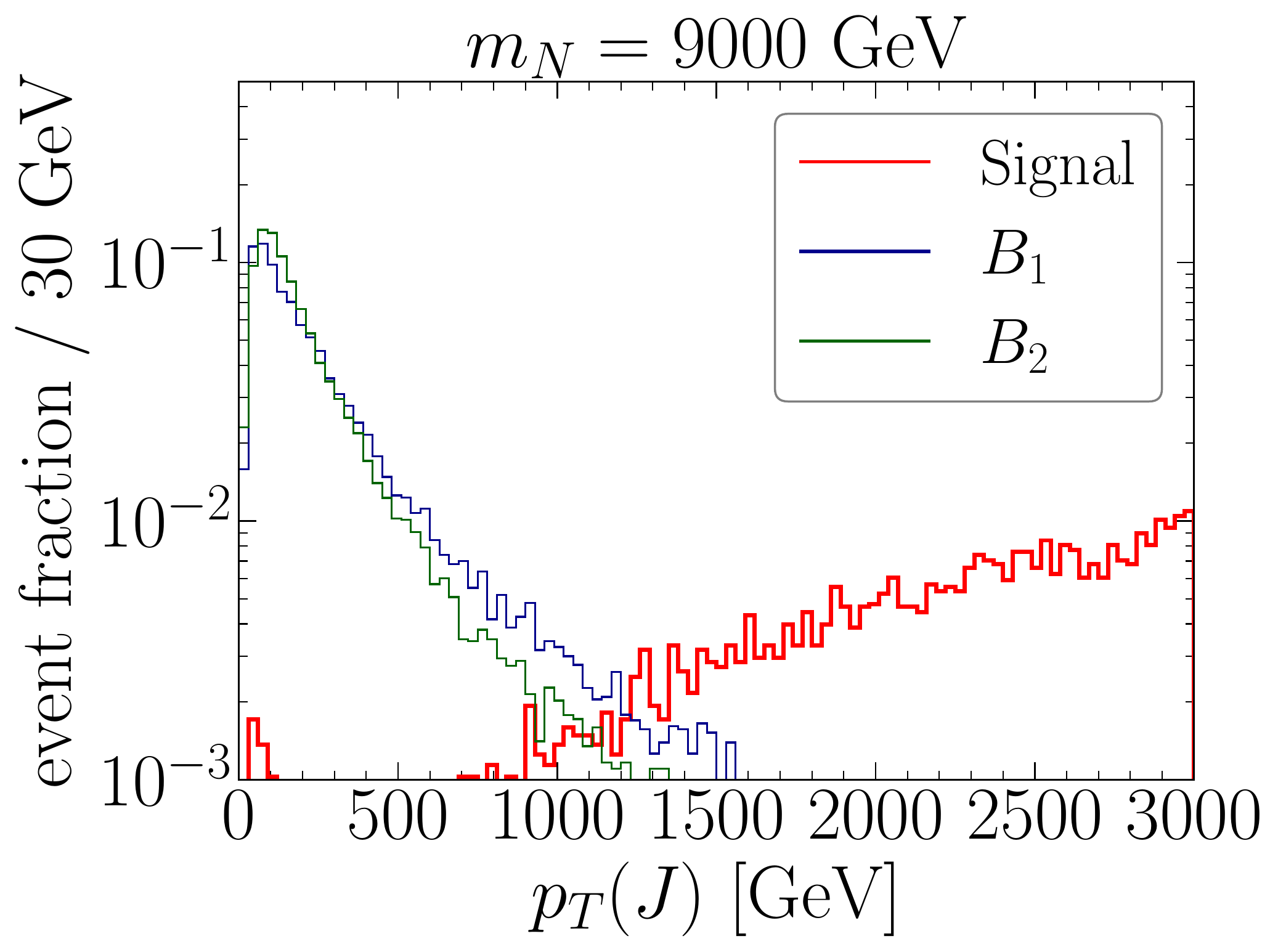}\\
      \minigraph{3.6cm}{-0.05in}{}{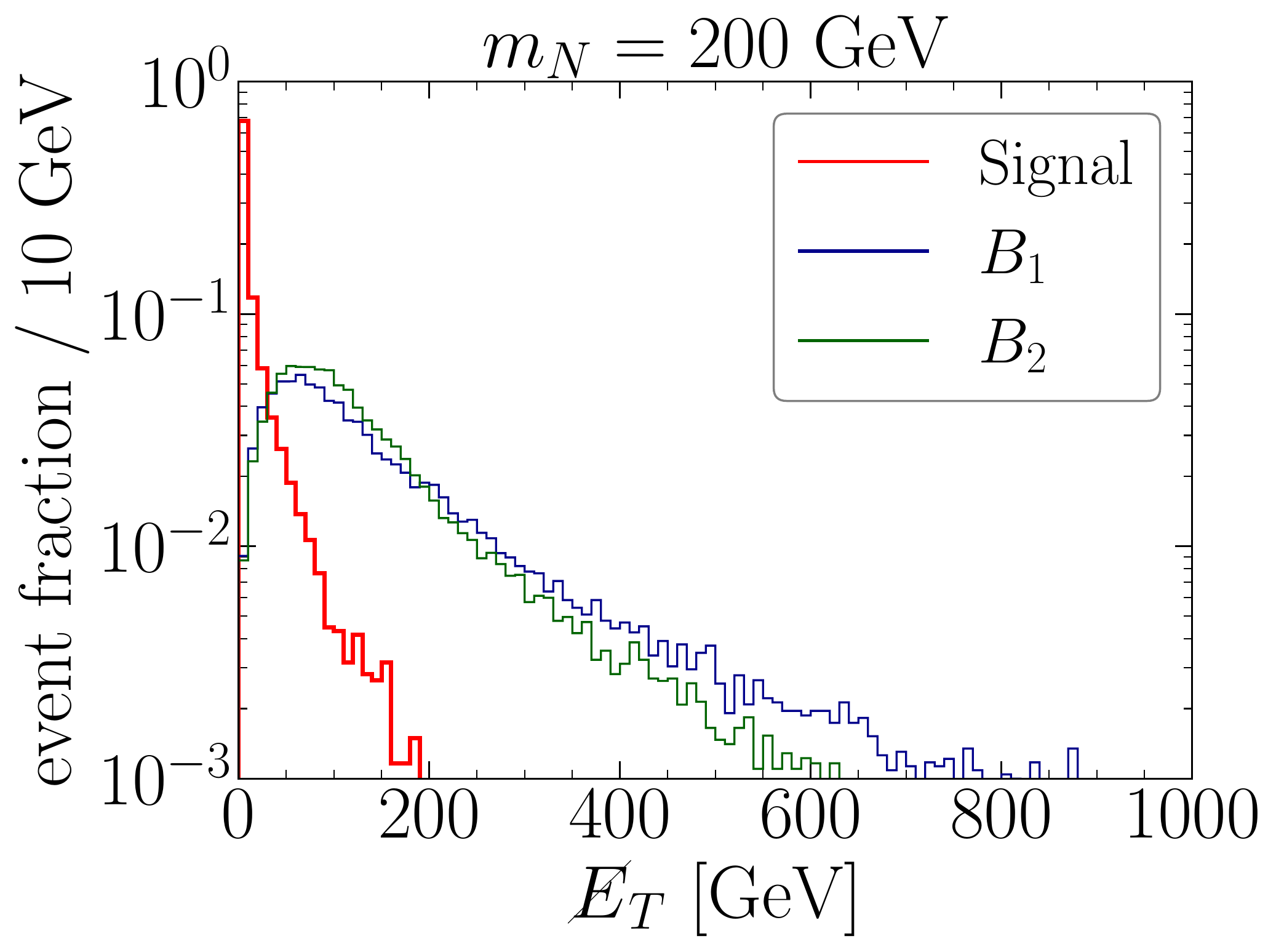}
      \minigraph{3.6cm}{-0.05in}{}{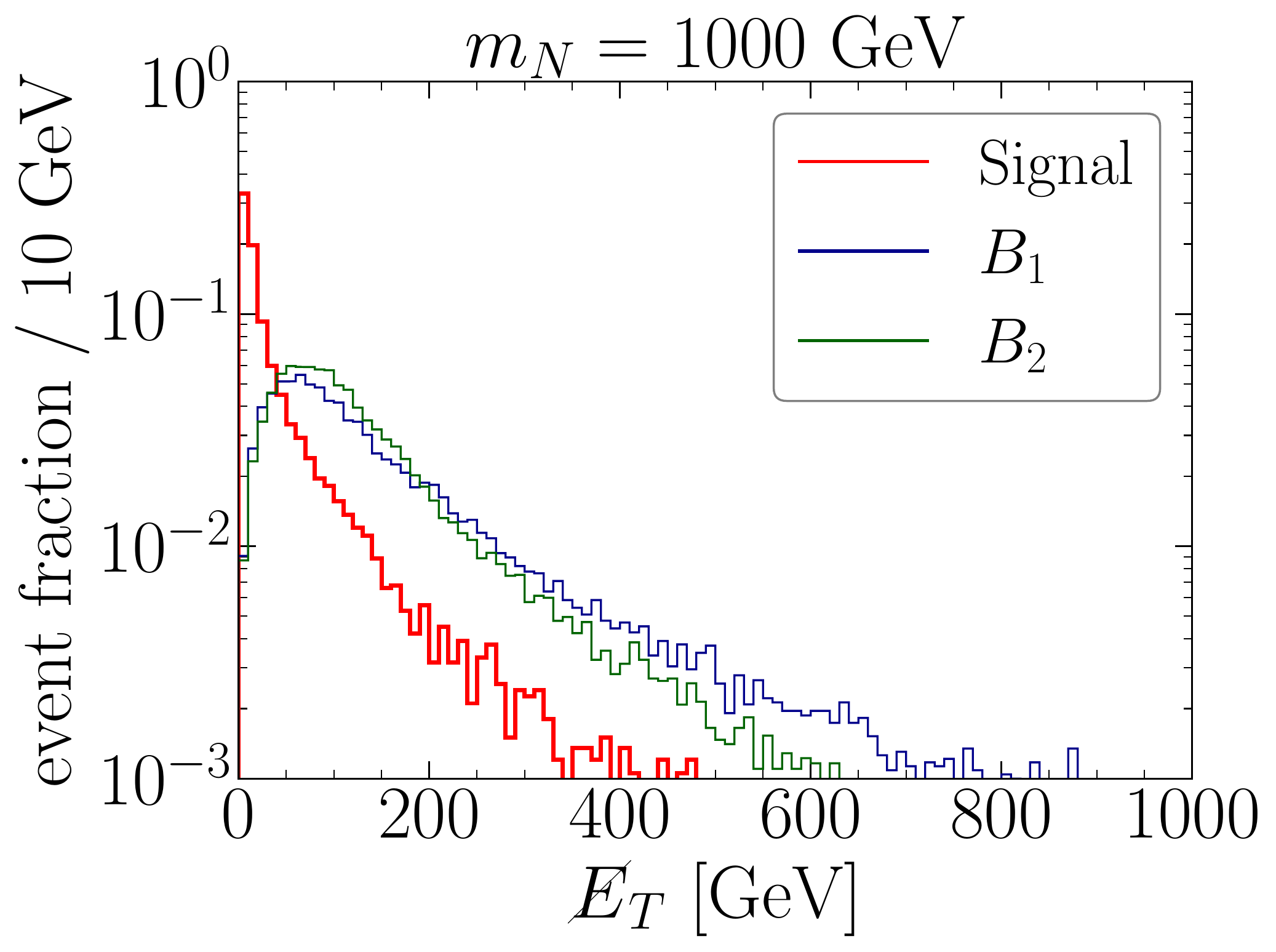}
      \minigraph{3.6cm}{-0.05in}{}{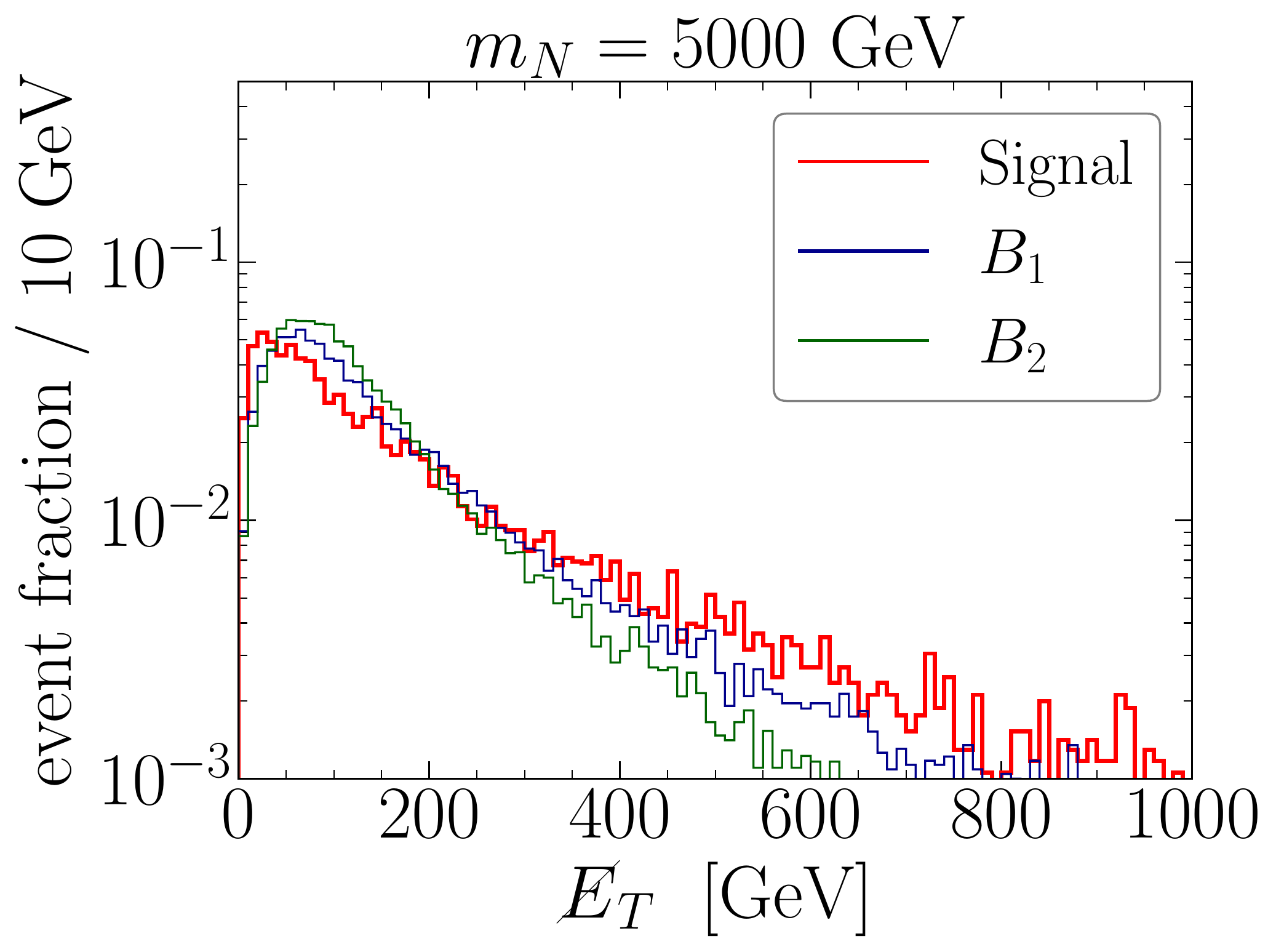}
      \minigraph{3.6cm}{-0.05in}{}{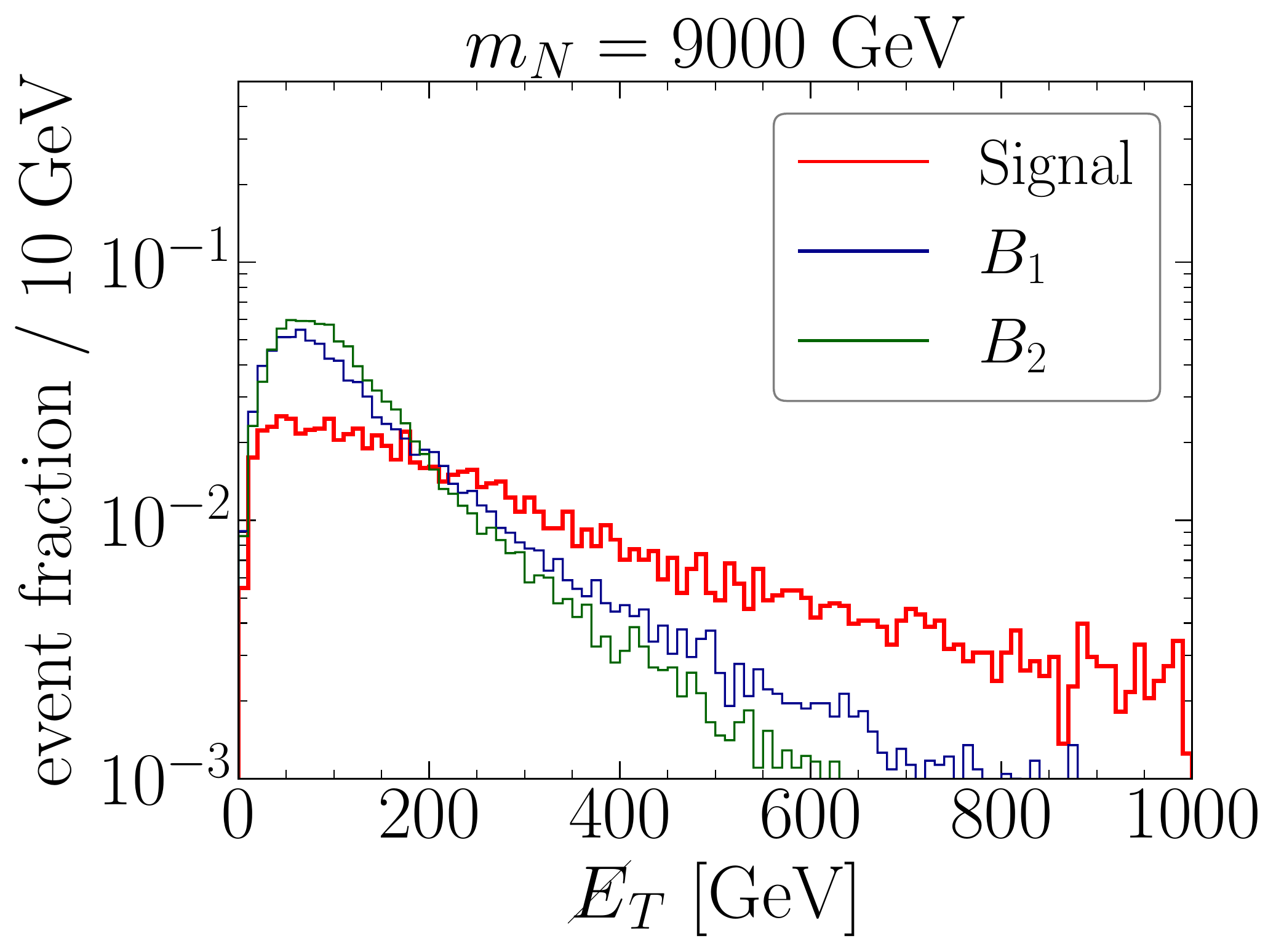}\\
      \minigraph{3.6cm}{-0.05in}{}{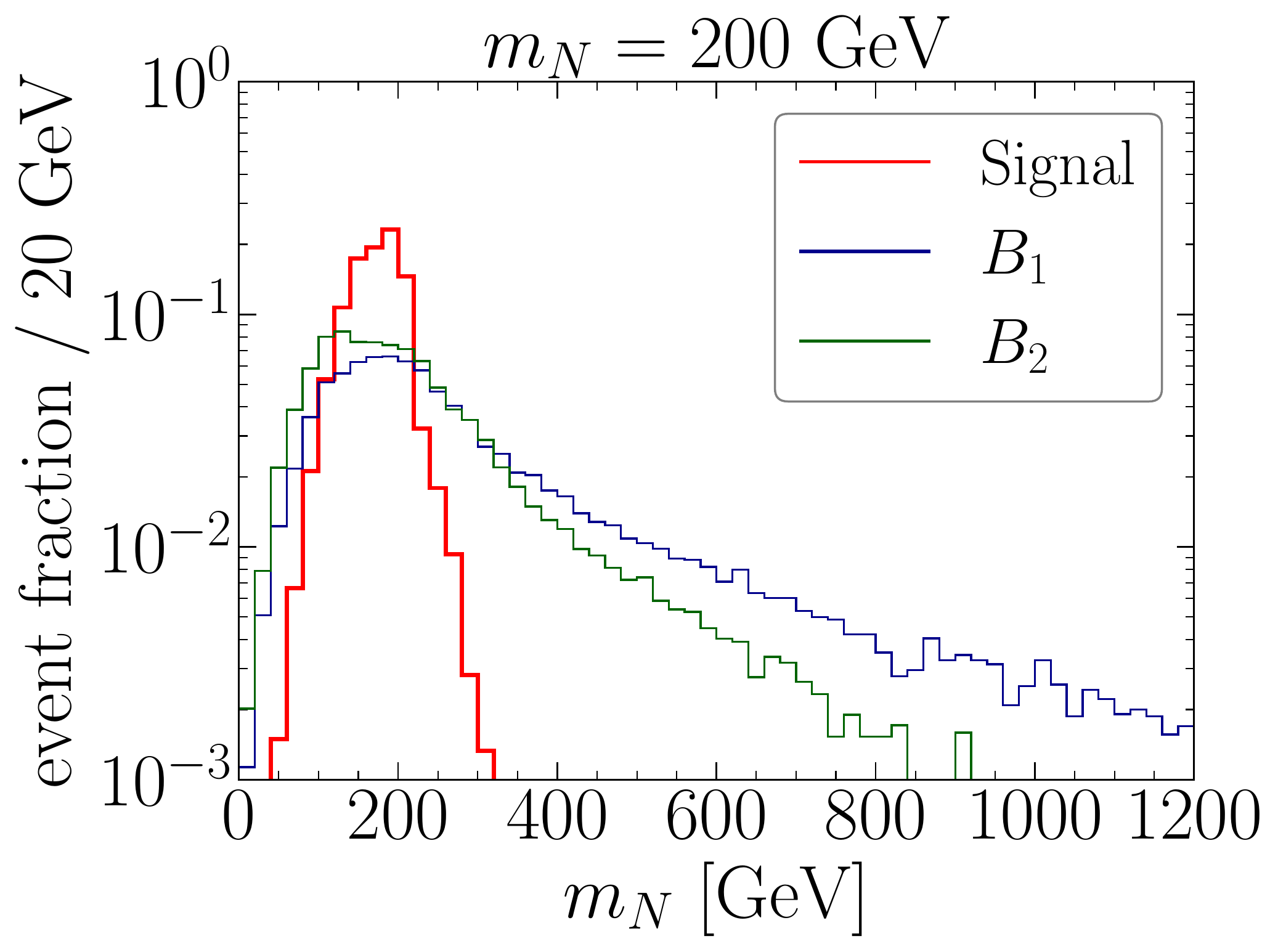}
      \minigraph{3.6cm}{-0.05in}{}{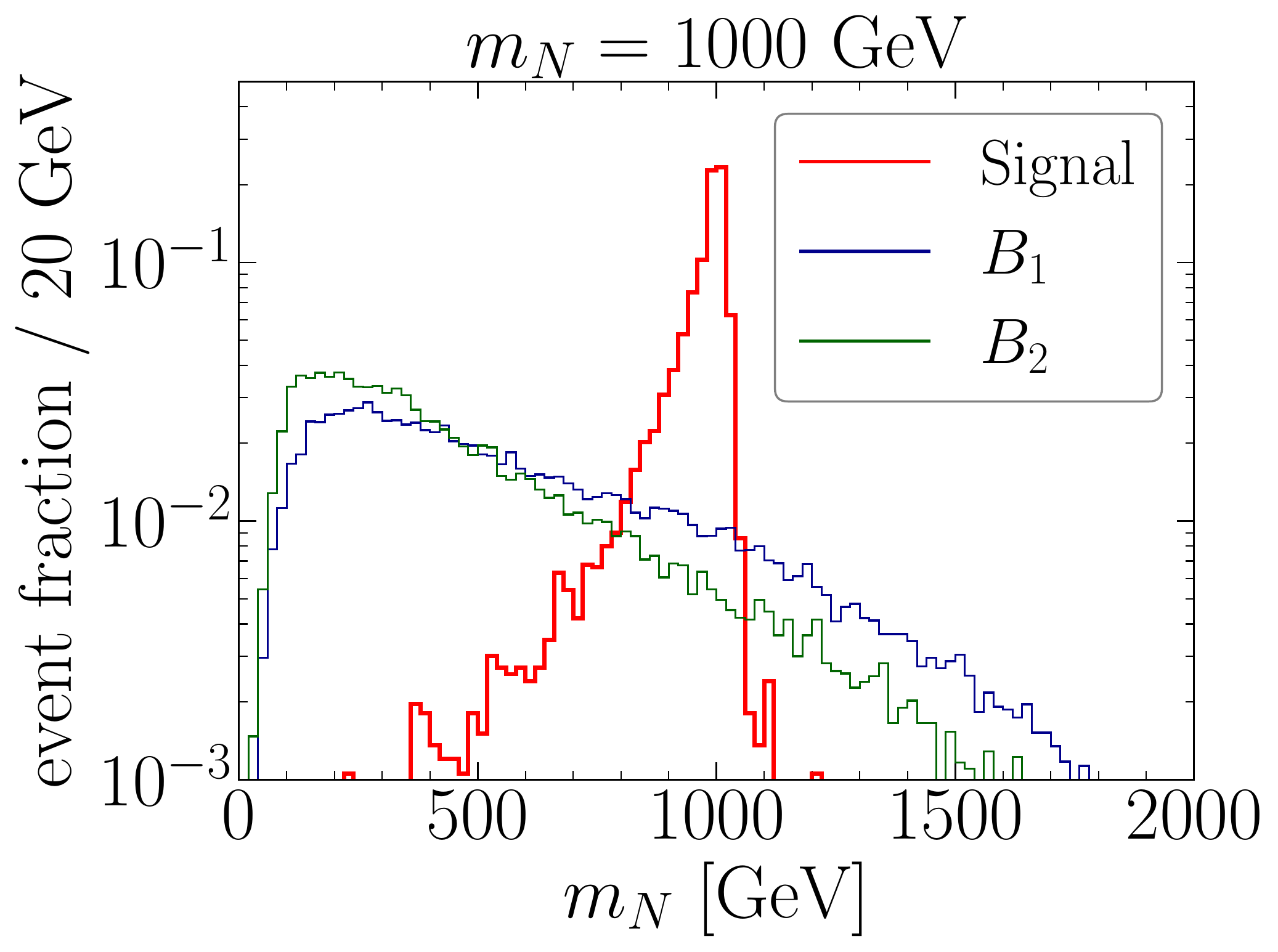}
      \minigraph{3.6cm}{-0.05in}{}{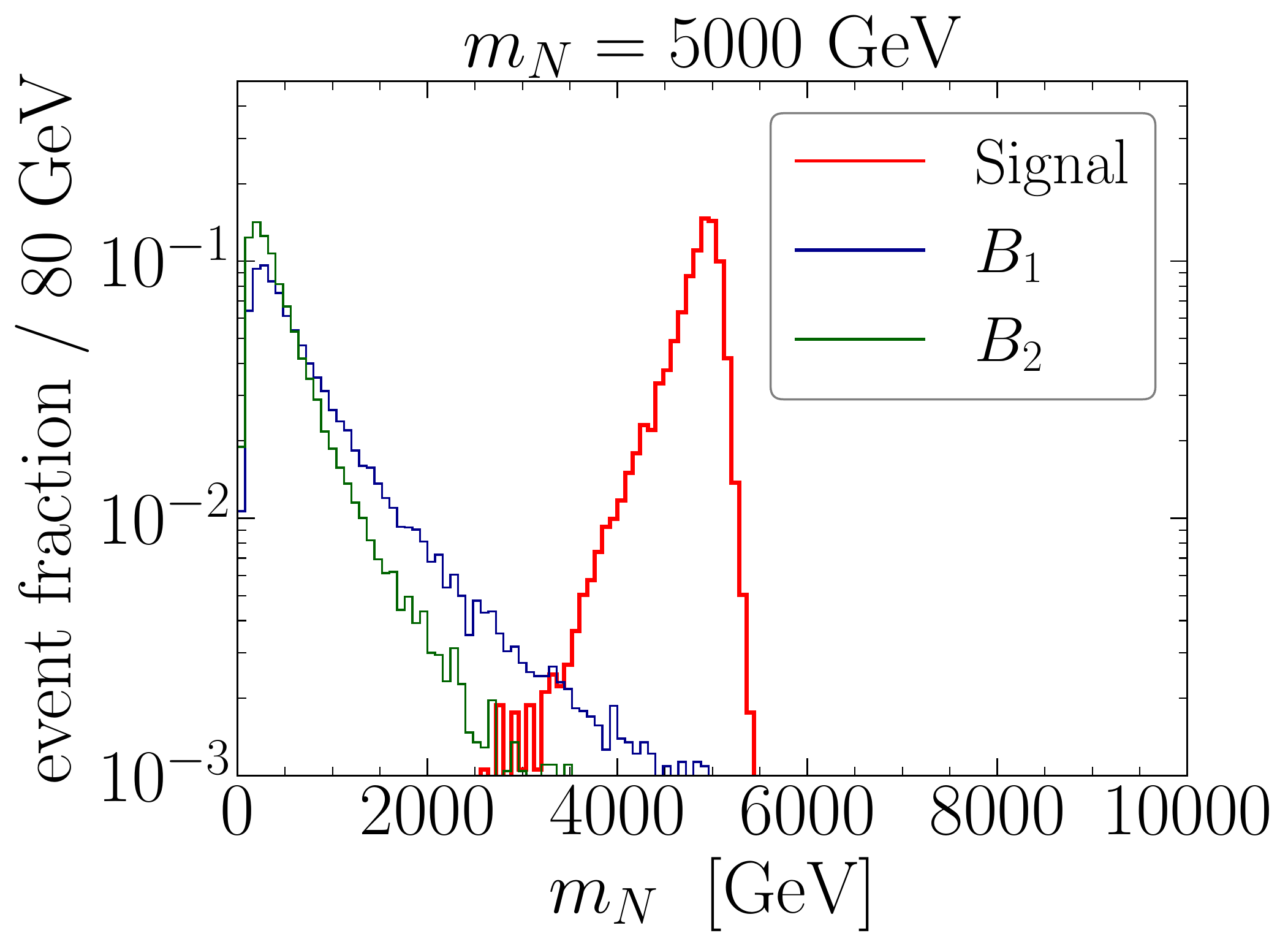}
      \minigraph{3.6cm}{-0.05in}{}{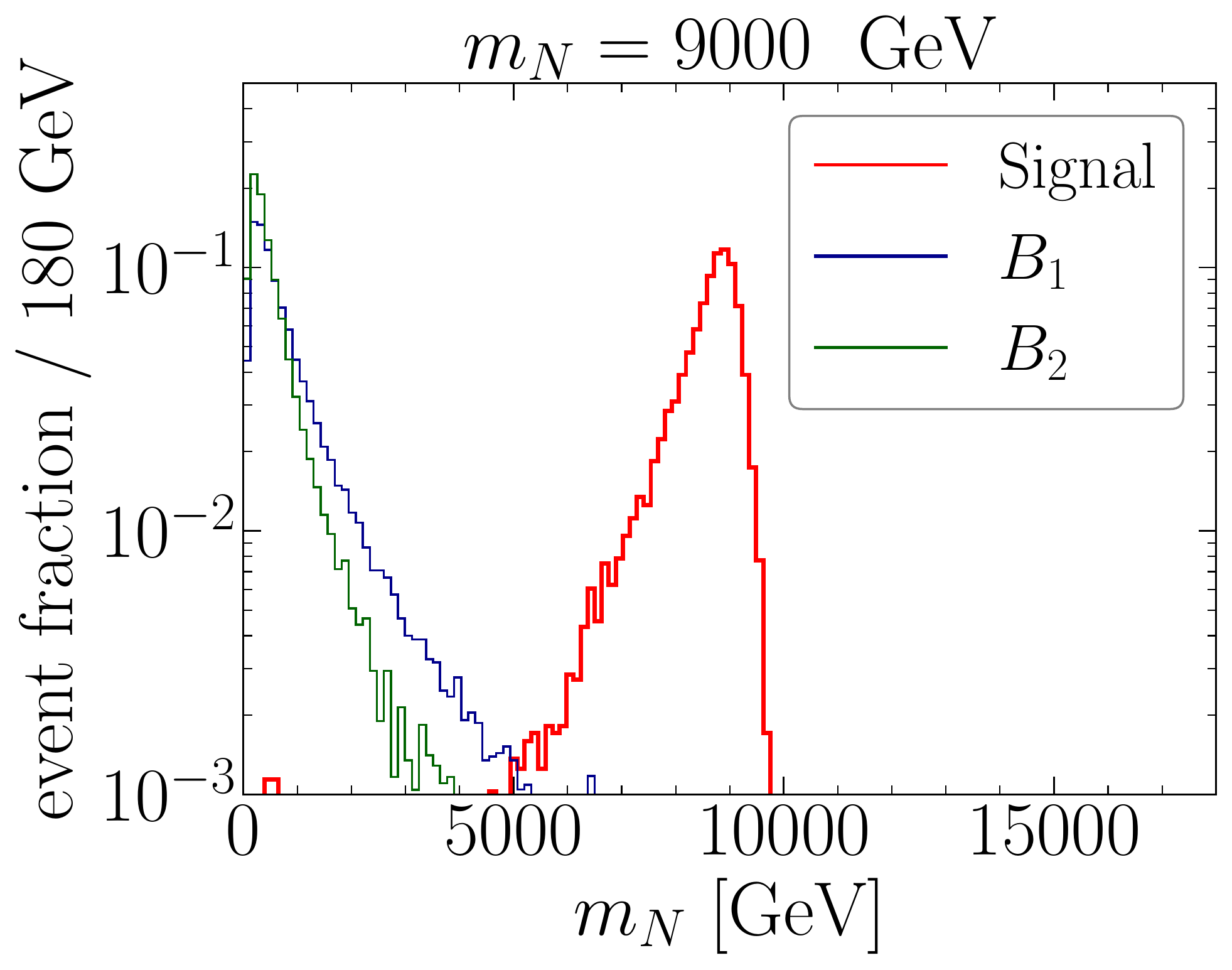}
\end{center}
\caption{The normalized distributions of $p_T(J)$ (top), missing energy $\cancel{E}_T$ (middle), and the invariant mass $m_N$ of HNL (bottom) are presented for the signal and the backgrounds for $\sqrt{s}=$ 30 TeV.
The benchmark masses of HNL from left to right are $m_N=200,~1000,~5000$ and $9000$ GeV. }
\label{fig-SB2}
\end{figure}

We take the case of $\ell_1=\ell_2=\mu$ in Eq.~(\ref{eqn-VV1}) to illustrate the kinematic observables of our signal and backgrounds in Fig.~\ref{fig-SB1} and Fig.~\ref{fig-SB2}.
The benchmark masses of HNL are considered to be $m_N=200,~1000,~5000$ and $9000$ GeV for $\sqrt{s}=30$ TeV.
From the decay of such heavy Majorana neutrino, the $W$ boson could be highly boosted and the produced dijet can be regarded as a single fat-jet $J$~\cite{Das:2017gke,Bhardwaj:2018lma,Das:2018usr} in the detector of muon collider. The fat-jet is reconstructed via the ``Valencia'' algorithm with $R = 1.2$.
We use $\mu_1$ to denote the muon produced in association with the HNL, and $\mu_2$ and $J$ (or $q\bar{q}'$) are the decay products of HNL. We will explain how we distinguish them in our analysis below.
The signal and backgrounds exhibit quite different kinematic distributions.
We employ a series of selection cuts based on the distributions to suppress the SM backgrounds and enhance the significance.

\begin{itemize}
\item
We first select the generated events by requiring the number of same-sign muons and fat-jet, and identify the fat-jet as a $W$ boson
\begin{align}
N_{\mu} \geq 2,~N_{J} \geq 1 \;;~~ 65~{\rm GeV} \leq m_J \leq 95~{\rm GeV}\;.
\label{eqn-cut0}
\end{align}
If no fat-jet is found in the event, we also request two leading ordinary jets to satisfy the invariant mass window $65~{\rm GeV} \leq m_{j_1 j_2} \leq 95~{\rm GeV}$. But we find that taking into account these events would not change our results much. Thus, we only deal with the events with at least one fat-jet in the following analysis.
\item We employ some basic cuts for the muons in final states
\begin{align}
& p_T(\mu_1) > 50~{\rm GeV},~~p_T(\mu_2) > m_N/4;~~|\eta(\mu_1)|,|\eta(\mu_2)|< 2.5;~~\Delta R_{\mu_1 \mu_2} > 0.5 \;.
\label{eqn-cut1}
\end{align}
The transverse momentum cut for $\mu_1$ is essential to avoid possible collinear divergence. As $\mu_2$ comes from the decay of $N$, it could be very energetic depending on $m_N$ as shown in Fig.~\ref{fig-SB1}. We thus tighten up the kinematic cut for the transverse momentum of $\mu_2$. It is worth mentioning that for the two muons in the final states, we separately combine them with $J$. The muon that generates an invariant mass closer to $m_N$ is considered as $\mu_2$.
\item For the final fat-jet $J$ coming from the decay of $N$, its $p_T$ distribution is shown in the upper panels of Fig.~\ref{fig-SB2}. We employ the following kinematic cuts
\begin{align}
p_T(J) > m_N/4,~~|\eta(J)| < 2.5 \;.
\label{eqn-cut2}
\end{align}
\item The presence of neutrinos in the final states of the backgrounds leads to missing energy, as shown in Eq.~\eqref{eqn-bkg1}. For our signal without neutrinos, due to the effect of the energy resolution of detector, there still exists a distribution of small missing energy as shown in the middle panels of Fig.~\ref{fig-SB2}. As $m_N$ increases, this effect becomes progressively more significant. Thus, we only employ the missing energy cut for $m_N\leq 1000$ GeV
\begin{align}
&\cancel{E}_T < 30~{\rm GeV} ~~~{\rm when}~~ m_N \leq 1000~{\rm GeV}\;.
\label{eqn-cut3}
\end{align}
\item We reconstruct Majorana HNL with $\mu_2$ and $J$ according to the invariant mass $m_{\mu_2 J}$ shown in the lower panels of Fig.~\ref{fig-SB2}. Based on the signal distribution of the $m_{\mu_2 J}$, we employ an asymmetric mass window cut
\begin{align}
0.8 \times m_N < ~m_{\mu_2 J}~< 1.1\times m_N \;.
\label{eqn-cut4}
\end{align}
\end{itemize}
The background events can be efficiently suppressed after employing the above cuts. For illustration, we take $\sqrt{s}=30$ TeV for muon collider to present the signal and background cross sections after these cuts, as shown in Table~\ref{tab-SB}.

\begin{table}[htb!]
\centering
\footnotesize
\begin{tabular}{|c|c|c|c|c|c|c|}
\hline
\hline
     \multirow{2}*{sig. and bkgs}   & \textcolor{blue}{$\sigma/S_{\mu\mu}$}  or  & $N_{\mu}$,~$N_{J}$,~$m_{J}$  & $\mu_1,~\mu_2$ & fat-jet $J$ &  $\cancel{E}_T$ & $m_{\mu_2 J}$\\
         & $\sigma_{\rm B_i}$ [fb]  & in Eq.~\eqref{eqn-cut0} & in Eq.~\eqref{eqn-cut1} & in Eq.~\eqref{eqn-cut2} &  in Eq.~\eqref{eqn-cut3} & in Eq.~\eqref{eqn-cut4}\\
    \hline
    \hline
    $m_N=200$ GeV &  \textcolor{blue}{7.67} &  \textcolor{blue}{1.17}&  \textcolor{blue}{0.90} &  \textcolor{blue}{0.89}&  \textcolor{blue}{0.80} &  \textcolor{blue}{0.80} \\
     $\rm B_1$& 0.65& 0.23& 0.13& 0.13 & 0.0081& 8.6$\times 10^{-4}$\\
     $\rm B_2$& 0.094& 0.0054& 0.0020& 0.0020 & 2.0$\times 10^{-4}$  & 4.0$\times 10^{-5}$ \\
    \hline
    \hline
    $m_N=1000$ GeV & \textcolor{blue}{9.05} &  \textcolor{blue}{4.54}&  \textcolor{blue}{4.11} & \textcolor{blue}{3.79} & \textcolor{blue}{2.80} & \textcolor{blue}{2.79}\\
     $\rm B_1$& 0.65& 0.23& 0.048& 0.036&0.0017 & 5.4$\times 10^{-4}$\\
     $\rm B_2$& 0.094& 0.0054& 4.4$\times 10^{-4}$& 2.6$\times 10^{-4}$ & 2.0$\times 10^{-5}$& 9.9$\times 10^{-6}$\\
    \hline
    \hline
    $m_N=5000$ GeV & \textcolor{blue}{4.56} &  \textcolor{blue}{2.70}&  \textcolor{blue}{2.41} &   \textcolor{blue}{2.19}& --&   \textcolor{blue}{2.17}\\
     $\rm B_1$& 0.65& 0.23& 0.0066 & 0.0035&--&0.0012\\
     $\rm B_2$& 0.094& 0.0054& 1.2$\times 10^{-5}$& 0.0 & --& 0.0\\
    \hline
    \hline
    $m_N=9000$ GeV & \textcolor{blue}{2.11} &  \textcolor{blue}{1.17}&  \textcolor{blue}{1.03} &   \textcolor{blue}{0.95}&  --&  \textcolor{blue}{0.94}\\
    $\rm B_1$& 0.65& 0.23& 0.0020& 8.8$\times 10^{-4}$&--&3.2$\times 10^{-4}$\\
    $\rm B_2$& 0.094& 0.0054& 5.1$\times 10^{-6}$ & 0.0 & -- & 0.0\\
\hline
\hline
\end{tabular}
\caption{The representative signal cross section $\sigma/S_{\mu\mu}$ (in blue) and those for SM backgrounds (in black) at muon collider after selection cuts.
For illustration, four benchmarks $m_N=200,~1000,~5000$ and $9000$ GeV are considered at muon collider with $\sqrt{s}=30$ TeV.
}
\label{tab-SB}
\end{table}

\begin{figure}[h!]
\begin{center}
\minigraph{7.4cm}{-0.05in}{}{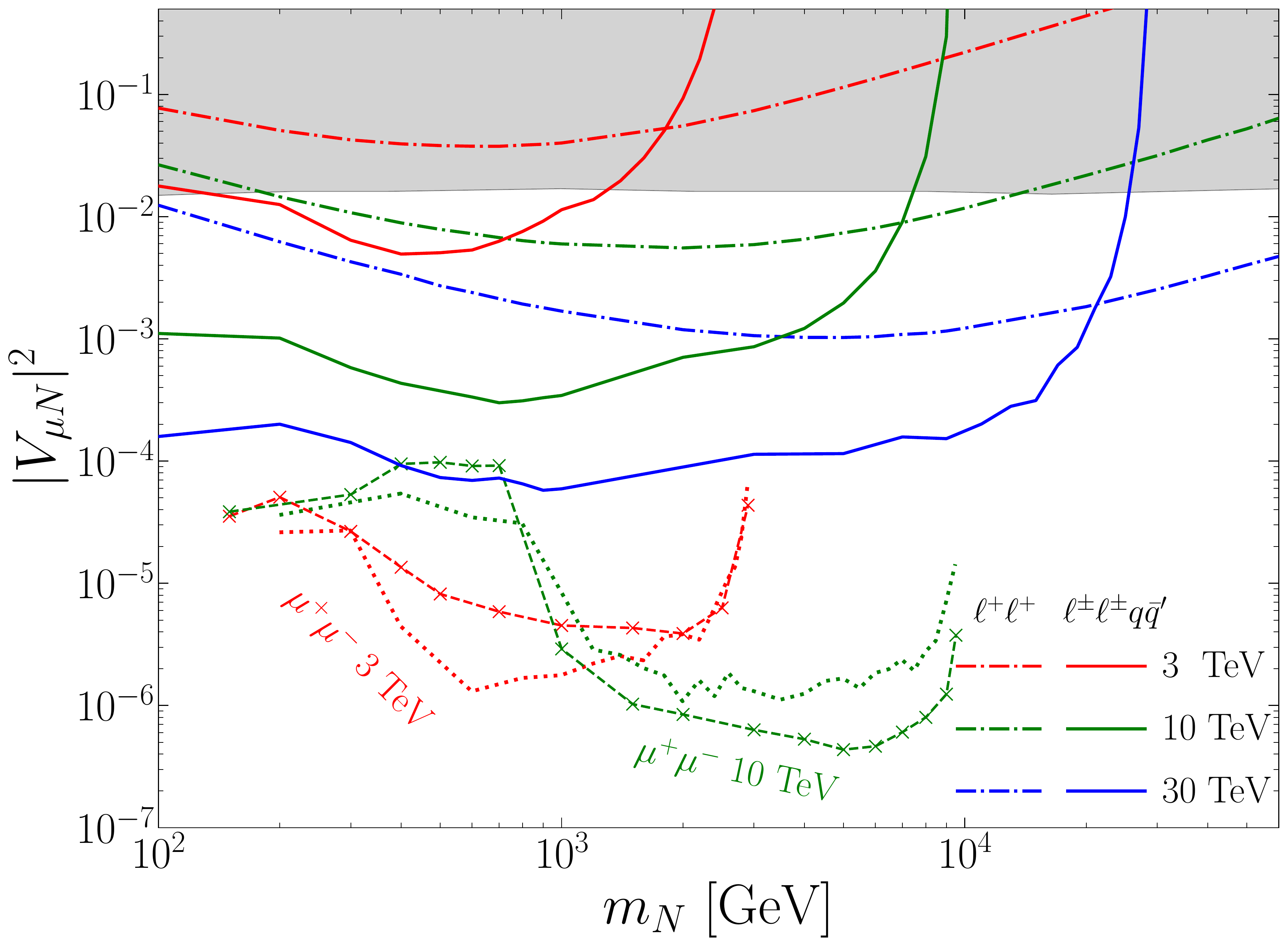}\\
\minigraph{7.4cm}{-0.05in}{}{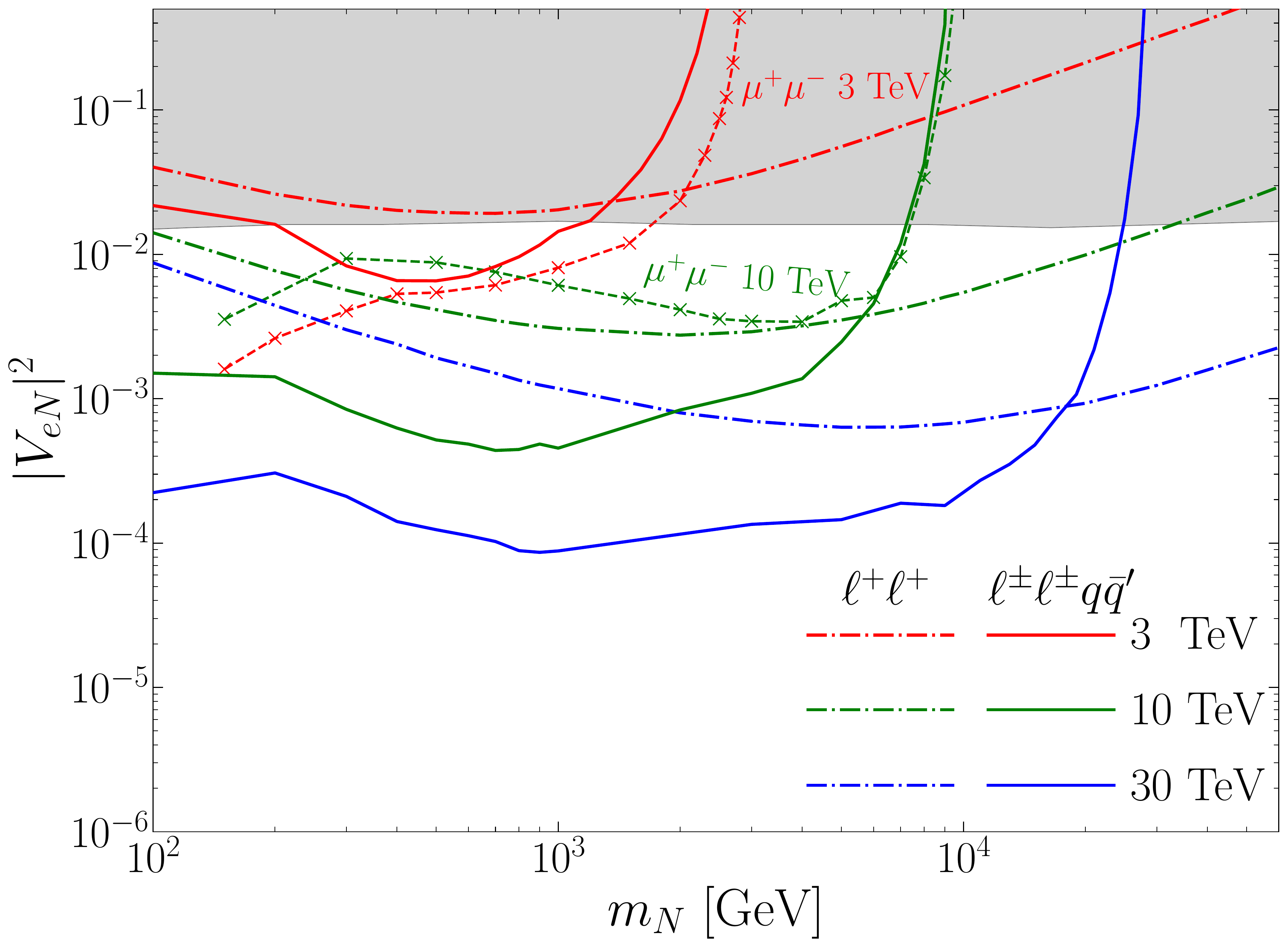}
\minigraph{7.4cm}{-0.05in}{}{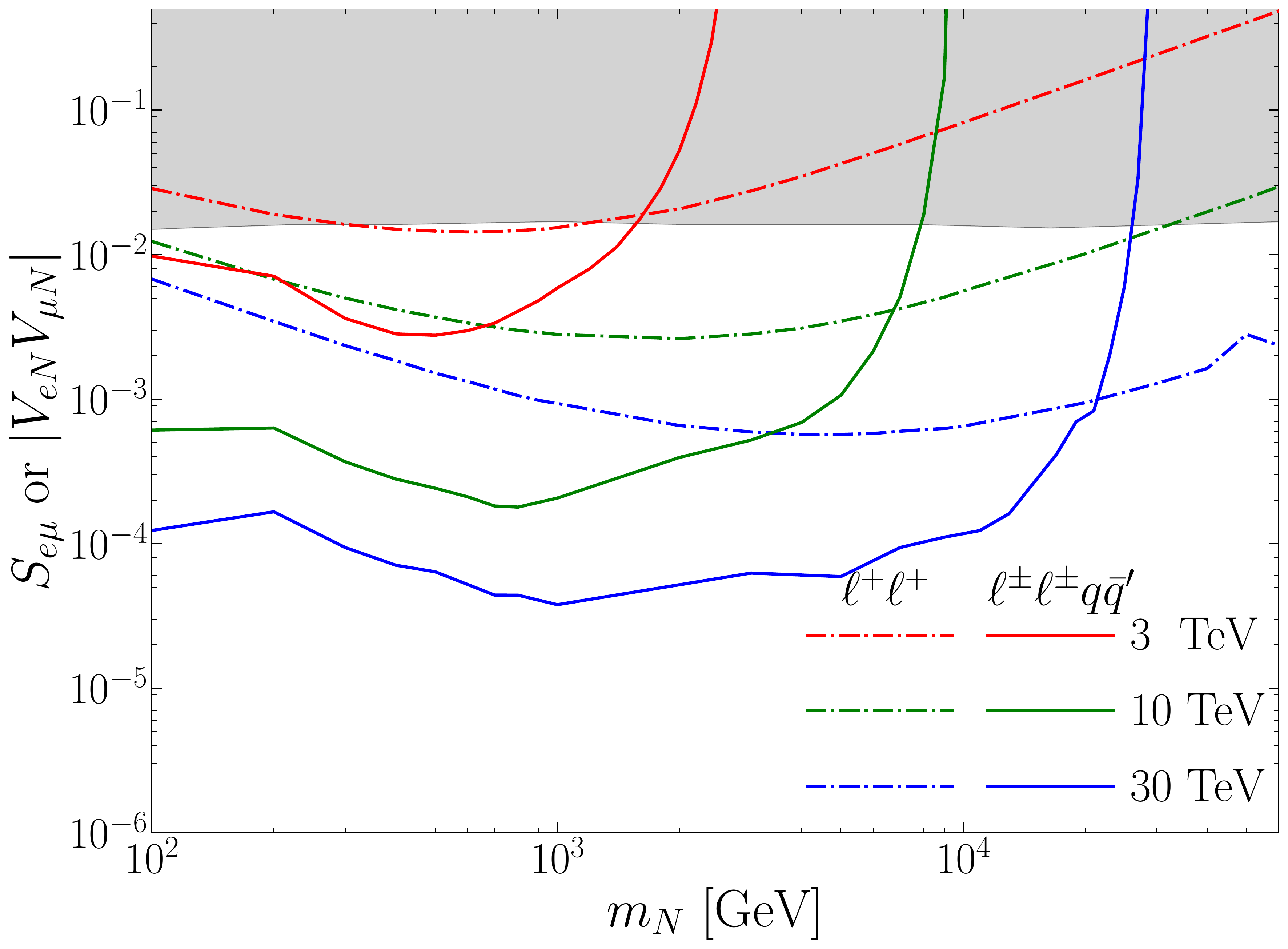}
\end{center}
\caption{The 2$\sigma$ exclusion limits for $|V_{\mu N}|^2$ (upper panel), $|V_{eN}|^2$ (lower-left panel) and $S_{e\mu}$ or $|V_{eN}V_{\mu N}|$ (lower-right panel) as a function of $m_N$ at muon colliders with $\sqrt{s}=3$ TeV (red), 10 TeV (green) and 30 TeV (blue). The solid and dash-dotted lines correspond to $V_i V_j\to \ell_1^\pm\ell_2^\pm + {\rm jet(s)}$ and $W^+W^+\to \ell_1^+\ell_2^+$ channels, respectively. The frame label of the lower-right panel is $S_{e\mu}$ for $\ell_1^\pm\ell_2^\pm + {\rm jet(s)}$ channel or $|V_{eN}V_{\mu N}|$ for $\ell_1^+\ell_2^+$. The grey area shows the region excluded by a global scan~\cite{Chrzaszcz:2019inj}. The recent results from two groups' simulations of $\mu^+\mu^-\to N\nu$ are also added for comparison, including the dotted lines~\cite{Kwok:2023dck} and the dashed lines~\cite{Li:2023tbx} for $\sqrt{s}=3$ TeV (red) and 10 TeV (green).
}
\label{fig-VmN2}
\end{figure}

After implementing the above analysis, we evaluate the sensitivity to the mixing constant $S_{\ell_1\ell_2}$ for Majorana HNL at muon colliders. For the significance, we use the following formula
\begin{equation}
\mathcal{S} = \frac{N_{\rm S}}{\sqrt{N_{\rm S}+N_{\rm B}}}\;,
\label{eqn:significance}
\end{equation}
where $N_{\rm S}$ and $N_{\rm B}=N_{\rm B_1}+N_{\rm B_2}$ are the event numbers of signal and backgrounds, respectively. They are given by
\begin{eqnarray}
N_{\rm S}= \sigma_0 ~S_{\ell_1\ell_2} \times \epsilon_{\rm S} \times  \mathcal{L}\;,~
N_{\rm B_1}= \sigma_{{\rm B}_1} \times \epsilon_{{\rm B}_1} \times  \mathcal{L}\;,~
N_{\rm B_2}= \sigma_{{\rm B}_2} \times \epsilon_{{\rm B}_2} \times  \mathcal{L}\;,
\end{eqnarray}
where $\epsilon_{\rm S,B_{1(2)}}$ represent the efficiencies of the above cuts and $\mathcal{L}$ denotes the integrated luminosity. From Eq.~\eqref{eq:lumi}, we can get the corresponding integrated luminosity of the muon collider for different c.m. energies
\begin{align}
\sqrt{s} &= ~3,~10,~~30~\left[{\rm TeV}\right] \;,\nonumber\\
\mathcal{L} &= ~1,~10,~~90~\left[{\rm ab}^{-1}\right] \;.
\label{eqn-lumi}
\end{align}
We obtain the 2$\sigma$ exclusion limits to $S_{\ell_1,\ell_2}$ with $\ell_1,\ell_2=e,\mu$ versus $m_N$ at muon colliders with $\sqrt{s}=3$ (red), 10 (green) and 30 (blue) TeV, as shown in solid lines in Fig.~\ref{fig-VmN2}.

The quantity $S_{\ell\ell}$ is equal to the commonly used $|V_{\ell N}|^2$ under the assumption of only one non-vanishing flavor element for the matrix $V_{\ell N}$. Under this assumption, we use $|V_{\mu N}|^2$ or $|V_{eN}|^2$ as the frame labels of the y-axis in the upper and lower-left panels of Fig.~\ref{fig-VmN2}.
One can see that, for $m_N$ well below $\sqrt{s}$, the value of $|V_{\mu N}|^2$ can be probed as low as $4\times 10^{-3}$ ($3\times 10^{-4}$) [$5\times 10^{-5}$] for $\sqrt{s}=3~(10)~[30]$ TeV. This potential is worse than
that through $\mu^+\mu^-\to N_\mu \bar{\nu}_\mu$ channel~\cite{Kwok:2023dck,Li:2023tbx}, because the latter happens via a t-channel exchange of a $W$ boson with huge production cross section. The projected sensitivity bound of $|V_{e N}|^2$ is similar to that of $|V_{\mu N}|^2$. The difference comes from the efficiency of identifying $e$ or $\mu$ from the collider simulation. However, this $|V_{e N}|^2$ exclusion
limit is stronger than that through $\mu^+\mu^-\to N_e \bar{\nu}_e$ channel for $\sqrt{s}=10$ TeV. This is because the electron HNL can only be produced from $\mu^+\mu^-$ annihilation via s-channel and its production cross section is exceeded by VBS production at high energies. Moreover, this LNV process through VBS can provide a clean signal of $e^\pm\mu^\pm$ and an exclusion limit on quantity $S_{e\mu}$ as shown in the lower-right panel of Fig.~\ref{fig-VmN2}.

\section{Lepton number violation through VBS at $\mu^+\mu^+$ collider}
\label{sec:HNLt}

\subsection{The probe of heavy neutral lepton}

\begin{figure}[h!]
\begin{center}
\includegraphics[width=4.5cm]{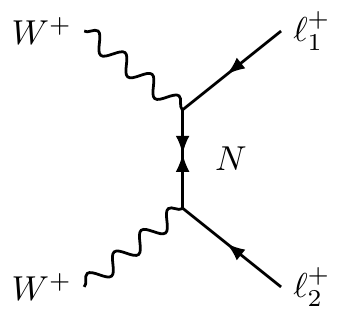}
\end{center}
\caption{The representative Feynman diagram of process $W^+W^+ \to \ell_1^+\ell_2^+$ at the $\mu^+\mu^+$ collider. There is another u-channel diagram by exchanging the initial $W$ bosons.
}
\label{fig-FD2}
\end{figure}

In this section we consider the LNV signature through VBS at same-sign muon collider
\begin{eqnarray}
W^+W^+\to \ell^+\ell^+\;.
\label{eqn-ll}
\end{eqnarray}
This VBS process is mediated by either light neutrino mass eigenstates $\nu_m$ or the HNL $N_{m'}$ in t-channel and is analogous to neutrinoless double-$\beta$ decay. Their amplitudes are proportional to $m_{\nu_m}$ and $m_{N_{m'}}$, respectively. We thus only consider the process mediated by one HNL $N$ in t-channel as shown by the Feynman diagram in Fig.~\ref{fig-FD2}.

\begin{figure}[h!]
\begin{center}
\minigraph{7.4cm}{-0.05in}{}{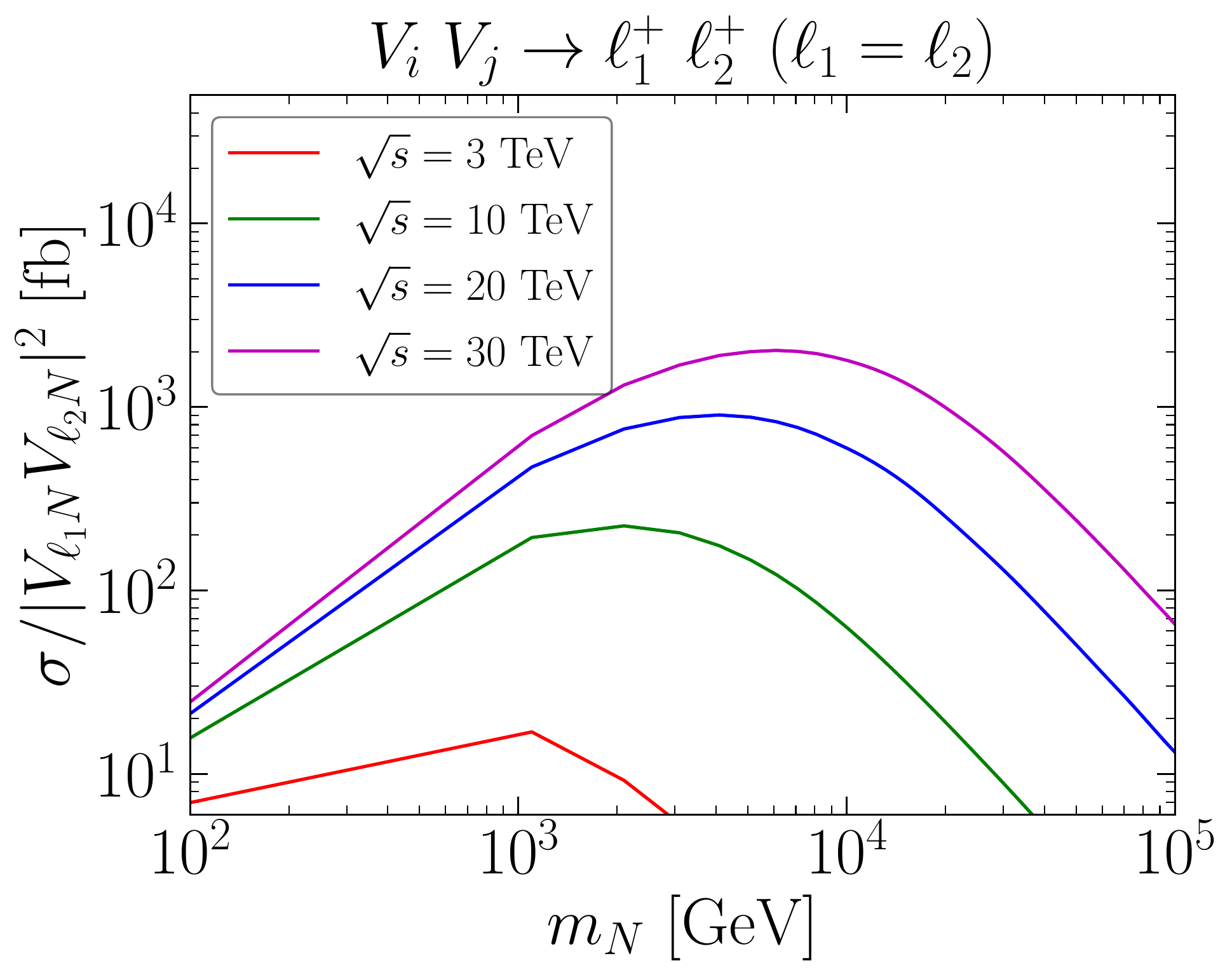}
\minigraph{7.4cm}{-0.05in}{}{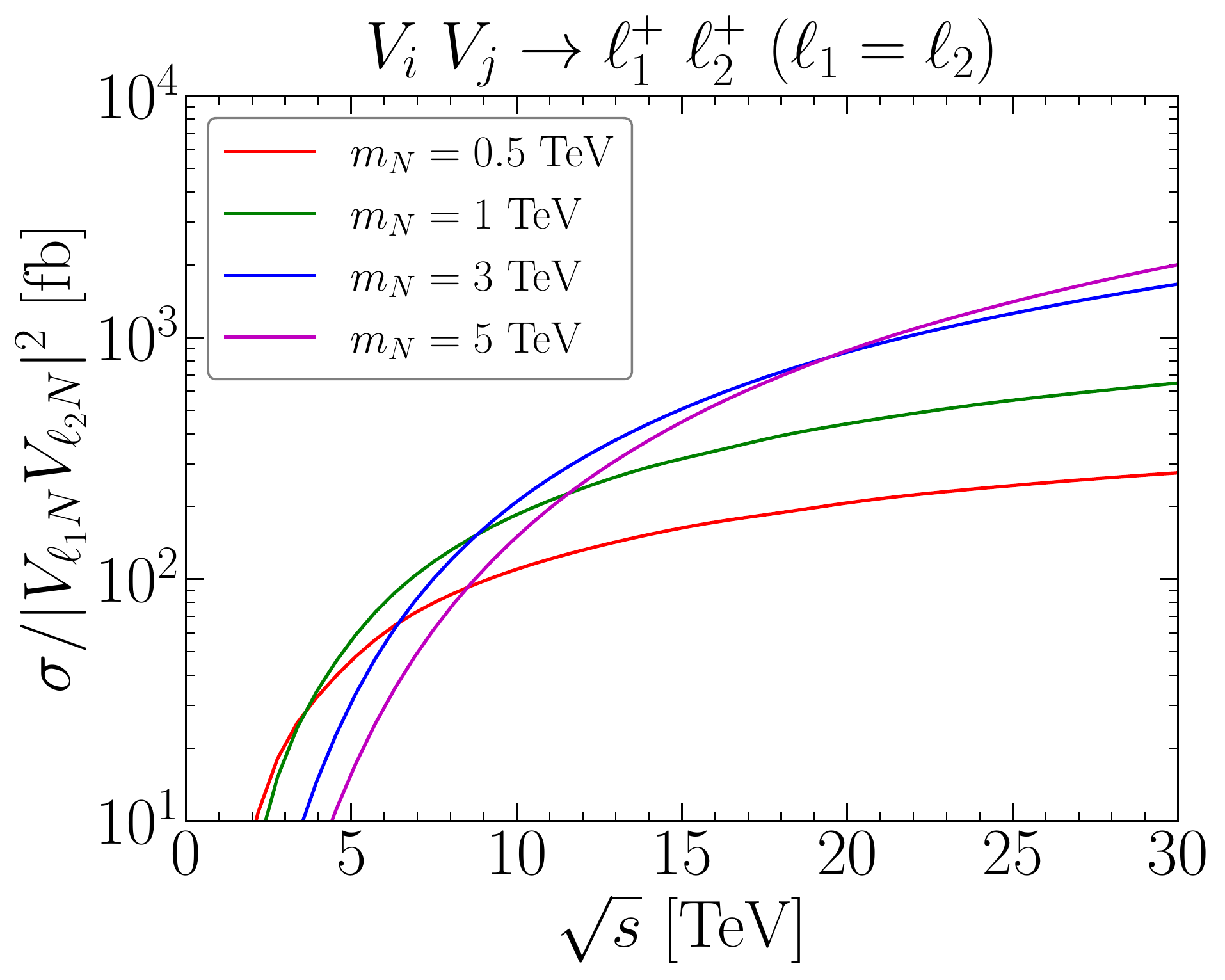}
\end{center}
\caption{The parameter-independent cross section $\sigma_0=\sigma/|V_{\ell_1 N} V_{\ell_2 N}|^2$ for the VBS process $W^+ W^+ \to \ell_1^+\ell_2^+$ with $\ell_1=\ell_2$, as a function of the HNL mass $m_N$ (left) and c.m. energy $\sqrt{s}$ (right) at muon colliders. The benchmark c.m. energy of the muon collider are $\sqrt{s}=3,~10,~20$ and 30 TeV, the benchmark HNL masses are $m_N=0.5,~1,~3$ and 5 TeV.
}
\label{fig-xsc2}
\end{figure}

\begin{figure}[h!]
\begin{center}
      \minigraph{3.6cm}{-0.05in}{}{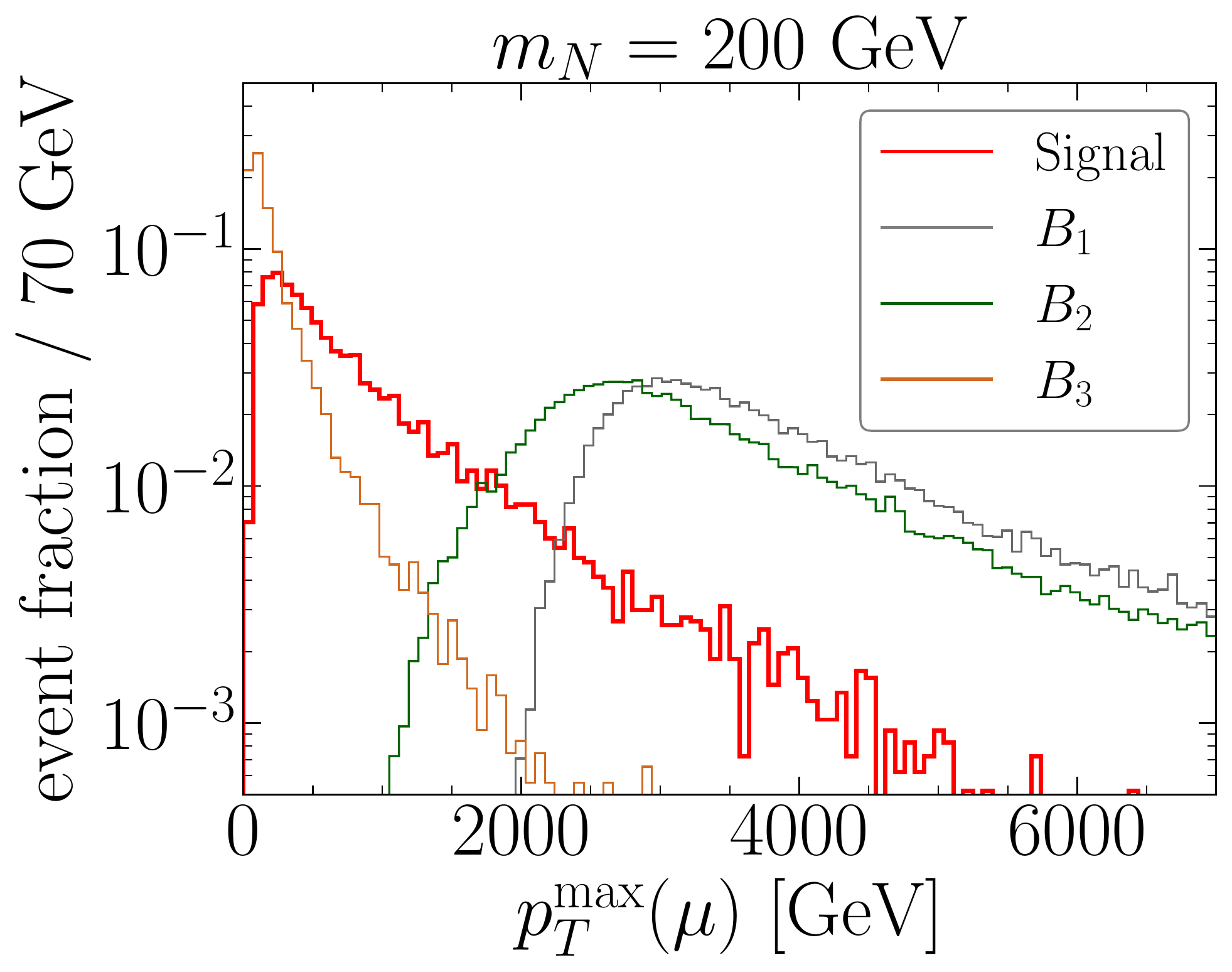}
      \minigraph{3.6cm}{-0.05in}{}{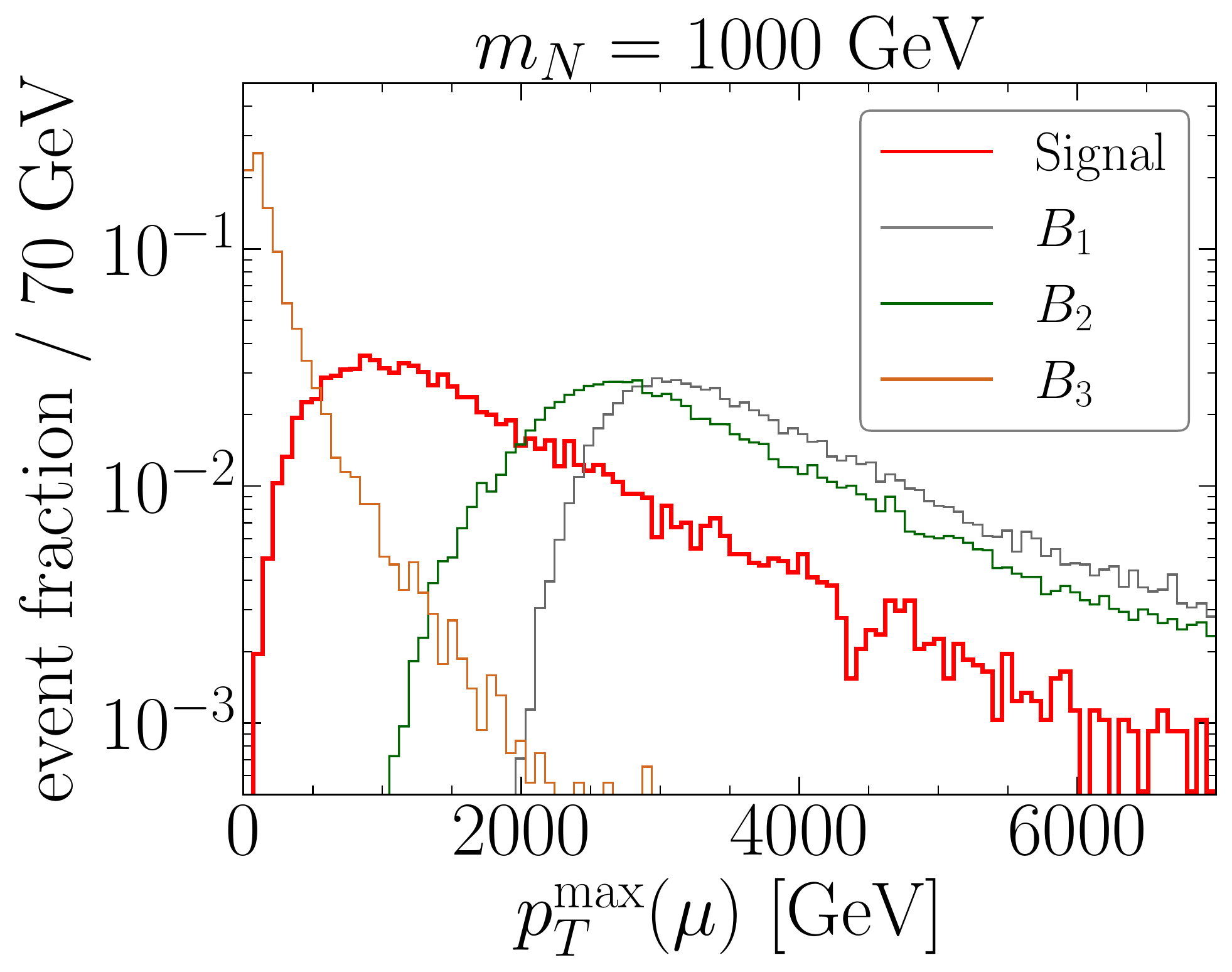}
      \minigraph{3.6cm}{-0.05in}{}{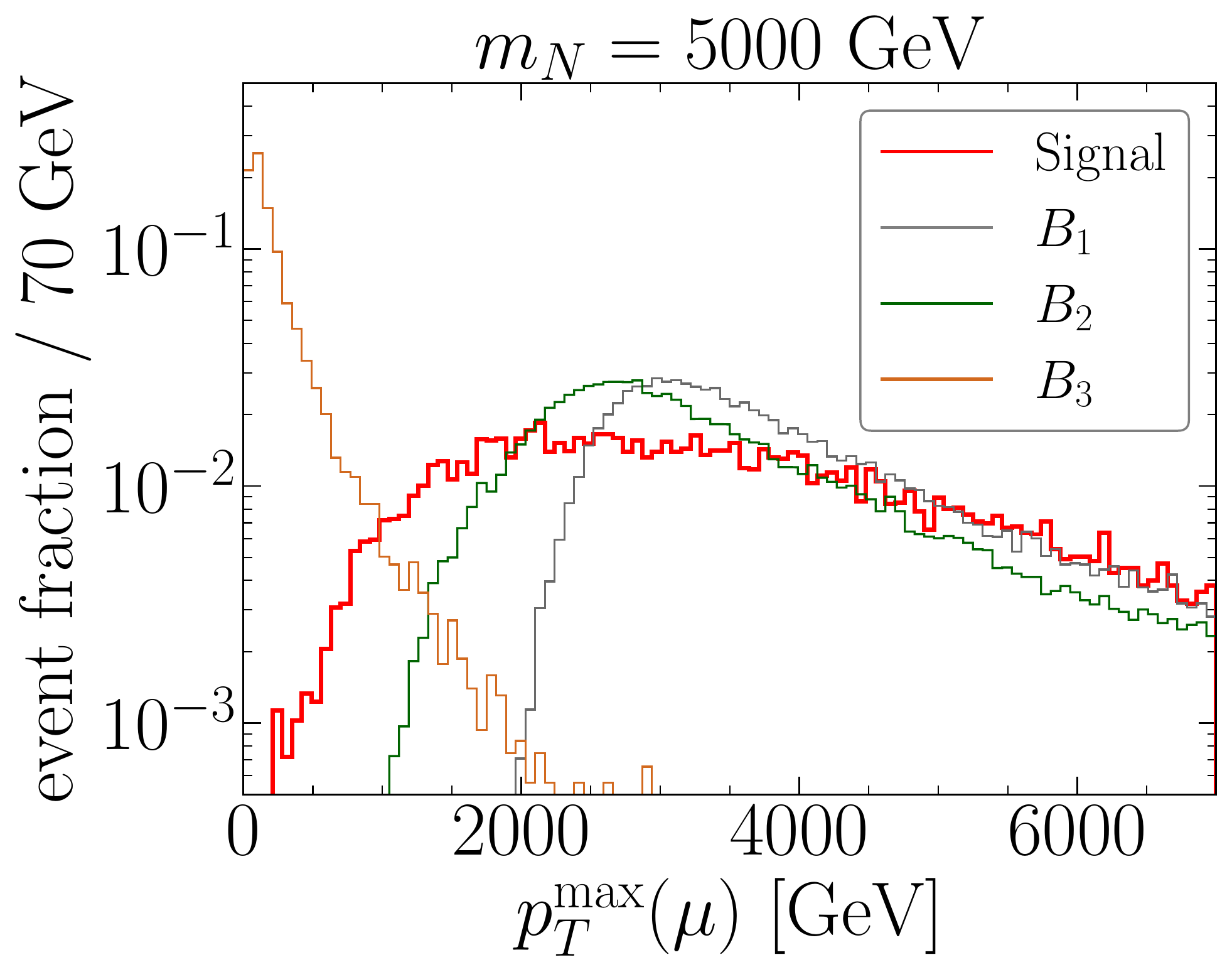}
      \minigraph{3.6cm}{-0.05in}{}{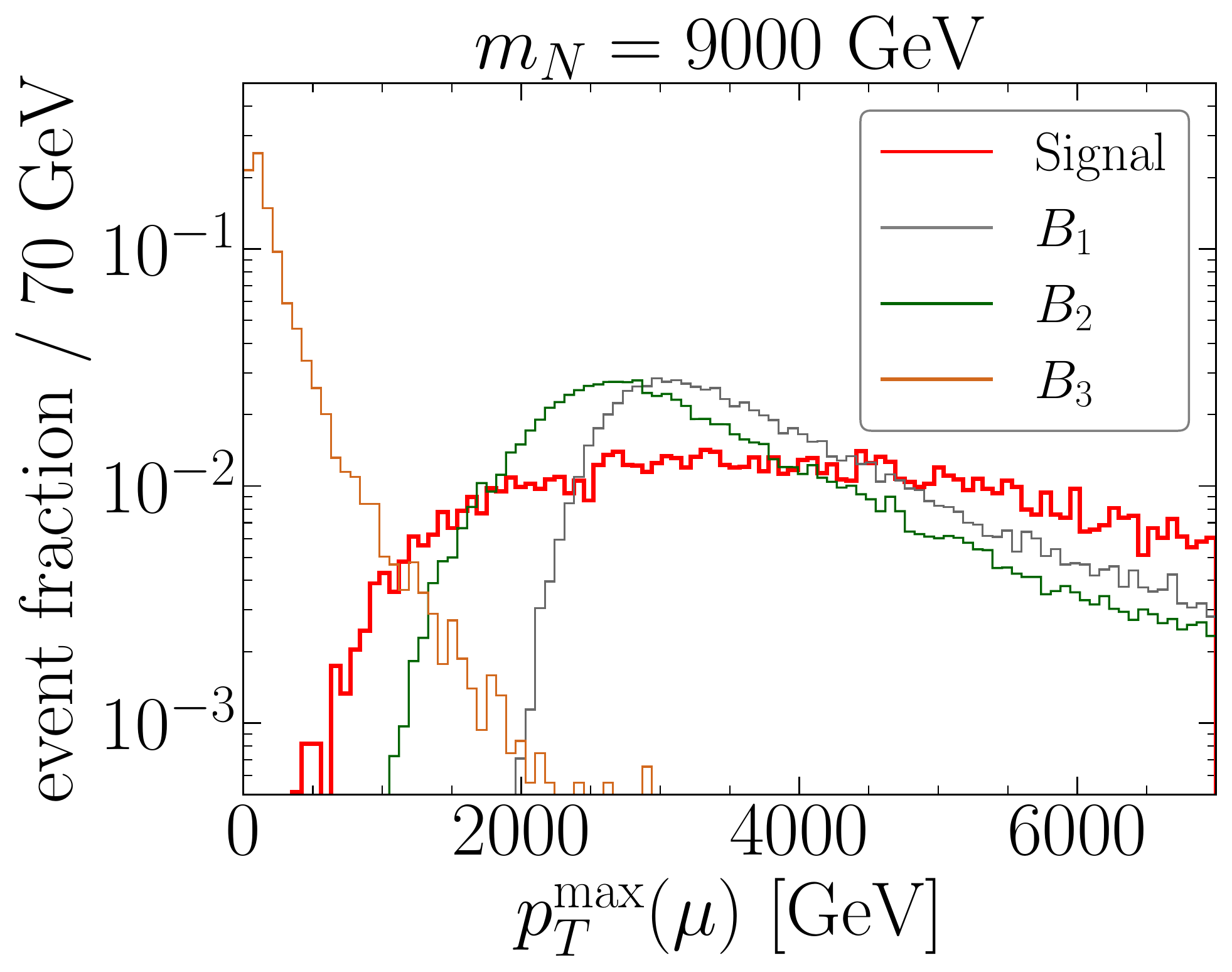}\\
      \minigraph{3.6cm}{-0.05in}{}{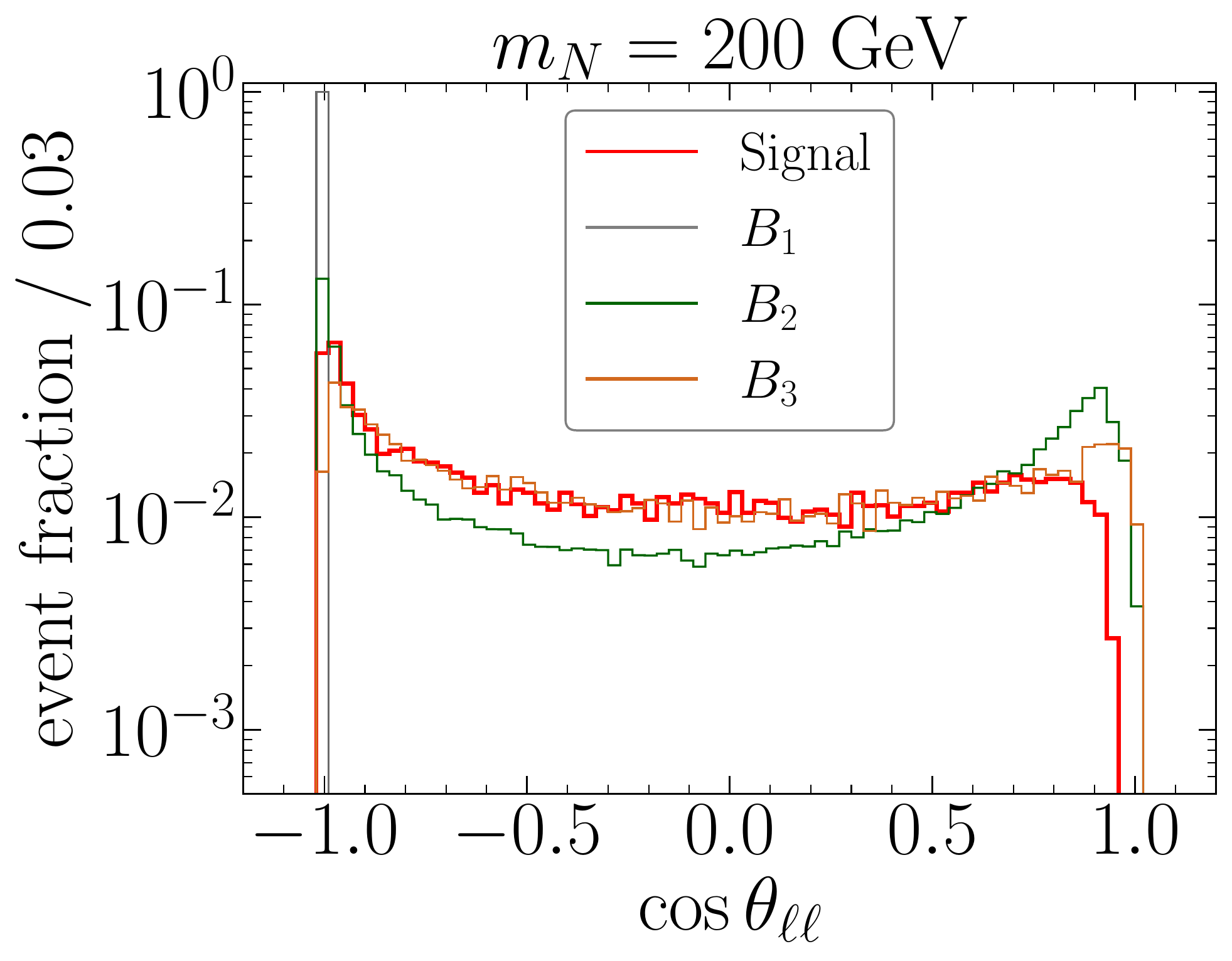}
      \minigraph{3.6cm}{-0.05in}{}{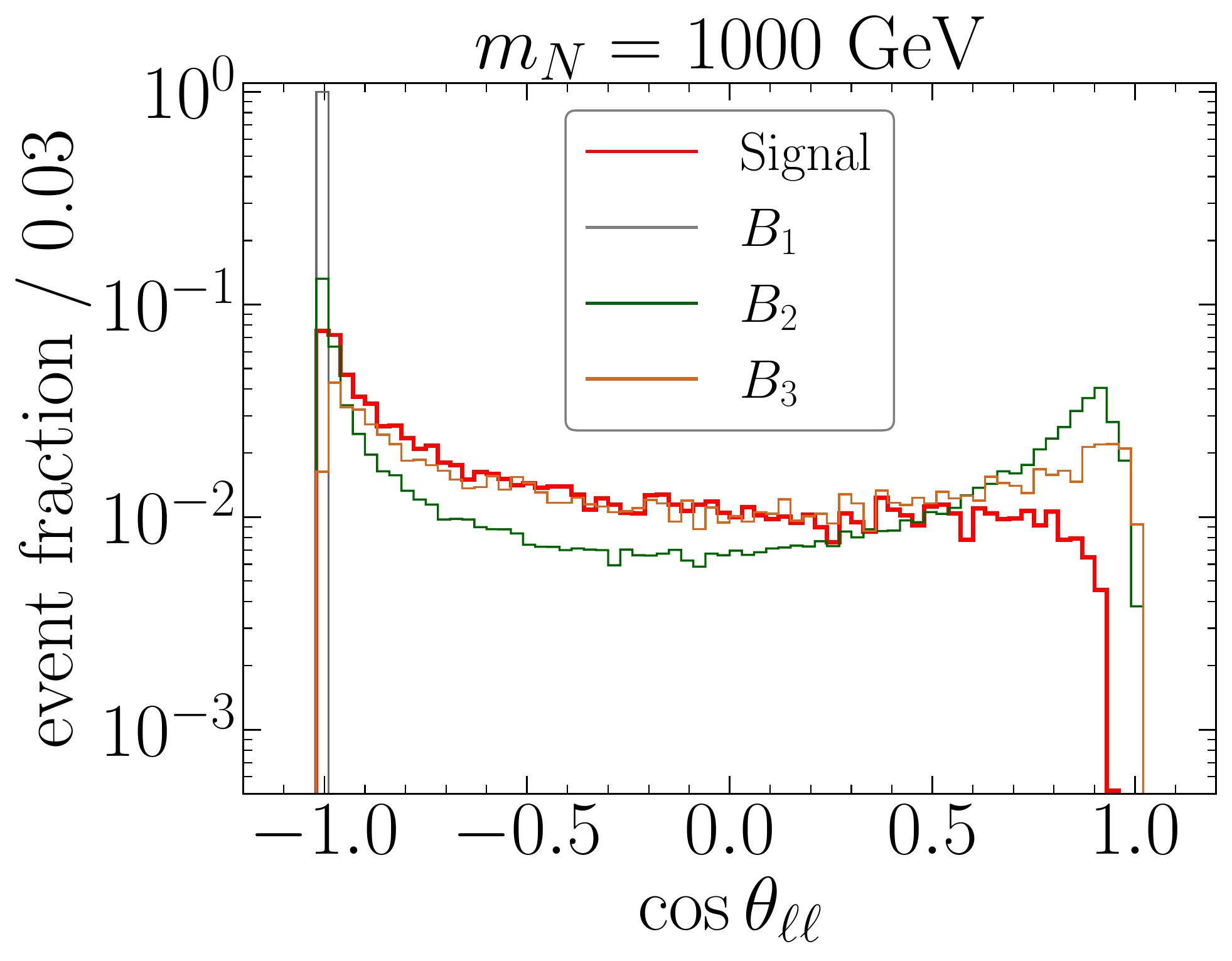}
      \minigraph{3.6cm}{-0.05in}{}{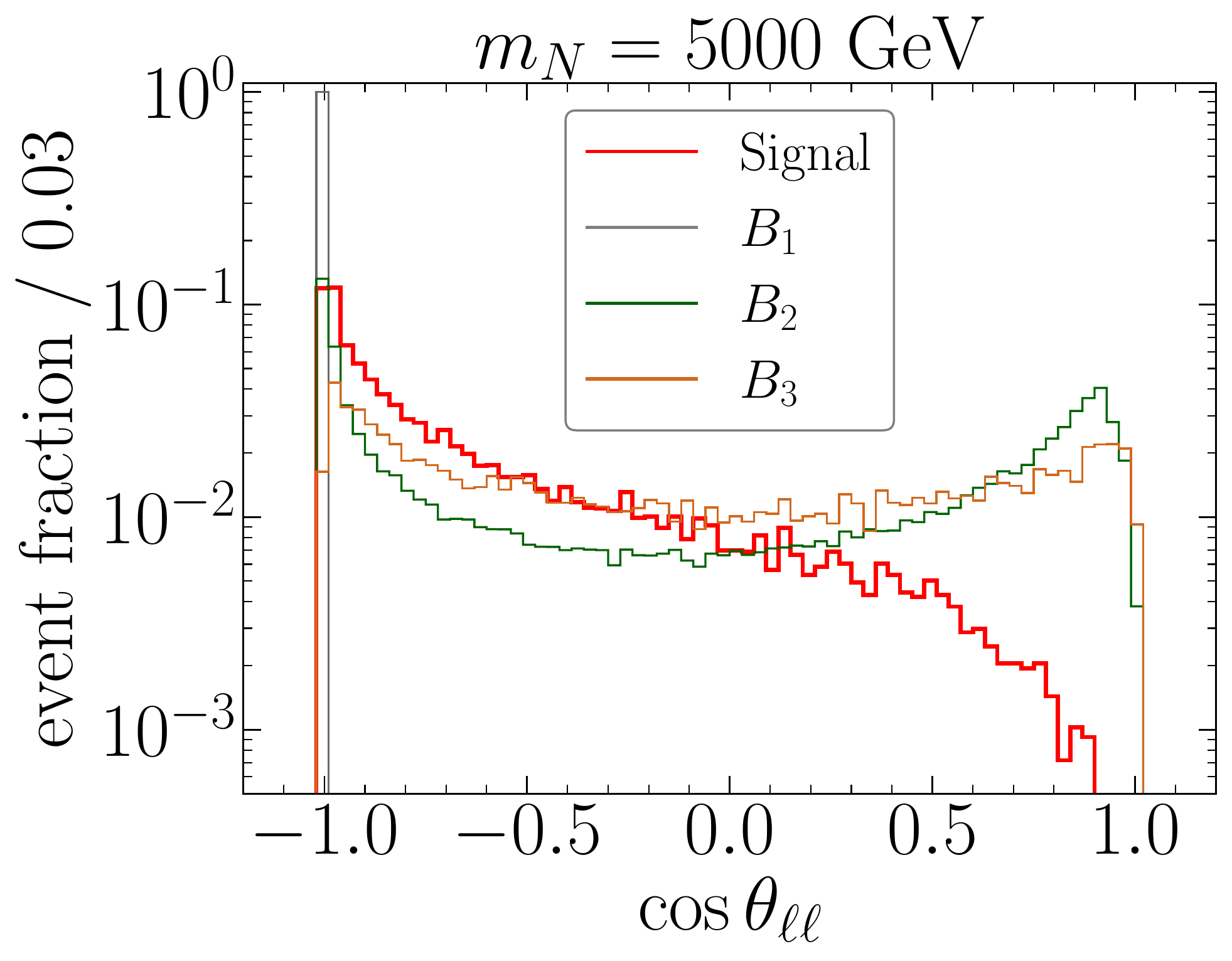}
      \minigraph{3.6cm}{-0.05in}{}{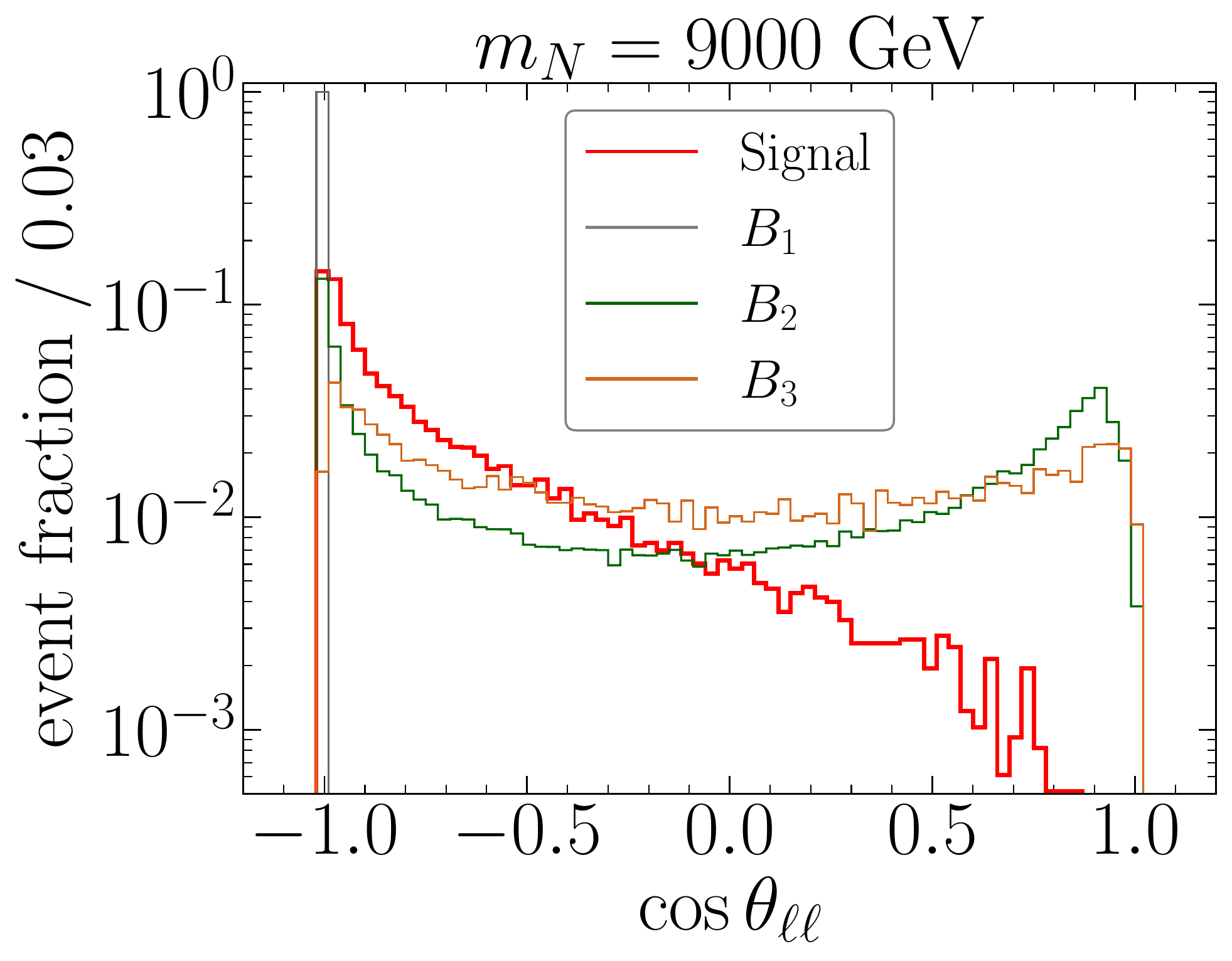}\\
      \minigraph{3.6cm}{-0.05in}{}{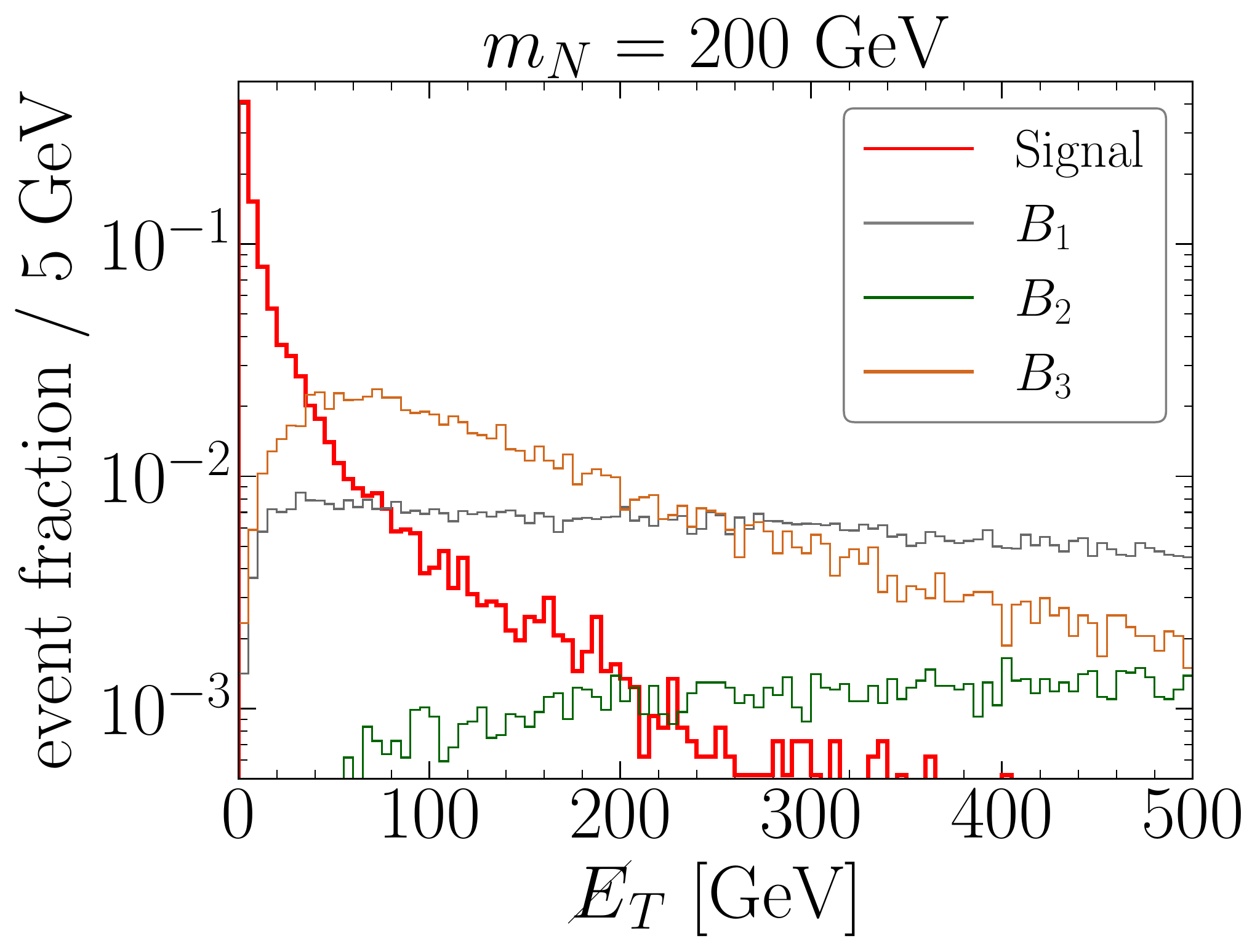}
      \minigraph{3.6cm}{-0.05in}{}{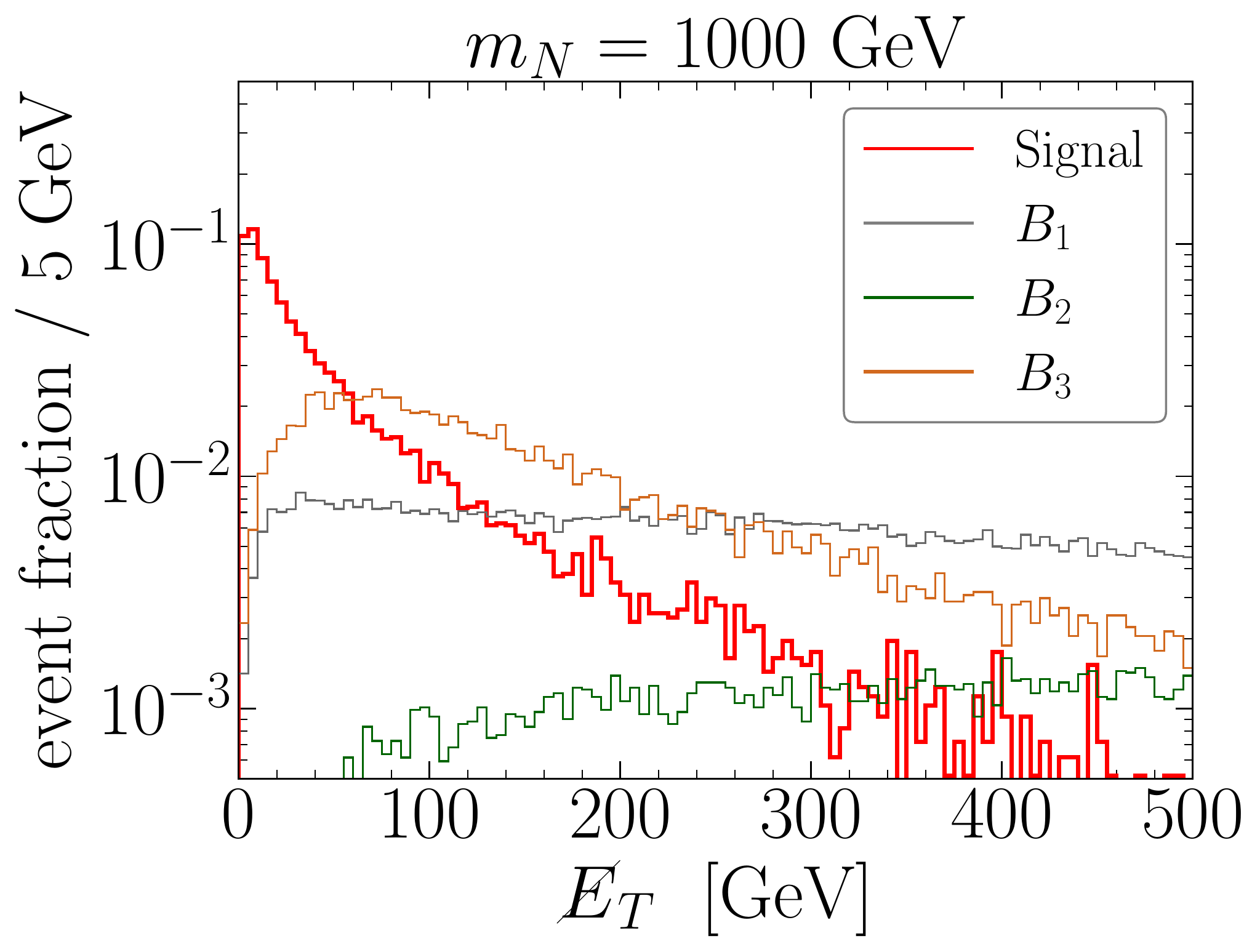}
      \minigraph{3.6cm}{-0.05in}{}{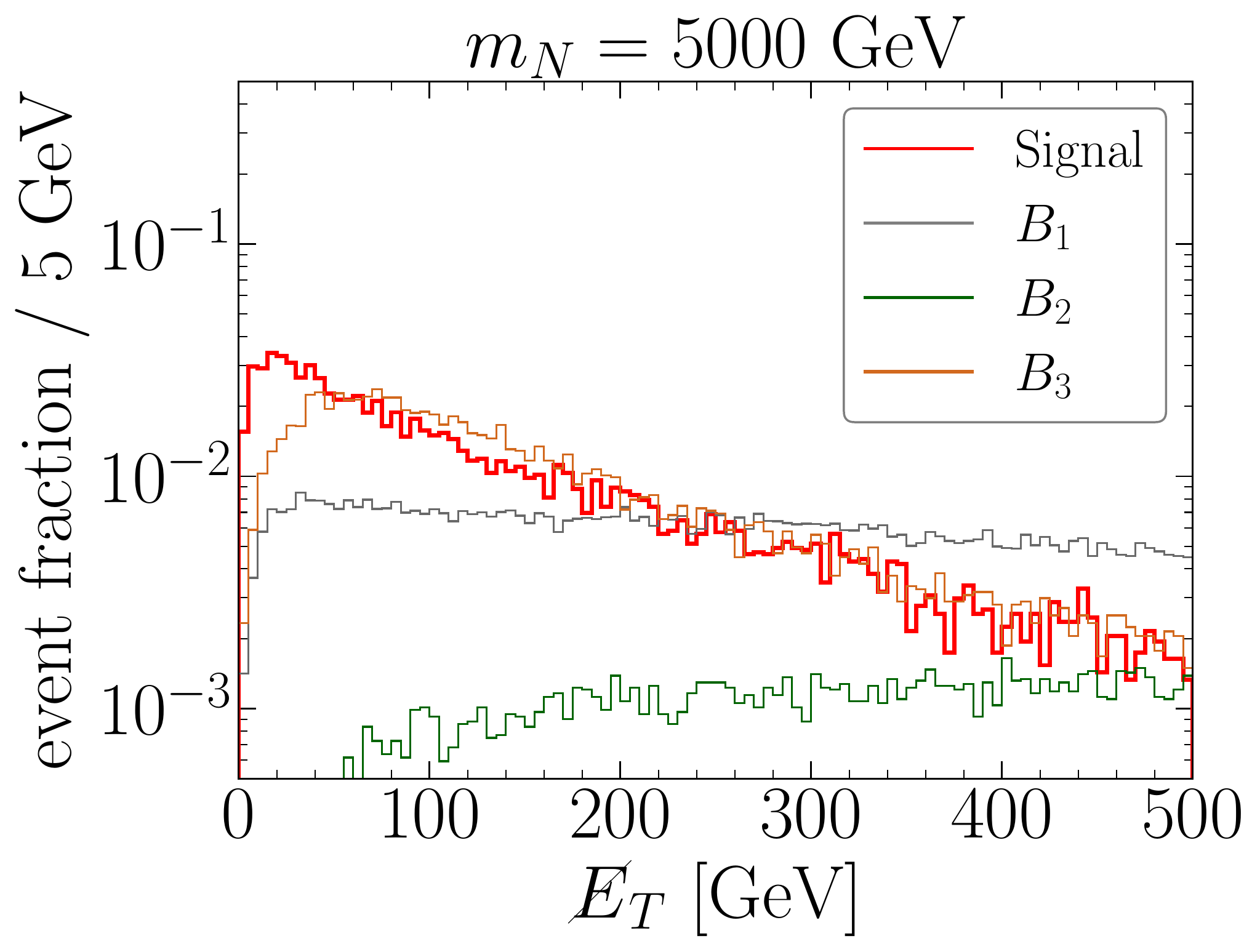}
      \minigraph{3.6cm}{-0.05in}{}{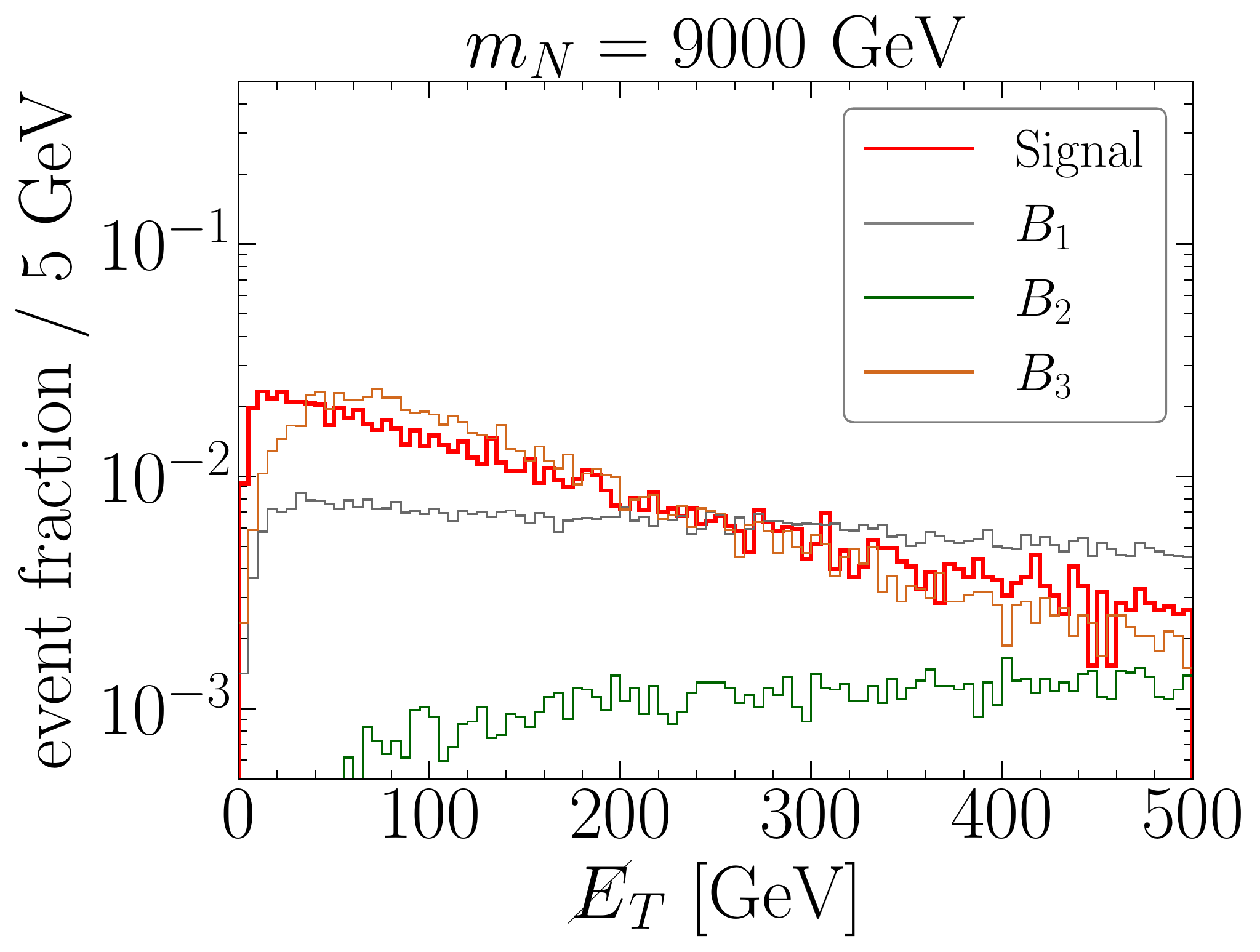}\\
      \minigraph{3.6cm}{-0.05in}{}{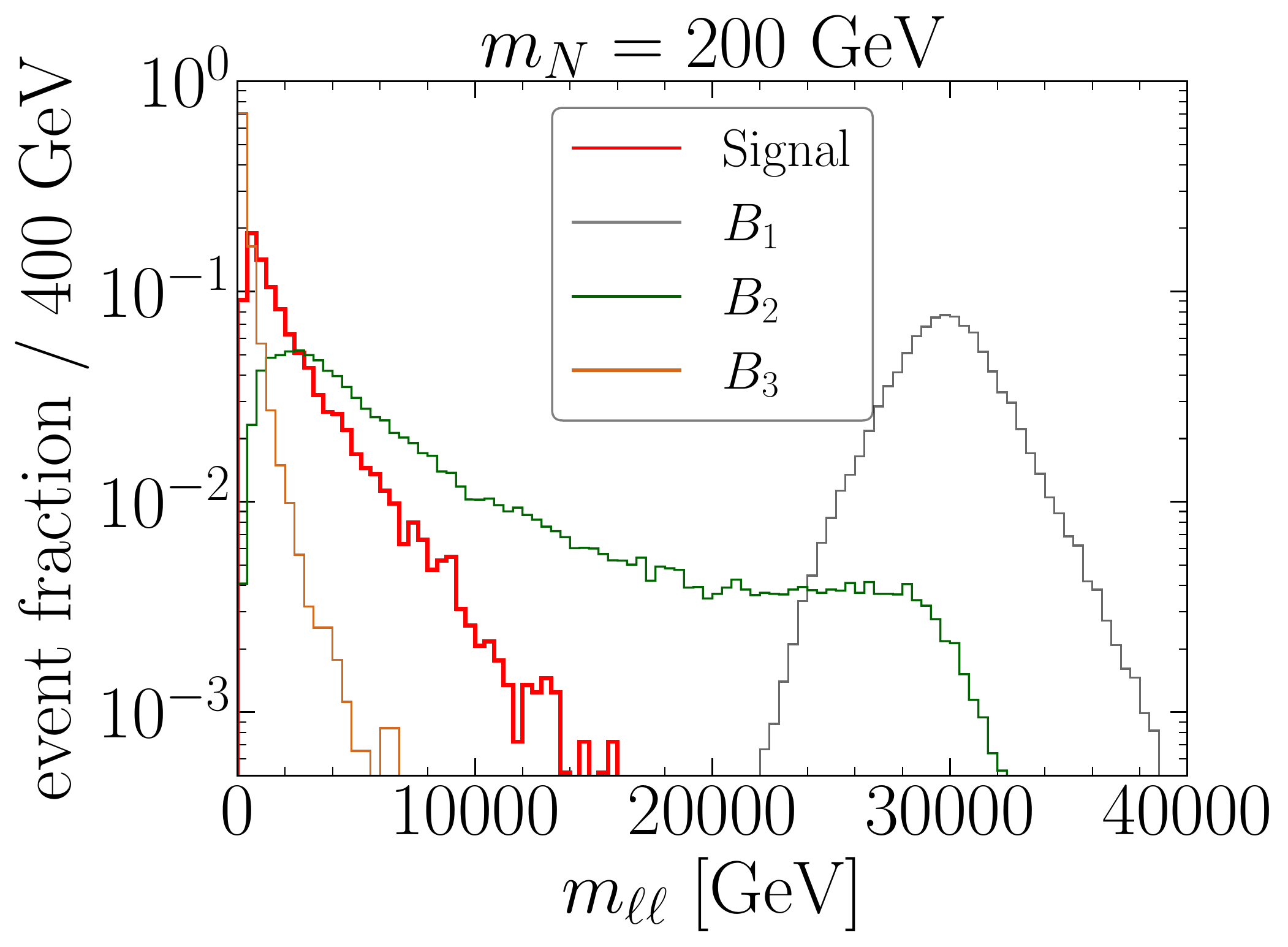}
      \minigraph{3.6cm}{-0.05in}{}{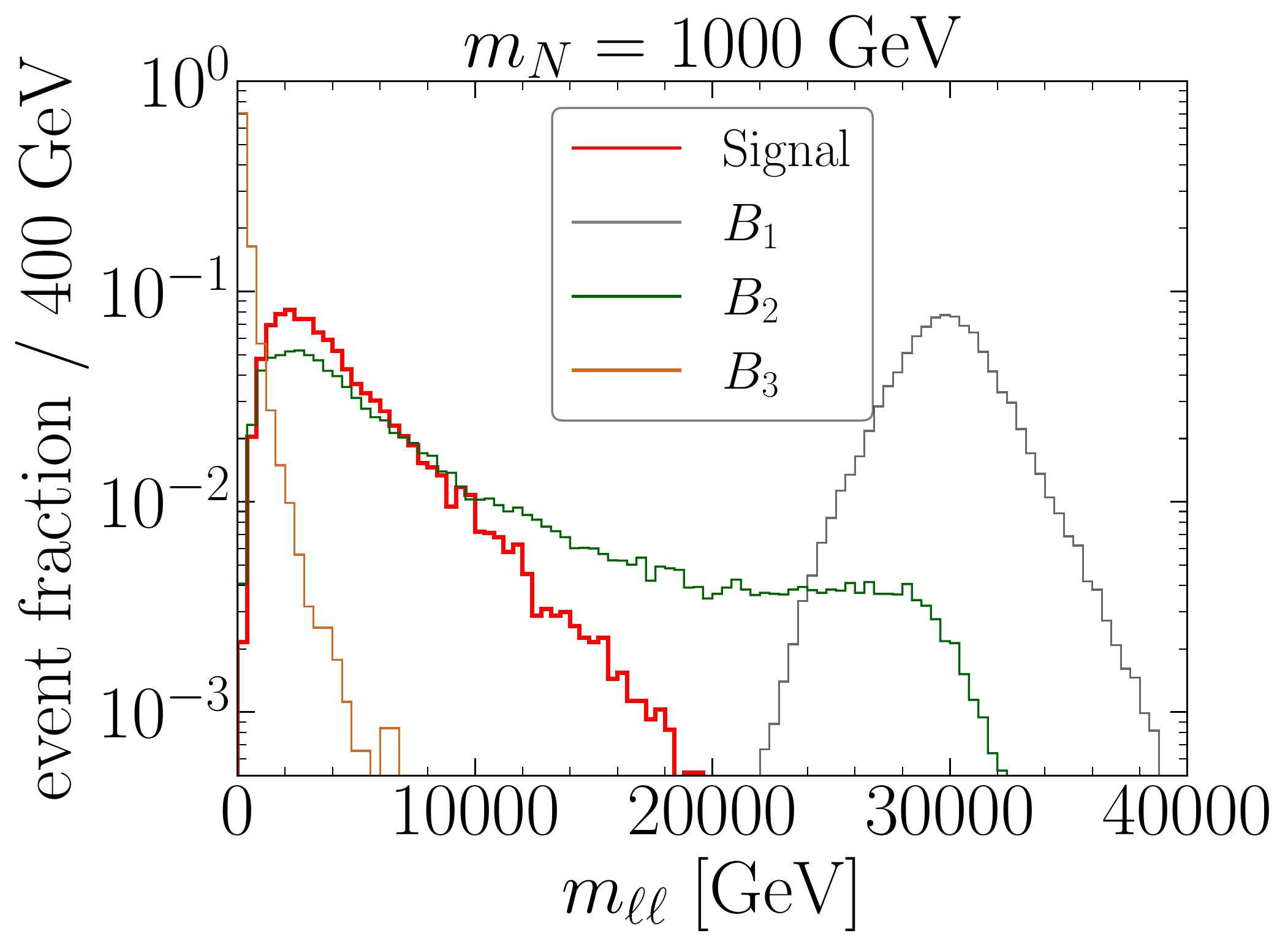}
      \minigraph{3.6cm}{-0.05in}{}{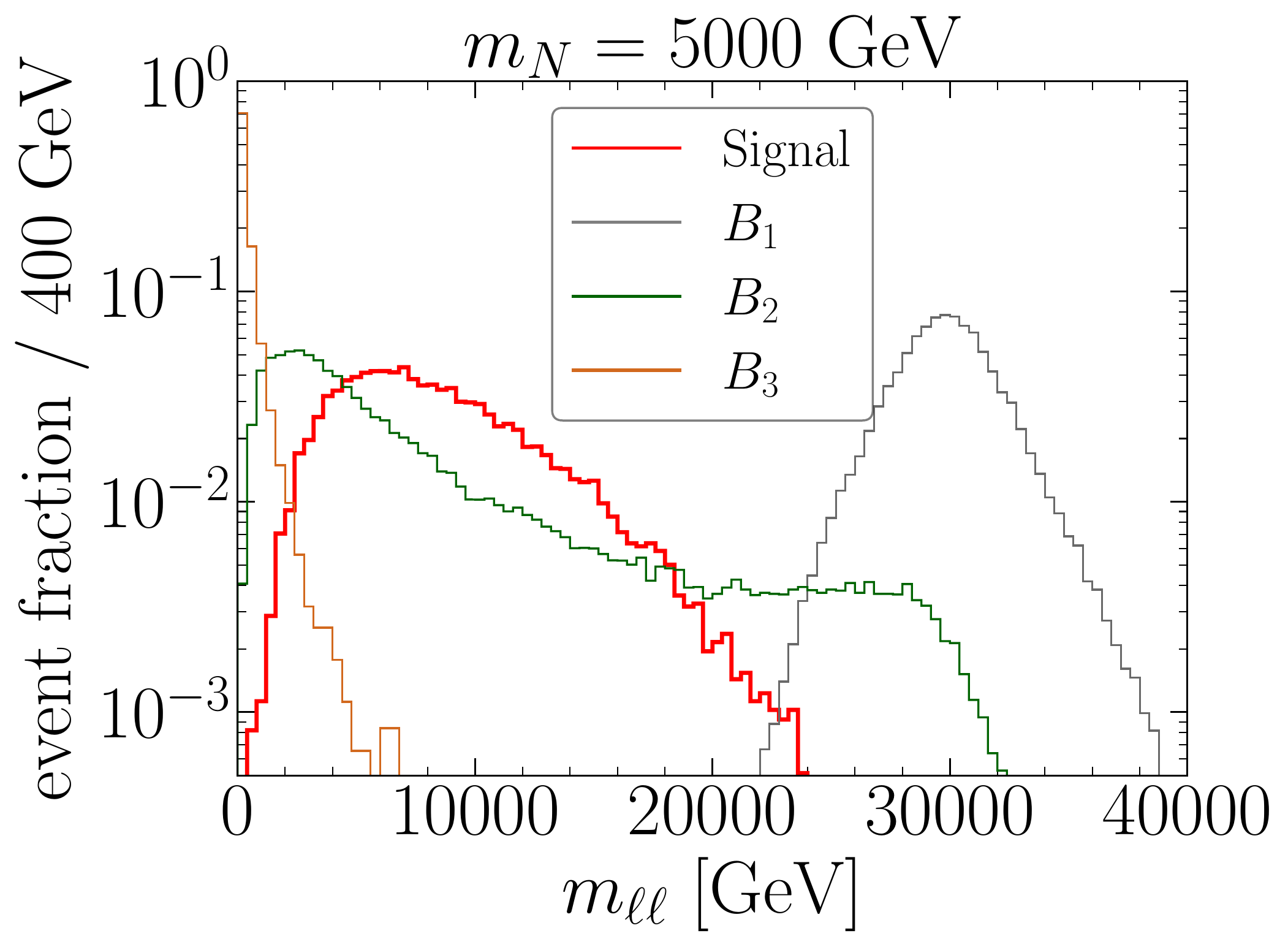}
      \minigraph{3.6cm}{-0.05in}{}{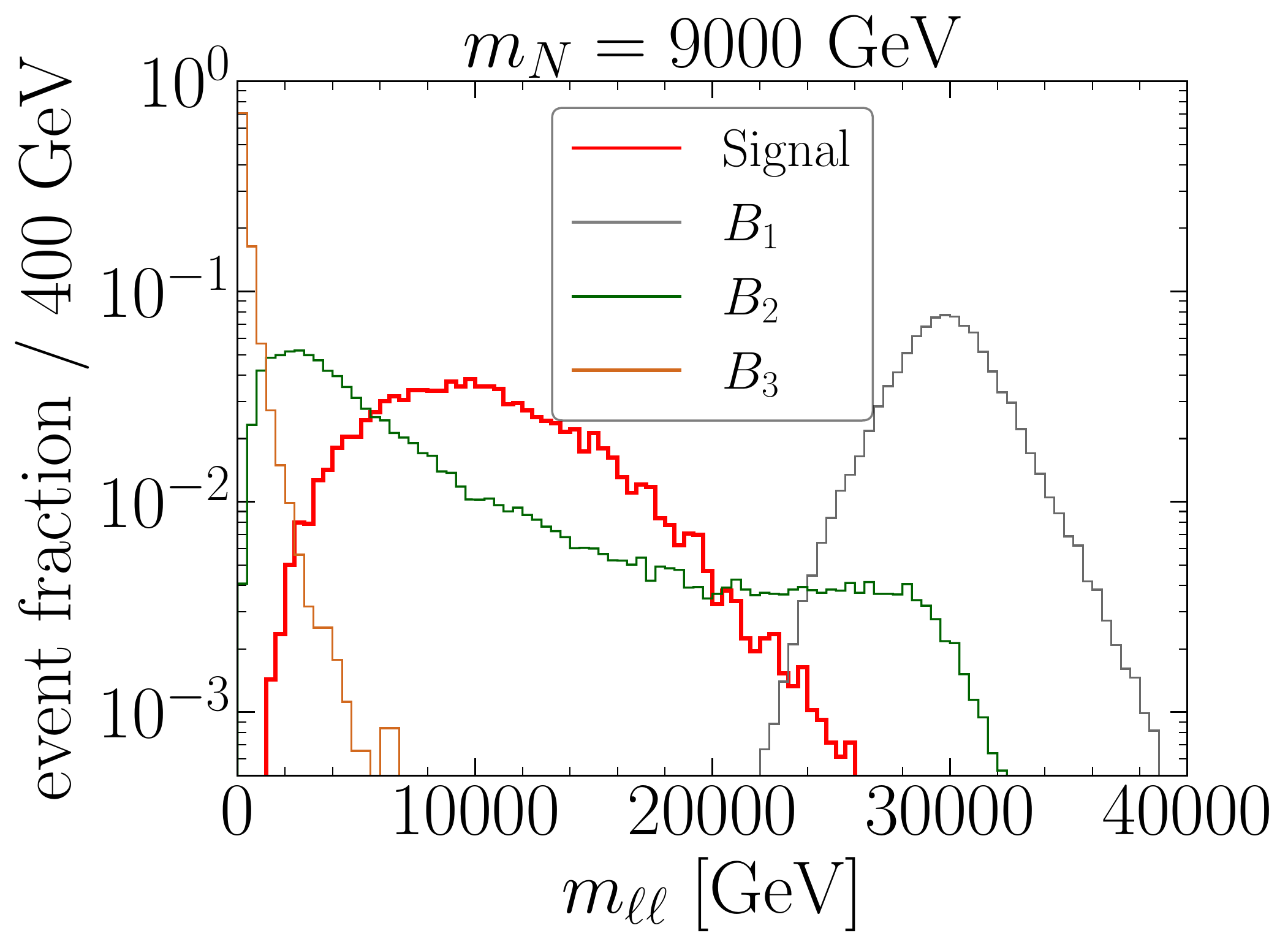}
\end{center}
\caption{The normalized distributions of the transverse momentum $p_T^{\rm max}(\mu)$ (row 1), $\cos\theta_{\ell\ell}$ (row 2), missing energy $\cancel{E}_T$ (row 3) and the invariant mass $m_{\ell\ell}$ (row 4) for the signal and the backgrounds.
The benchmark masses of HNL are $m_N=200,~1000,~5000$ and $9000$ GeV from left to right for $\sqrt{s}=$ 30 TeV. }
\label{fig-SB3}
\end{figure}

The cross section of this VBS process can be expressed as
\begin{equation}
\sigma ~(W^+W^+ \to \ell_1^+ \ell_2^+  ) \equiv |V_{\ell_1 N} V_{\ell_2 N}|^2 \times \sigma_0  \;,
\label{eqn-VBSxesc2}
\end{equation}
where $\sigma_0$ is the ``bare'' cross section independent of the mixing parameters. The values of cross section $\sigma_0$ for $\ell_1=\ell_2$ are shown in Fig.~\ref{fig-xsc2} which are one-half of those for $\ell_1\neq \ell_2$ because of the average of identical final states.
This LNV signal is purely composed of two same-sign charged leptons. We consider three primary SM backgrounds
\begin{align}
& {\rm B_1}:~~ \mu^+ \mu^+ \to \mu^{+} \mu^{+}  \nonumber \;,\\
& {\rm B_2}:~~ \mu^+ \mu^+ \to \mu^{+} \mu^{+}  \nu~ \bar{\nu} \nonumber \;, \\
& {\rm B_3}:~~ \mu^+\mu^+\to W^+W^+ \bar{\nu}_\mu~ \bar{\nu}_\mu \to \ell^+ \ell^+  \nu_{\ell} ~\nu_{\ell}~ \bar{\nu}_\mu~ \bar{\nu}_\mu \nonumber \;,
\label{eqn-bkg2}
\end{align}
where $\nu~(\Bar{\nu})$ donates the summation of neutrinos (anti-neutrinos) with three flavors.
The backgrounds ${\rm B_1}$ and ${\rm B_2}$ only appear for the situation with $\ell_1=\ell_2=\mu$, and we also take this case as an example to illustrate the distributions of signal and backgrounds in Fig.~\ref{fig-SB3}. Corresponding to other flavor combinations of charged leptons in our signal, only background ${\rm B_3}$ applies. According to the different kinematic distributions of signal and backgrounds, we adopt a series of selection cuts to suppress the backgrounds and enhance the significance.

\begin{itemize}
\item Firstly, we require the number of same-sign muons in the generated events, and employ some basic cuts for the muons in final states
\begin{equation}
N_{\mu} \geq 2, ~~p_T(\mu_1,\mu_2) > 50 ~{\rm GeV},~~|\eta(\mu_1)|,|\eta(\mu_2)| < 2.5, ~~\Delta R_{\mu_1 \mu_2} > 0.5\;.
\label{eqn-cut5}
\end{equation}
\item The leading transverse momentum $p_T^{\rm max}(\mu)$ of the final muons is shown in the first row of Fig.~\ref{fig-SB3}. We employ some relatively conservative kinematic cuts
\begin{equation}
p_T^{\rm max}(\mu) ~ >~ 100~(200)~[500]~ {\rm GeV}\;,
\label{eqn-cut6}
\end{equation}
for $\sqrt{s}=3~(10)~[30]$ TeV. They help to reduce the background ${\rm B_3}$ in which the charged leptons are from two $W$ bosons' leptonic decay.
\item The cosine of the angle between two muons in final states $\cos\theta_{\ell\ell}$ exhibits very different behaviors between signal and SM backgrounds, as shown in the second row of Fig.~\ref{fig-SB3}.
Thus, we employ the following cut
\begin{equation}
\cos\theta_{\ell\ell} < 0.2\;.
\label{eqn-cut7}
\end{equation}
\item For backgrounds ${\rm B_2}$ and ${\rm B_3}$, the presence of neutrinos leads to missing energy, as shown in the third row of Fig.~\ref{fig-SB3}. We can employ the cut for the missing energy to reduce the backgrounds~\footnote{The cut on the $p_T$ of the sub-leading charged lepton could also help to reduce the backgrounds at the LHC~\cite{ATLAS:2023tkz}. We find that it is worse than the missing energy cut here.}
\begin{equation}
\cancel{E}_T < 20~(50)~[150] ~{\rm GeV}\;,
\label{eqn-cut8}
\end{equation}
for $\sqrt{s}=3~(10)~[30]$ TeV.
\item The final muons in background ${\rm B_1}$ carry all the collision energy. Its invariant mass leads to a significantly different distribution from the signal and other backgrounds, as shown in the last row of Fig.~\ref{fig-SB3}. According to this feature, we can employ the following invariant mass cut to extremely reduce the background ${\rm B_1}$
\begin{align}
m_{\ell \ell}~ &<~ 2000~(5000)~[15000]~ {\rm GeV} \;,
\label{eqn-cut9}
\end{align}
for $\sqrt{s}=3~(10)~[30]$ TeV.
\end{itemize}
We take the case of $\sqrt{s}=30$ TeV for muon collider to illustrate the cross sections of the signal and backgrounds after the above series of cuts, as shown in Table~\ref{tab-SB3}.

\begin{table}[htb!]
\centering
\footnotesize
\begin{tabular}{|c|c|c|c|c|c|c|}
\hline
\hline
    \multirow{2}*{sig. and bkgs}   & \textcolor{blue}{$\sigma/|V_{\mu N}|^4$}  or  & $N_{\mu}$~and basic cuts  & $p_T^{\rm max}(\mu)$ & ${\rm cos}~ \theta_{\ell\ell}$ &  $\cancel{E}_T$ & $m_{\ell \ell}$ \\
    &  $\sigma_{\rm B_i}$ [pb]  & in Eq.~\eqref{eqn-cut5} & in Eq.~\eqref{eqn-cut6} & in Eq.~\eqref{eqn-cut7} &  in Eq.~\eqref{eqn-cut8} & in Eq.~\eqref{eqn-cut9}\\
    \hline
    \hline
    $m_N=200$ GeV &  \textcolor{blue}{0.072} &  \textcolor{blue}{0.070}&  \textcolor{blue}{0.041} &  \textcolor{blue}{0.030}&  \textcolor{blue}{0.027} & \textcolor{blue}{0.026} \\
    $m_N=1000$ GeV & \textcolor{blue}{0.64} &  \textcolor{blue}{0.62}&  \textcolor{blue}{0.57} & \textcolor{blue}{0.44} & \textcolor{blue}{0.36} & \textcolor{blue}{0.36} \\
    $m_N=5000$ GeV & \textcolor{blue}{1.99} &  \textcolor{blue}{1.95}&  \textcolor{blue}{1.94} &   \textcolor{blue}{1.77}& \textcolor{blue}{1.01} & \textcolor{blue}{0.97} \\
    $m_N=9000$ GeV & \textcolor{blue}{1.84} &  \textcolor{blue}{1.84}&  \textcolor{blue}{1.83} &   \textcolor{blue}{1.75}&  \textcolor{blue}{0.83} &  \textcolor{blue}{0.76}\\
    \hline
    \hline
    $\rm B_1$& 0.055 & 0.051 & 0.051& 0.051 & 0.011 & 0.0\\
    $\rm B_2$& 0.0046 & 0.0042& 0.0042 & 0.0024 & 4.5$\times 10^{-5}$ & 1.3$\times 10^{-5}$ \\
    $\rm B_3$& 0.011  & 0.0018& 4.1$\times 10^{-4}$ & 2.8$\times 10^{-4}$ & 9.9$\times 10^{-6}$ & 9.9$\times 10^{-6}$ \\
\hline
\hline
\end{tabular}
\caption{The representative signal cross section $\sigma/|V_{\mu N}|^4$ (in blue) and those for SM backgrounds (in black) at muon collider after selection cuts.
For illustration, four benchmarks $m_N=200,~1000,~5000$ and $9000$ GeV are considered at muon collider with $\sqrt{s}=30$ TeV. }
\label{tab-SB3}
\end{table}

We adopt the same significance formula in Eq.~\eqref{eqn:significance} here with $N_{\rm B}=N_{\rm B_1}+N_{\rm B_2}+N_{\rm B_3}$. The $N_{\rm S}$ and $N_{{\rm B}_i}$ ($i=1,2,3$) are given by
\begin{eqnarray}
N_{\rm S}= \sigma_0 ~|V_{\ell_1 N}V_{\ell_2 N}|^2\times \epsilon_{\rm S} \times  \mathcal{L}\;,~
N_{\rm B_i}= \sigma_{{\rm B}_i} \times \epsilon_{{\rm B}_i} \times  \mathcal{L}\;.
\end{eqnarray}
The corresponding integrated luminosity of the muon collider for different c.m. energies are shown in Eq.~\eqref{eqn-lumi}.
We obtain the sensitivity of muon colliders to the mixing parameter $|V_{\ell_1 N}V_{\ell_2 N}|$ for Majorana HNL.
The $2\sigma$ exclusion limits to $|V_{\mu N}|^2$, $|V_{e N}|^2$ and $|V_{e N} V_{\mu N}|$ versus the mass of Majorana HNL are shown in dash-dotted lines in Fig.~\ref{fig-VmN2}.
In contrast to the $\ell_1^\pm \ell_2^\pm +{\rm jet(s)}$ signature losing sensitivity near the energy threshold, this t-channel LNV signal can probe much heavier HNL.

\subsection{The probe of Weinberg operator}

Finally, we comment on the probe of dimension-5 Weinberg operator in the absence of HNL.
The above signature of same-sign charged leptons can also be used to probe the Weinberg operator~\cite{Fuks:2020zbm,ATLAS:2023tkz}
\begin{eqnarray}
{C_5^{\ell \ell'}\over \Lambda} H\cdot \bar{\ell}_L^c \ell'_L\cdot H\;,
\end{eqnarray}
where $\ell_L^T=(\nu_\ell,\ell)$. After EW symmetry breaking and the SM Higgs gains its vev $v$, this Weinberg operator generates a Majorana neutrino mass in the flavor basis
\begin{eqnarray}
m_{\ell\ell'}={C_5^{\ell \ell'}v^2\over \Lambda}\,,
\end{eqnarray}
which appears as the mass of the intermediate fermion in the same-sign $W$ scattering. It was shown in Ref.~\cite{Fuks:2020zbm} that the parton-level cross section of this same-sign $W$ scattering process in this model is
\begin{eqnarray}
\sigma(W^+W^+\to \ell \ell')\approx {2-\delta_{\ell \ell'}\over 18\pi} \Big| {C_5^{\ell \ell'}\over \Lambda}\Big|^2\;.
\label{eq:Wxsec}
\end{eqnarray}
The cross section is proportional to an overall scaling factor of $| C_5^{\ell \ell'}/\Lambda|^2$.

Next we follow the approach in Ref.~\cite{Fuks:2020zbm} to simulate the probe of Weinberg operator for $\ell=\ell'=\mu$ at high-energy muon colliders. It was verified that a reasonable simulation using the UFO for Weinberg operator (called ``SMWeinberg'') has to be performed with $\Lambda/|C_5^{\ell \ell'}|\geq 200$ TeV~\cite{Fuks:2020zbm}.
We adopt the same selection strategies mentioned in last subsection and obtain the number of signal events denoted by $N_{\rm S}^{200}$ for a benchmark with $\Lambda=200$ TeV and $C_5^{\mu\mu}=1$. According to Eq.~(\ref{eq:Wxsec}), the number of signal events is given by
\begin{eqnarray}
N_{\rm S}=N_{\rm S}^{200} |C_5^{\mu\mu}|^2 \Big({200~{\rm TeV}\over \Lambda}\Big)^2\;.
\end{eqnarray}
By requiring $2\sigma$ significance, we can obtain the sensitivity to the scale $\Lambda/|C_5^{\mu \mu}|$. We find that the sensitivity bound of muon collider to the scale is
\begin{eqnarray}
\Lambda/|C_5^{\mu \mu}|\lesssim 9.3~(23.0)~[47.7]~{\rm TeV}\;,
\end{eqnarray}
for $\sqrt{s}=3~(10)~[30]$ TeV. This is equivalent to the limit of effective $\mu\mu$ Majorana mass as
\begin{eqnarray}
m_{\mu\mu}\gtrsim 6.5~(2.6)~[1.3]~{\rm GeV}\;.
\end{eqnarray}

\section{Conclusion}
\label{sec:Con}

In this work, we propose the search for Majorana HNLs and LNV signatures through VBS processes at high-energy muon colliders. At high energies, the muon beams radiate substantial EW gauge bosons and can play as ``gauge boson colliders''. The VBS processes lead to clear LNV signatures so as to tell the nature of Majorana HNLs and thus provide more advantageous benefits than direct $\mu\mu$ annihilation. We use the method of EW PDFs to calculate the VBS production of HNL at both $\mu^+\mu^-$ and $\mu^+\mu^+$ colliders, and analyze the produced LNV signatures. At $\mu^+\mu^-$ collider, the scattering of gauge bosons $W^\pm Z/\gamma$ induces the associated production of HNL $N$ and charged lepton $\ell^\pm$. At same-sign muon collider, the $W^+W^+$ scattering can result in a $0\nu\beta\beta$-like LNV process $W^+W^+\to \ell^+\ell^+$ with the HNL in t-channel.

We perform the detector simulation of LNV signal and SM backgrounds. The search potentials on heavy Majorana neutrino are analyzed and the exclusion limits on the mixing of HNL $N$ and charged lepton $V_{\ell N}$ are obtained. We find the following conclusions:
\begin{itemize}
\item The VBS processes can provide smoking-gun LNV signatures $\ell_1^\pm \ell_2^\pm+{\rm jet(s)}$ and $\ell_1^+\ell_2^+$ at $\mu^+\mu^-$ collider and same-sign muon collider, respectively.
\item The probing potential of $|V_{\mu N}|^2$ through $W^\pm Z/\gamma\to \mu^\pm \mu^\pm+{\rm jet(s)}$ signature is worse than that from annihilation channel $\mu^+\mu^-\to N_\mu \bar{\nu}_\mu$. However, the exclusion limit on $|V_{eN}|^2$ from VBS channel is stronger than that through the $\mu^+\mu^-$ annihilation for $\sqrt{s}=10$ TeV or above.
\item At same-sign muon collider, compared with the $\ell^\pm \ell^\pm +{\rm jet(s)}$ signature, the $0\nu\beta\beta$-like LNV signal can help to probe much heavier HNL. The mixings other than $|V_{\mu N}|$ can be probed through VBS as well, which is contrary to the process $\mu^+\mu^+\to W^+W^+$. Based on this signature, we also estimate the sensitivity of muon collider to the scale of Weinberg operator.
\item The above LNV signatures can both be employed to look for the combinations of different charged lepton flavors and probe the mixing $|V_{\ell_1 N} V_{\ell_2 N}|$ with $\ell_1\neq \ell_2$.
\end{itemize}

\acknowledgments
We would like to thank Richard Ruiz for useful discussion.
T.L. is supported by the National Natural Science Foundation of China (Grants No. 11975129, 12035008) and ``the Fundamental Research Funds for the Central Universities'', Nankai University (Grant No. 63196013). C.Y.Y. is supported in part by the Grants No. NSFC-11975130, No. NSFC-12035008, No. NSFC-12047533, the Helmholtz-OCPC International Postdoctoral Exchange Fellowship Program, the National Key Research and Development Program of China under Grant No. 2017YFA0402200, the China Postdoctoral Science Foundation under Grant No. 2018M641621, and the Deutsche Forschungsgemeinschaft (DFG, German Research Foundation) under Germany's Excellence Strategy --- EXC 2121 ``Quantum Universe'' --- 390833306.

\bibliographystyle{JHEP}
\bibliography{refs}

\end{document}